\documentclass[aps, superscriptaddress, preprintnumbers,reprint]{revtex4-1}

\usepackage{amsmath,amssymb}
\usepackage{graphicx}

\usepackage{cases,bm}
\usepackage{xcolor}

\usepackage{framed}
\usepackage{enumitem}

\begin{document}                  

\title{ Partially Coherent Ptychography by Gradient Decomposition of the Probe}

\ifx \aff \undefined

\author{Huibin Chang}{}
\affiliation{School of Math. Sci., Tianjin Normal University}
\affiliation{Computational Research Division, Lawrence Berkeley National Laboratory}
\author{Pablo Enfedaque}{}
\affiliation{Computational Research Division, Lawrence Berkeley National Laboratory}
\author{Yifei Lou}{}{}
\affiliation{Dept. of Math,  University of Texas, Dallas}
\author{Stefano Marchesini}
\email{smarchesini@lbl.gov}
\affiliation{Computational Research Division, Lawrence Berkeley National Laboratory}


\else
\author[a,b]{Huibin Chang}{}
\author[b]{Pablo Enfedaque}{}
\author[c]{Yifei Lou}{}{}
\cauthor[b]{Stefano}{ Marchesini}{smarchesini@lbl.gov}

\aff[a]{School of Math. Sci., Tianjin Normal University}
\aff[b]{Computational Research Division, Lawrence Berkeley National Laboratory }
\aff[c]{Dept. of Math,  University of Texas, Dallas}
\fi

\date{\today}
\begin{abstract}
  Coherent ptychographic imaging experiments often discard over
  99.9 \% of the flux from a light source  to define the
  coherence of an illumination.  Even when coherent flux is
  sufficient, the stability required during an exposure is another
  important limiting factor. Partial coherence analysis can
  considerably reduce these limitations.  A partially coherent
  illumination can often be written as the superposition of a single
  coherent illumination convolved with a separable translational
  kernel. In this paper we propose the Gradient Decomposition of the
  Probe (\emph{GDP}), a model that exploits translational kernel
  separability, coupling the variances of the kernel with the
  transverse coherence.  We describe an efficient
  first-order splitting algorithm \emph{GDP-ADMM} to solve the
  proposed nonlinear optimization problem.  Numerical experiments
  demonstrate the effectiveness of the proposed method with Gaussian
  and binary kernel functions in {\textit{fly-scan}} measurements.
  Remarkably, \emph{GDP-ADMM} produces satisfactory results even when
  the ratio between kernel width and beam size is more than one, or
  when the distance between successive acquisitions is twice as large as
  the beam width.
\end{abstract}

\maketitle

\section{Introduction}

\begin{figure}
\begin{center}
\includegraphics[width=.95\columnwidth]{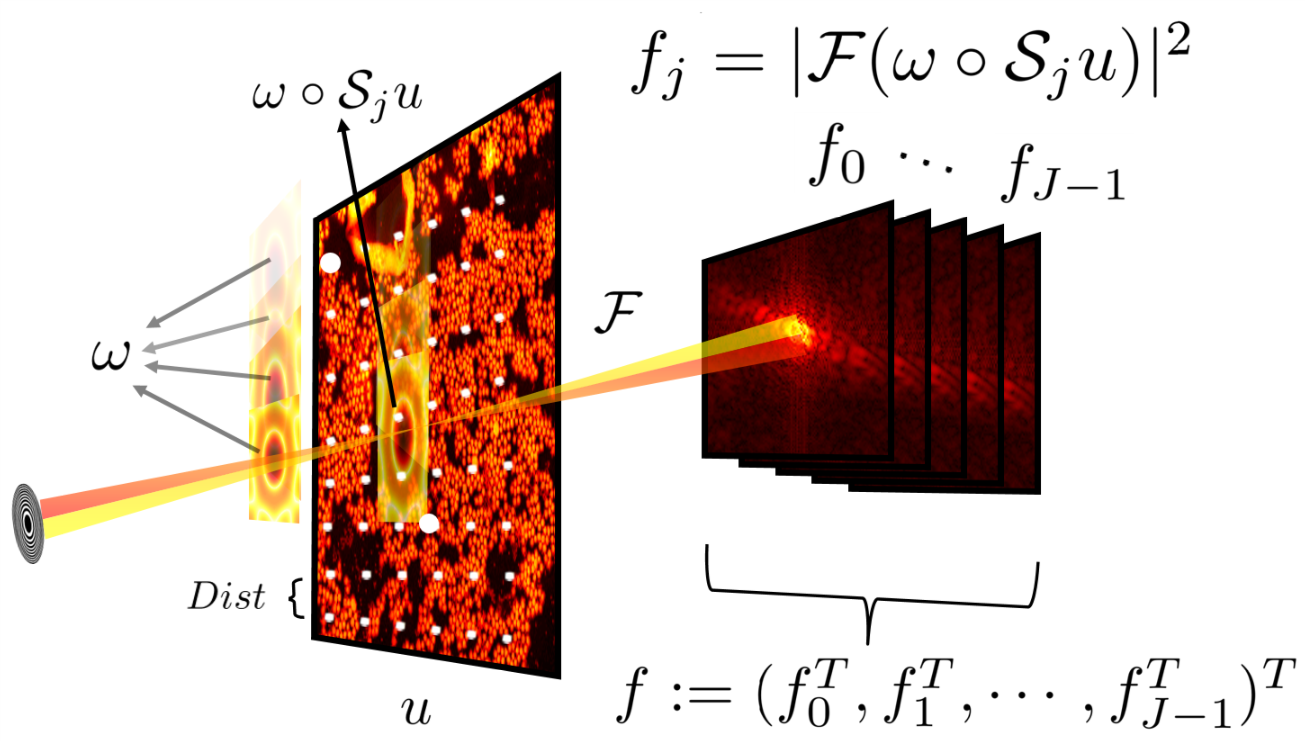}
\end{center}
\caption{Ptychographic  experiment (Far-field): A stack of  intensities  $f_j=|\mathcal F(\omega\circ\mathcal S_j u)|^2$ are collected, with
 $\omega$ being the localized coherent probe and $u$ being the image of interest (or specimen).
 The white dots on the image represent the center of probe or scanning lattice points with $Dist$ denoting the sliding distance between centers of two successive frames.}
\label{fig1}
\vskip -.15in
\end{figure}

Ptychography is a popular imaging technique in scientific
fields as diverse as condensed matter physics \cite{shi2016soft}, cell
biology \cite{giewekemeyer2010quantitative}, materials science
\cite{shapiro2014chemical} and electronics \cite{holler2017high},
among others.  In a coherent ptychography experiment (Figure
\ref{fig1}), a localized coherent X-ray probe (or illumination)
$\omega$ scans through an specimen $u$, while the detector collects a
sequence $J$ of phaseless intensities $f$ in the far field.  The goal is
to obtain a high resolution reconstruction of the specimen $u$ from
the sequence of intensity measurements. In a discrete setting,
$u\in\mathbb C^n$ is a 2D image with $\sqrt{n}\times\sqrt{n}$ pixels,
$\omega\in\mathbb C^{\bar m}$ is a localized 2D probe with $\sqrt{\bar
  m}\times \sqrt{\bar m}$ pixels ($u$ and $\omega$ are both written as
a vector by a lexicographical order), and $f_j=|\mathcal F(\omega\circ
\mathcal S_j u)|^2 $ is a stack of phaseless measurements ~$f_j\in
\mathbb R_+^{\bar m}~\forall 0\leq j\leq J-1$.  Here $|\cdot|$
represents the element-wise absolute value of a
vector, 
$\circ$ denotes the element-wise multiplication, and $\mathcal F$
denotes the normalized 2 dimensional discrete Fourier transform. Each
$\mathcal S_j\in \mathbb R^{\bar m\times n}$ is a binary matrix that
crops a region $j$ of size $\bar m$ from the image $u$.

In practice, as the probe is almost never completely known, one has to
solve a blind ptychographic phase retrieval problem or probe
retrieval \cite{thibault2009probe}, as follows:

\begin{equation}\label{PtychoPR}
{\text{ To find ~}\omega \in \mathbb C^{\bar m}\text{~and~} u\in \mathbb C^n,~~} s.t.~~|\mathcal A(\omega,u)|^2= f,
\end{equation}
where  bilinear operators  $\mathcal A:\mathbb C^{\bar m}\times \mathbb C^{n}\rightarrow \mathbb C^{m}$ and
 $\mathcal A_j:\mathbb C^{\bar m}\times \mathbb C^{n}\rightarrow \mathbb C^{\bar m}~\forall 0\leq j\leq J-1$,  are denoted as follows:
  \[
\begin{split}
 \mathcal A(\omega,u):=&(\mathcal A_0^T (\omega,u), \mathcal A_1^T(\omega,u),\cdots, \mathcal A_{J-1}^T(\omega,u))^T,\\
 \mathcal A_j(\omega,u):=&\mathcal F(\omega\circ \mathcal S_j u),
 \end{split}
 \]
 and $f:=(f^T_0, f^T_1, \cdots, f^T_{J-1})^T\in \mathbb R^m_+.$

Coherent ptychographic imaging experiments often rely on apertures to
define a coherent illumination. Research institutions around the world
are investing considerable resources to produce brighter x-ray sources
in order to overcome this limitation. Meanwhile, most of the x-ray photons
generated are currently discarded by secondary apertures. Even when
there is enough coherent flux, the stability required during an
exposure is often another limiting factor. Both flux and stability
limitations can be reduced using partial coherence
analysis \cite{pelz2014fly,deng2015continuous,huang2015fly}.

First, we briefly review the existing algorithmic work for partial
coherence.  In \cite{Fienup:93} the authors applied a gradient descent
phase retrieval algorithm from incoherently averaged illuminations to
compute the aberration of the Hubble space telescope.  In
\cite{clark2011simultaneous}, the authors considered a constant
quasi-monochromatic illumination on the sample (beam much larger than
the sample), and derived a convolution based model using the mutual
optical intensity as:
\begin{equation}\label{modelPC}
f_{pc}=f\ast \kappa,
\end{equation}
where $f_{pc}$ is the measured partial coherent intensity, $f$ is the
coherent intensity, $\ast$ denotes the convolution operator, and
$\kappa$ is the kernel function (Fourier transform of the complex
coherence function).  Physically, the convolution kernel represents
the combination of the detector response function and the angular
spread of the illumination.  {It was solved by the alternating
  projection algorithm in \cite{clark2011simultaneous}, when $\kappa$
  is a Gaussian kernel function with free horizontal and vertical
  coherence length parameters.  Burdet et al.
  \cite{burdet2015evaluation} applied the above model to 
  ptychographic imaging and proposed the Douglas-Rachford algorithm
  with a known Gaussian kernel function.}

A more general description of a partially coherent wave field
illuminating a specimen in a ptychography experiment was considered in
\cite{thibault2013reconstructing}, where the authors employed an
orthogonal decomposition of the mutual coherence function
\cite{wolf1982new} to describe partially coherent measurement as
follows:
\begin{equation}\label{modelMD}
f_{pc, j}=\sum\nolimits_{l=0}^{L-1} |\mathcal A_j(\omega_l,u)|^2, ~\forall 0\leq j\leq J-1,
\end{equation}
with $L$ orthogonal probes $\{\omega_l\}$, where
$f_{pc}:=(f^T_{pc,0},\cdots,$ $f^T_{pc,J-1})^T$ is the measured
intensity.  The extended ptychographic iterative engine
\cite{maiden2009improved} was adopted to solve such model
\cite{cao2016Modal}.  Experimental `{\textit{fly-scan}}' data with
translational blur was successfully reconstructed using
\eqref{modelMD} by
\cite{pelz2014fly,deng2015continuous,huang2015fly}.  However, it is
important to note that such blur is a special case of a model that has
many more degrees of freedom.  Moreover, the physical interpretation
of the multiple modes is unclear, and relationship with the coherence
function is indirect.  

  Other existing studies focus mostly on
\eqref{modelPC}, which only addressed blurring coherent intensities
at the detector. In practice, the blur is dominated by the source
dimensions and by the translation of the probe with respect to the
image during an exposure \cite{pelz2014fly,deng2015continuous}.  To
the best of our knowledge, there is no algorithm in the literature to
jointly recover the unknown image, probe, and coherence kernel
function that exploits such property.

In this paper, we propose {\emph{G}}radient {\emph{D}}ecomposition of
the {\emph{P}}robe (\emph{GDP}), a new forward model to characterize the
partially coherent ptychography problem.  The new model is based on
coupling the experimental coherence widths with the variances of the
kernel functions using the second order Taylor expansion to translate
the probe. In the second part of this work we present
\emph{GDP-ADMM}: a novel fast iterative solver that jointly optimizes
the image, the probe and the variance of the kernel function.
 The main benefits of the algorithm are listed below: 
\setlist{nolistsep}
\begin{enumerate}[noitemsep]
\item The approximation accuracy for a general partial
  coherent source is increased, while providing its coherent properties.
\item Subproblems can be solved using low computational and
  memory resources, and usually have close form.
\item It is insensitive to the coherence kernel and scanning step sizes, and 
achieves high SNR even when the data is contaminated by Poisson noise.
\end{enumerate}
Numerical experiments show that satisfactory results with
\emph{GDP-ADMM} can be obtained even when the ratio between kernel 
and beam widths is more than one, or when the distance between
successive acquisitions is almost twice as large as the beam width
(Full Width Half Maximum-FWHM).

\section{Approximate forward model for partial coherence}\label{sec2}

Partially coherent illumination from standard microscopes can be
written as the superposition of a single quasi-monochromatic coherent
illumination convolved with separable angular and translational
kernels \cite{GORI1978185,kim1986brightness,coissonwalker86,coisson1997gauss}.
Translational convolution of
the illumination is equivalent to translational motion during a scan
while exposing the detector \cite{deng2015continuous}.  In other
words, an extended incoherent source upstream of the lens can be
viewed as the superposition, photon-by-photon, of a rapidly moving
source demagnified by the lens onto the image plane. The demagnified
source defines the degree of coherence of the probe, or the blurring
kernel. { Coherence and vibrations kernels can be combined into
  one, such that partially coherent ptychography imaging with
  coherence kernel function $\kappa$ in a continuous setting (same
  notations as in the discrete setting) is formulated} as:
\begin{equation}\label{PCF}
\int |\mathcal F_{x\rightarrow q}\left(\mathcal S_j u\left (x \right ) \omega\left (x-y\right )\right)|^2 \kappa(y)\mathrm{d}y=f_{pc,j}(q),
\end{equation}
where $f_{pc}$ is the measured partial coherent intensity and  $\mathcal F_{x\rightarrow q}$ is the normalized Fourier transform. {
When  setting $\kappa$ to a binary function,  the above model is exactly the same as {\textit{fly-scan}} ptychography \cite{pelz2014fly,deng2015continuous,huang2015fly}};
Setting $\kappa$ to the Dirac delta function reduces it to the  coherent model \eqref{PtychoPR}.
We remark that \eqref{PCF} is quite different from \eqref{modelPC}, since  \eqref{PCF} illustrates the effects of blurring of images with respect to the probe,  while \eqref{modelPC} can be interpreted as blurring or binning multiple pixels at the detector.

Generally speaking, solving Eq. \eqref{PCF}  is a non-linear ill-posed problem with an unknown kernel and there is no fast method to { even compute  the integral on} the right hand side with known kernel, probe and images.
{In \cite{marchesini2013augmented,tian2014multiplexed},  the authors considered  $L $ probes translated with weights $\{w_l\}$:
\begin{equation}\label{modelMX}
f_{pc,j}=\sum\limits_{l=0}^{L-1} w_l | {\mathcal A_{j}} (\mathcal T_l\omega,u)|^2, ~\forall 0\leq j\leq J-1,
\end{equation}
with the translation operators $\{\mathcal T_l\},$ and partially
coherent intensity $f_{pc}.$ However, those methods in
\cite{marchesini2013augmented,tian2014multiplexed} cannot be directly
applied to \eqref{PCF}, since either the weights
\cite{tian2014multiplexed} or the probe \cite{marchesini2013augmented}
are assumed to be known in advance. Moreover, as coherence degrades,
or the number of probes increases, their computation and memory costs
would increase dramatically.  Hence instead of solving \eqref{PCF}
directly as in \cite{marchesini2013augmented}, { in the following
  sections we will reformulate the above model \eqref{PCF}, and solve
  the related nonlinear optimization problem.}  }

\subsection{Gradient Decomposition of the Probe (\emph{GDP})}
Following \eqref{PCF}  and using the Taylor expansion of $\omega$, one has:
\[
\begin{split}
f_{pc,j}(q)=&\int \left |\mathcal F_{x\rightarrow q} \left (\mathcal S_j u(x) \left ( \omega(x)-y^T\nabla \omega(x)\right . \right . \right .\\ &\left . \left . \qquad\qquad +\tfrac{1}{2}y^T\nabla^2\omega(x)y +\mathcal O(|y|^3)) \right ) \right |^2 \kappa(y)\mathrm{d}y,
\end{split}
\]
where we assume that the third order derivatives are uniformly bounded, e.g.
 \begin{equation}\label{eq:bounded_derivative}
 \max_{\{\alpha\in\mathbb N^2:~|\alpha|=3\}}\max_{x\in\mathbb C^2}|\partial^{\alpha} \omega(x)|\leq C,
 \end{equation}
  with a positive constant $C.$ 
{ It is easy to verify that this condition is satisfied
when the illumination is generated by a small lens (with small aperture).}

Consider a kernel function characterized by its moment expansion $m_{i_1,i_1}=\int \kappa(y) y_{1}^{i_1} y_{2}^{i_2}\mathrm{d}y$,  normalized with $m_{0,0}=1$,  center of mass $(m_{1,0},m_{0,1})=(0,0)$, and  second order moments $\sigma_{1}^2$, $\sigma_{2}^2$, and
$\sigma_{12}$ ($\sigma_1^2 \sigma_2^2-\sigma_{12}^2\geq 0$). We  further assume that:
\begin{equation}\label{thirdMZ}
m_{i_1,i_2}=0, \text{if~} \mathrm{mod}(i_1+i_2,2)={ 1}.
\end{equation}
The above relation holds if the kernel function is centro-symmetric with respect to  the origin, i.e. $\kappa(-y_1,-y_2)=\kappa(y_1,y_2)$. For higher order moments, we also assume that
\begin{equation}\label{assumpMoment}
\int |y|^{k}\kappa(y)=\mathcal O(\max\{\sigma_1,\sigma_2\}^{k}), \, k\geq 3.
\end{equation}
Therefore, one has:
\begin{equation}
\small{
\begin{split}
f_{pc,j}(q)=&|\mathcal A_j(\omega+\tfrac{1}{2}(\sigma_1^2\nabla_{11}\omega+\sigma_2^2\nabla_{22}\omega+2\sigma_{12}\nabla_{12}\omega),u)|^2\\
&+\sigma_1^2|\mathcal A_j(\nabla_1 \omega, u)|^2
+\sigma_2^2|\mathcal A_j(\nabla_2 \omega, u)|^2\\
&+\mathcal O(\int |y|^3\kappa(y)\mathrm{d}y).
\end{split}
}\label{eqM-1}
\end{equation}
More details of the above derivation can be found  in Appendix \ref{apdx-1}.
 In order to further simplify the partial coherence approximation we introduce a new variable  $\tilde \omega$ (variance adjusted probe):
 \[
 \tilde \omega:=\omega+\tfrac{1}{2}(\sigma_1^2\nabla_{11}\omega+\sigma_2^2\nabla_{22}\omega+{2\sigma_{12}\nabla_{12}\omega) },
\]
and  nonlinear operator $\mathcal{G}_j:\mathbb C^{\bar m}\times \mathbb C^{n} \times \mathbb R^{2} \rightarrow \mathbb R^{\bar m}_{+}$:
\begin{equation}\label{eq:G}
\begin{split}
\mathcal{G}_{j} (\omega,u,\bm \sigma)&:=
|\mathcal A_j(\omega,u)|^2\\&
+ \sigma_1^2|\mathcal A_j(\nabla_1 \omega, u)|^2+
\sigma_2^2|\mathcal A_j(\nabla_2 \omega, u)|^2,
\end{split}
\end{equation}
where $\bm \sigma:=(\sigma_1,\sigma_2)$. 
Neglecting high order terms $\mathcal O( \| \bm \sigma\|^k),$ $k \geq 3$, following \eqref{eqM-1}, 
 we obtain the approximate forward model:
\begin{equation}\label{PCM-1st}
\small{\begin{split}
\mathrm{GDP:~~}f_{pc,j}\approx &\mathcal{G}_j(\tilde \omega, u,\bm \sigma).
\end{split}
}
\end{equation} 
{\it
\paragraph{Remarks:}
\begin{itemize}
\item[$\bullet$] The above assumptions hold if $\kappa$ is a Gaussian
  kernel function, which well approximates standard light sources such as
  synchrotrons \cite{kim1986brightness,coissonwalker86,coisson1997gauss},
  SASE FELs \cite{saldin2008coherence}, and others. 
   When $\bm \sigma=(0,0),$ the above formula
  \eqref{PCM-1st} reduces to the coherent model in \eqref{PtychoPR}.
\item The variance adjusted probe $\tilde \omega$ incorporates second order effects (such as the covariance $\sigma_{12}$) into the illumination, while reducing the cost of computing the intensities using the forward model \eqref{PCM-1st} compared to  \eqref{modelMX}.  
\item[$\bullet$]  If the probe $\omega$ has a support by a lens of finite size,   $(\mathcal F\omega)(q)=0~\forall~q\in \Omega_0\subset\mathbb R^2,$
then  the same support applies to the variance adjusted probe 
$\forall~q\in\Omega_0,$ $(\mathcal F\tilde\omega)(q)=0$
 due to the Fourier transform relationship:
\[
\begin{split}
(\mathcal F\tilde\omega)(q)&=
 \mathcal F(( 1+\tfrac{1}{2} \sigma_1^2\nabla_{11}+\tfrac{1}{2}\sigma_2^2\nabla_{22}+{\sigma_{12}\nabla_{12})\omega)(q) }\\
&= \left (1-\tfrac{1}{2}\sigma_1^2q_1^2-\tfrac12\sigma_2^2q_2^2-\sigma_{12}q_1q_2 \right )(\mathcal F\omega)(q), \\
&\forall~q=(q_1,q_2)\in \mathbb R^2.
\end{split}
\]

\item[$\bullet$] Similarly to the decomposition model \eqref{modelMD},  \emph{GDP}   has three different modes $\{\tilde\omega, \nabla_1\tilde\omega, \nabla_2\tilde\omega\}$. The difference between them is that in the \emph{GDP}  model,  two modes $\nabla_l \omega$ can be expressed by the first mode $\tilde \omega$ explicitly, while the multiple modes in the decomposition model  \eqref{modelMD} are only assumed to be orthogonal to each other  \cite{wolf1982new,thibault2013reconstructing}. Such orthogonal constraint  is essentially  nonconvex, which is more difficult to handle and  leads to more local minima as well.
\end{itemize}
}

\section{Fast iterative algorithm: \emph{GDP-ADMM} }\label{sec3}

The amplitude based nonlinear optimization model can be established as:
\begin{equation}\label{modelPCM}
\!\!\!\min\limits_{\tilde\omega,u,\bm \sigma} \tfrac12
\sum\limits_{j=0}^{J-1}\big\|\sqrt{f_{pc,j}}-
\sqrt{\mathcal{G}_j\left (\tilde\omega,u,\bm \sigma \right)}
\big\|^2,
\end{equation}
 where $\|\cdot\|$ denotes the standard $L^2$ norm in Euclidean space.
We remark that by introducing the new variable $\tilde \omega$, it is much easier to solve the subproblems of the nonlinear optimization model \eqref{modelPCM} with respect to $\bm \sigma$  than using the original formula \eqref{eqM-1}.

The \emph{GDP} based nonlinear optimization model \eqref{modelPCM} is
nonconvex and non-differentiable. We are interested in designing a
fast first order algorithm, whose subproblems can be easily
implemented.  The Alternating Direction Method of Multipliers (ADMM)
\cite{Glowinski1989} has been successfully applied to large scale
nonlinear and non-differential optimization problems arising from
machine learning or computer vision, among other areas.  The
connection between ADMM, the Douglas-Rachford algorithm, and the
popular Hybrid Input Output \cite{Fienup1982} for classical phase
retrieval was discussed in \cite{Bauschke2003,Wen2012}.  By
introducing auxiliary variables $z_{l,j}\in \mathbb C^{\bar m}$, one
has:
\begin{equation}
\label{ModelS-PC}
\begin{split}
\min\limits_{\tilde\omega, u,\bm \sigma, z_{l,j}}&\tfrac12\sum\limits_{j=0}^{J}\left\|\sqrt{f_{pc,j}}- \sqrt{\sum\limits_{l=0}^2| z_{l,j}|^2}\right\|^2,\\
s.t.\qquad&z_{0,j}=\mathcal A_j(\tilde\omega,u), z_{1,j}=\sigma_1\mathcal A_j(\nabla_1\tilde\omega,u),\\& \,  z_{2,j}=\sigma_2\mathcal A_j(\nabla_2\tilde\omega,u),~0\leq j\leq J-1.
\end{split}
\end{equation}
The corresponding augmented Lagrangian reads as:
\begin{equation}
\label{AL-S-PC}
\begin{split}
&\mathcal L_{r}(\tilde\omega,u,\{z_{l}\}_{l=0}^2,\bm {\sigma},\{\Lambda_{l}\}_{l=0}^2):=\tfrac12\sum\limits_j\left\|\sqrt{f_{pc,j}}- \tiny{\sqrt{\sum\limits_l | z_{l,j}|^2}}\right\|^2\\
&+r\Re(\langle z_{0}-\mathcal A(\tilde\omega,u), \Lambda_{0}\rangle)+\tfrac{r}{2}\|z_{0}-\mathcal A(\tilde\omega,u)\|^2\\
&+r\Re(\langle z_{1}-\mathcal A(\sigma_1\nabla_1\tilde\omega,u), \Lambda_{1}\rangle)+\tfrac{r}{2}\|z_{1}-\mathcal A(\sigma_1\nabla_1\tilde\omega,u)\|^2\\
&+r\Re(\langle z_{2}-\mathcal A(\sigma_2\nabla_2\tilde\omega,u), \Lambda_{2}\rangle)+\tfrac{r}{2}\|z_{2}-\mathcal A(\sigma_2\nabla_2\tilde\omega,u)\|^2,
\end{split}
\end{equation}
with multipliers
$\Lambda_l=(\Lambda^T_{l,0},\cdots,\Lambda^T_{l,J-1})^T\in{\mathbb
  C}^m$, $z_l=(z_{l,0}^T,\cdots,z_{l,J-1}^T)^T\in \mathbb C^m$, and
the positive parameter $r,$ where $\langle\cdot,\cdot\rangle$ denotes
the standard inner product in complex Euclidean space, and
$\Re(\cdot)$ denotes the real part of a complex number.  Briefly, the
\emph{GDP-ADMM} algorithm alternates minimization of the above
augmented Lagrangian with respect to the variables $\omega, u,
\sigma_1, \sigma_2$, updates the multipliers $\Lambda_l, l=0,1,2$, and
repeats.  In our previous work, ADMM was applied to coherent
ptychographic imaging in \cite{Wen2012,chang2017blind}.  Here we
propose a new, more robust variant of ADMM. It employs an additional
proximal term added to the augmented Lagrangian:
$\tfrac{\tau}{2}\|u-u^{k}\|^2_{M^{k}}$ to avoid division by zeros when
minimizing with respect to $u$.  The detailed description of
\emph{GDP-ADMM} can be found in Algorithm 1 (further details in Appendix
\ref{apdx-2}). 
The numerical results can be found in the following
section. For simplicity, the gradient operator $\nabla_1, \nabla_2$ in the numerical section 
is considered in a discrete setting, using the forward and  finite difference operator with respect to $x,y$ directions.

Theoretical analysis about the convergence properties of this blind
algorithm (to refine the probe, the image, and the coherence function
variances during iterations) is likely to be very challenging.
However, if the variances are known or fixed, convergence to the
stationary point of the nonlinear optimization model \eqref{modelPCM}
can be guaranteed by assuming that the iterative sequence is bounded
and the parameter $r$ is sufficiently large \cite{chang2017blind}.
\newline
\newline
\newline
\newpage

\begin{framed}
\begin{center}
{\bf{Algorithm 1}}: \emph{GDP-ADMM}

\noindent\hrulefill
\end{center}
\noindent
{\bf{Initialization: }} Set $\tilde\omega^0=\mathcal
F^*(\tfrac{1}{J}\sum \sqrt{f_{pc,j}}),$ $u^0=1, z_0^0=\mathcal
A(\tilde\omega^0, u^0),$ $\sigma^0_1=\sigma^0_2=1,$ $z^0_0=\mathcal
A(\tilde\omega^0, u^0),$ $z^0_1=\sigma^0_1\mathcal
A(\nabla_1\tilde\omega^0, u^0),$ $z^0_2=\sigma^0_2\mathcal
A(\nabla_2\tilde\omega^0, u^0),$ $\Lambda_l^0=0, k:=0,$ maximum
iteration number $Iter_{Max},$ parameters $r$ and $\tau$.
\newline
\newline
\noindent
{\bf {output:}}x
$u^\star:=u^{Iter_{Max}}$ and $\tilde\omega^\star:=\tilde\omega^{Iter_{Max}}.$\vskip .1in
\noindent
{\bf{for}}
{$k=0$ to $Iter_{Max}-1$}
\noindent
\begin{enumerate}
\item
Compute $\tilde\omega^{k+1}$ by solving the linear system:
\begin{equation}
\label{eqSubMask}
\begin{split}
&\tilde\omega^{k+1}=\left ({N_1^{k}}\right )^{-1}\times c^k,
\end{split}
\end{equation}
using conjugate gradient method,
where $N_1^k$, and $c^k$ are given in Eq. \eqref{eqK}  (Appendix \ref{apdx-2}).

\item
Compute $u^{k+1}$ as
\begin{equation}
{\label{eqSubU}
\begin{split}
u^{k+1}=\left ( N_2^{k} +\tfrac{\tau}{r} M^{k}\right )^{-1}
\times \left( b^k+\tfrac{\tau}{r} M^{k} u^{k}\right).
\end{split}
}
\end{equation}
with diagonal matrices $N_2^{k}$,  $M^{k}$ and vector $b^k$ defined in Eq. 
 \eqref{eqH}, and \eqref{eq:MK} (see Appendix \ref{apdx-2}). 
\item Compute $z_l^{k+1}$ as:
\begin{equation}
\label{eqSubZ}
z^{k+1}_{l,j}=\dfrac{\sqrt{f_{pc,j}}+r\sqrt{\sum_{d=0}^2 |\tilde z^{k+1}_{d,j}|^2 } }{1+r}\circ\mathrm{sign}(\tilde  z^{k+1}_{l,j}),
\end{equation}
with
\[
\begin{split}
\mathrm{sign}(z_{l,j}):=&\tfrac{z_{l,j}}{\sqrt{\sum_{d=0}^2 |z_{d,j}|^2 }},\\
\tilde z^{k+1}_{0,j}=&\mathcal A(\tilde\omega^{k+1},u^{k+1})-\Lambda^{k}_0,\\ 
\tilde z^{k+1}_{1,j}=&\sigma^{k}_1\mathcal A(\nabla_1\tilde\omega^{k+1},u^{k+1})-\Lambda^{k}_1,\\
\tilde z^{k+1}_{2,j}=&\sigma^{k}_2\mathcal A(\nabla_2\tilde\omega^{k+1},u^{k+1})-\Lambda^{k+1}_2.
\end{split}
\]

\item
 Compute $\sigma_l^{k+1}$ as:
\[
\begin{split}
&\sigma^{k+1}_l=\frac{\Re(\langle z^{k+1}_l+\Lambda^{k}_l,\mathcal A(\nabla_l \tilde\omega^{k+1},u^{k+1}) \rangle)}{\|\mathcal A(\nabla_l \tilde\omega^{k+1},u^{k+1})\|^2},~l=1,2.\\
\end{split}
\]
\item Update the multipliers $\{\Lambda_l^{k+1}\}$ as
\[
\begin{split}
&\Lambda_{0}^{k+1}=\Lambda_{0}^{k}+(z_{0}^{k+1}-\mathcal A(\tilde\omega^{k+1},u^{k+1}));\\
&\Lambda_{1}^{k+1}=\Lambda_{1}^{k}+(z_{1}^{k+1}-\mathcal A(\sigma^{k+1}_1\nabla_1\tilde\omega^{k+1},u^{k+1}));\\
&\Lambda_{2}^{k+1}=\Lambda_{2}^{k}+(z_{2}^{k+1}-\mathcal A(\sigma^{k+1}_2\nabla_2\tilde\omega^{k+1},u^{k+1})).
\end{split}
\]
\end{enumerate}
\end{framed}

\newpage
\section{Numerical experiments}\label{sec4}

\begin{figure}
\begin{center}
{{\includegraphics[width=\columnwidth,height=.45\columnwidth,trim={4.2cm 1.2cm 3cm 0cm},clip]{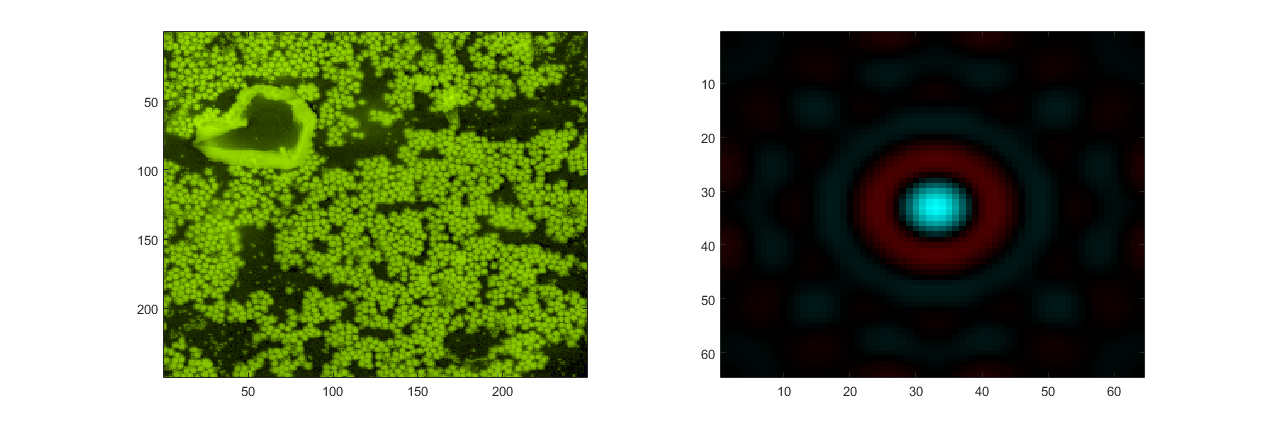}}}    
\end{center}
\caption{True image ``Goldballs'' \cite{marchesini2016sharp}  $249\times 249$ (left)  and true probe (right) $64\times 64$, with ``beam diameter'' $D=7$ pixels  (FWHM).
}
\label{fig2}
\end{figure}

The experimental setup of this section is introduced next. The
reference specimen used is a complex valued image
``Goldballs'' \cite{marchesini2016sharp}, of size $249 \times 249$ pixels
(Figure \ref{fig2}). The probe is an Airy disk of size $64\times
64$ pixels (Figure \ref{fig2}). 

%

We compare the proposed \emph{GDP-ADMM} (Appendix \ref{apdx-2}) with
the ``fully-coherent'' model by ADMM (\emph{FC-ADMM}
\cite{chang2017blind}).  Both algorithms are executed with a maximum
iteration number of 300.  
We measure the quality of recovery images by
relative residuals as:
 \[
{
 \begin{split}
 e_{fc}&=\tfrac{\left\|\sqrt{f}-|\mathcal A(\omega^{k},u^{k})|\right\|_1}{\big\|\sqrt{f}\big\|_1},\\
 \!\!e_{pc}&=
 \tfrac{\left\|\sqrt{f}-\sqrt{\mathcal{G}\left (\tilde \omega^{k},u^{k},\bm \sigma^{{k}} \right )}
 \right\|_1}{\left\|\sqrt{f}\right\|_1},\\
 \end{split}
}
\]
for \emph{FC-ADMM} and \emph{GDP-ADMM} respectively, where
$\omega^{k},\tilde\omega^{k}, u^{k}, \bm \sigma^{k}$ are the iterative
solutions, and $f$ is the measured intensity. The signal-to-noise
ratio (SNR) is also measured:
\[
\begin{split}
&\mathrm{SNR}(u^{k},u_{true})=-20\min_{c\in\mathbb C}      \log\frac{\|u^{k}- c~u_{true}\|}{\|u^{k}\|},\\
\end{split}
\]
where $u_{true}$  is the ground truth of the image.

We show the performance of \emph{GDP-ADMM} with the following Gaussian kernel
function: 
\begin{equation}\label{eq:gaussian}
\kappa(y):=\tfrac{1}{2\pi\sigma_1\sigma_2}\exp(-\tfrac{y_1^2}{2\sigma_1^2}-\tfrac{y_2^2}{2\sigma_2^2}).
\end{equation}
The discrete truncated Gaussian matrix is generated by
\textsl{fspecial} in \textsc{Matlab} with variances $\sigma_1,
\sigma_2$ and support sizes $(4\sigma_1+1,4\sigma_2+1)$. 
We remark that the kernel width is  $2\sqrt{2 \log (2)}\times(\sigma_1,\sigma_2)\approx 2.35 \times(\sigma_1,\sigma_2)$.

Simulated data is generated using raster grid scanning with sliding
distance $\mathrm{Dist}=8$ pixels (slightly bigger than the beam
width) as a default, and we incorporate the support constraint of the
lens as in \cite{marchesini2016sharp}. The parameter $r$ is selected
manually with default value 0.2, and $\tau=1.01 r$.

\paragraph{Noiseless data.}
Following \eqref{PCF}, the partially coherent intensity in a discrete
setting is generated as
\begin{equation}\label{intGen}
f_{pc,j}=\sum_{i} \kappa_i\big|\mathcal F((\mathcal T_{i} \omega)\circ\mathcal S_ju)\big|^2,
\end{equation}
with translation operator $\mathcal T_{i}$, discrete Gaussian weights
$\{\kappa_i\}$, and periodical boundary condition for the probe.  We
conduct the first numerical experiment to explore the performance of
the proposed algorithm with respect to different degrees of partial
coherence by varying the variances $\sigma_1, \sigma_2$, while keeping
the beam width constant (the smaller the variances, the more coherent
the data).

\begin{figure}
\begin{center}
\begin{tabular}{cccc}
\tiny{$\bm \sigma=(2,2)$}& 
\tiny{$(3,3)$}& 
\tiny{$(4,4)$}& 
\tiny{$(5,5)$}
\\
\includegraphics[width=.2\columnwidth,height=.2\columnwidth,trim={.6cm .4cm 0cm 0cm},clip]{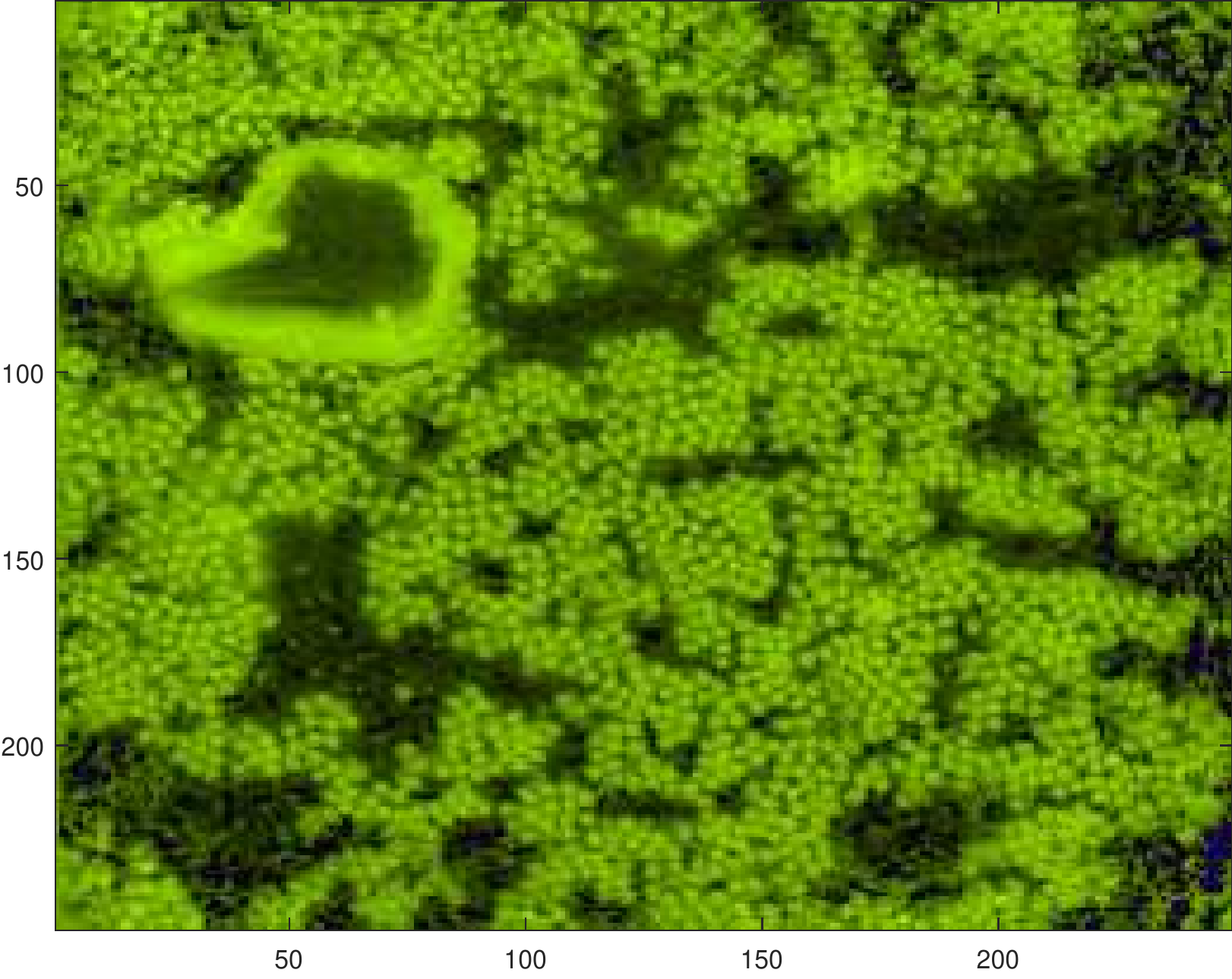}&
\includegraphics[width=.2\columnwidth,height=.2\columnwidth,trim={.6cm .4cm 0cm 0cm},clip]{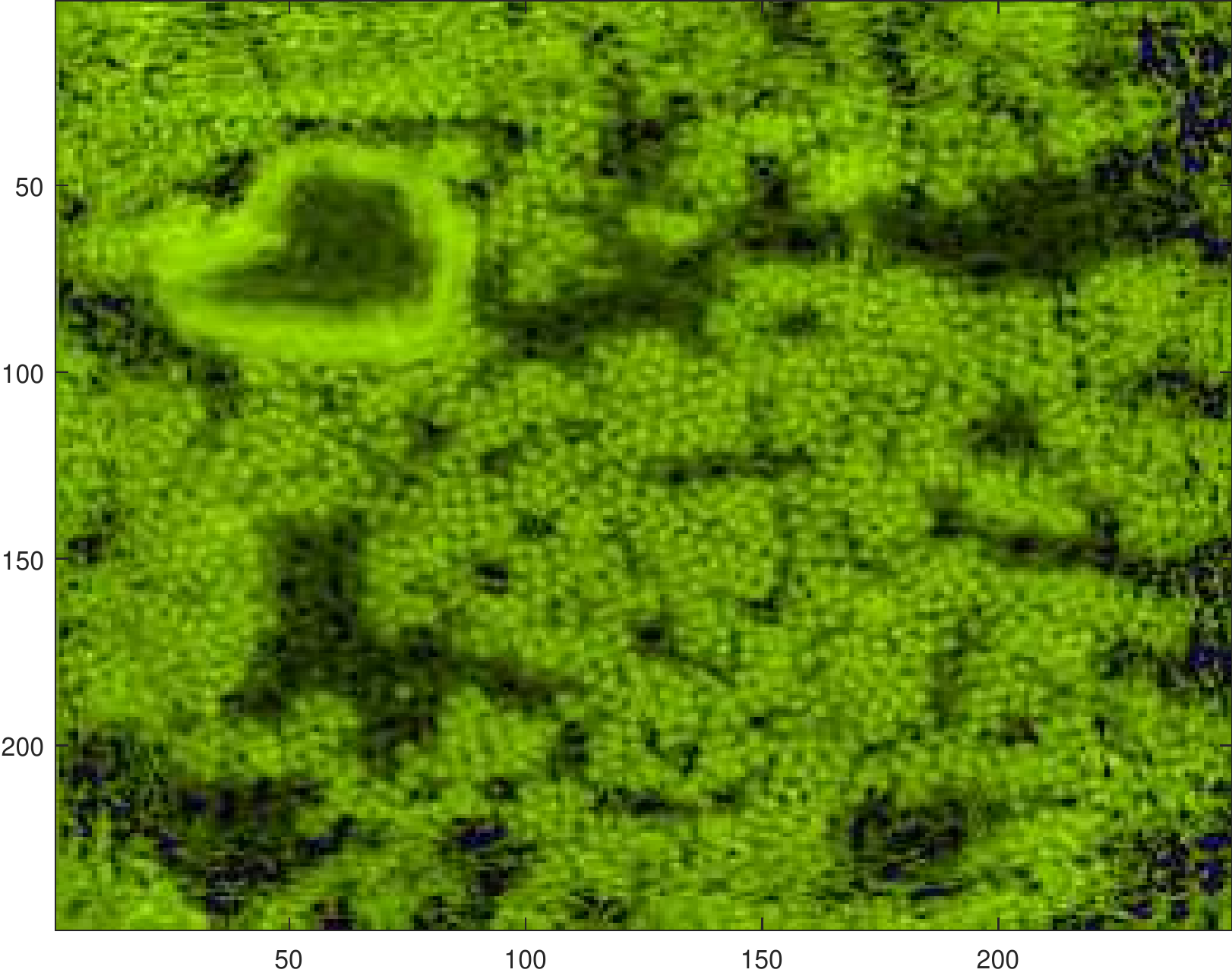}&
\includegraphics[width=.2\columnwidth,height=.2\columnwidth,trim={.6cm .4cm 0cm 0cm},clip]{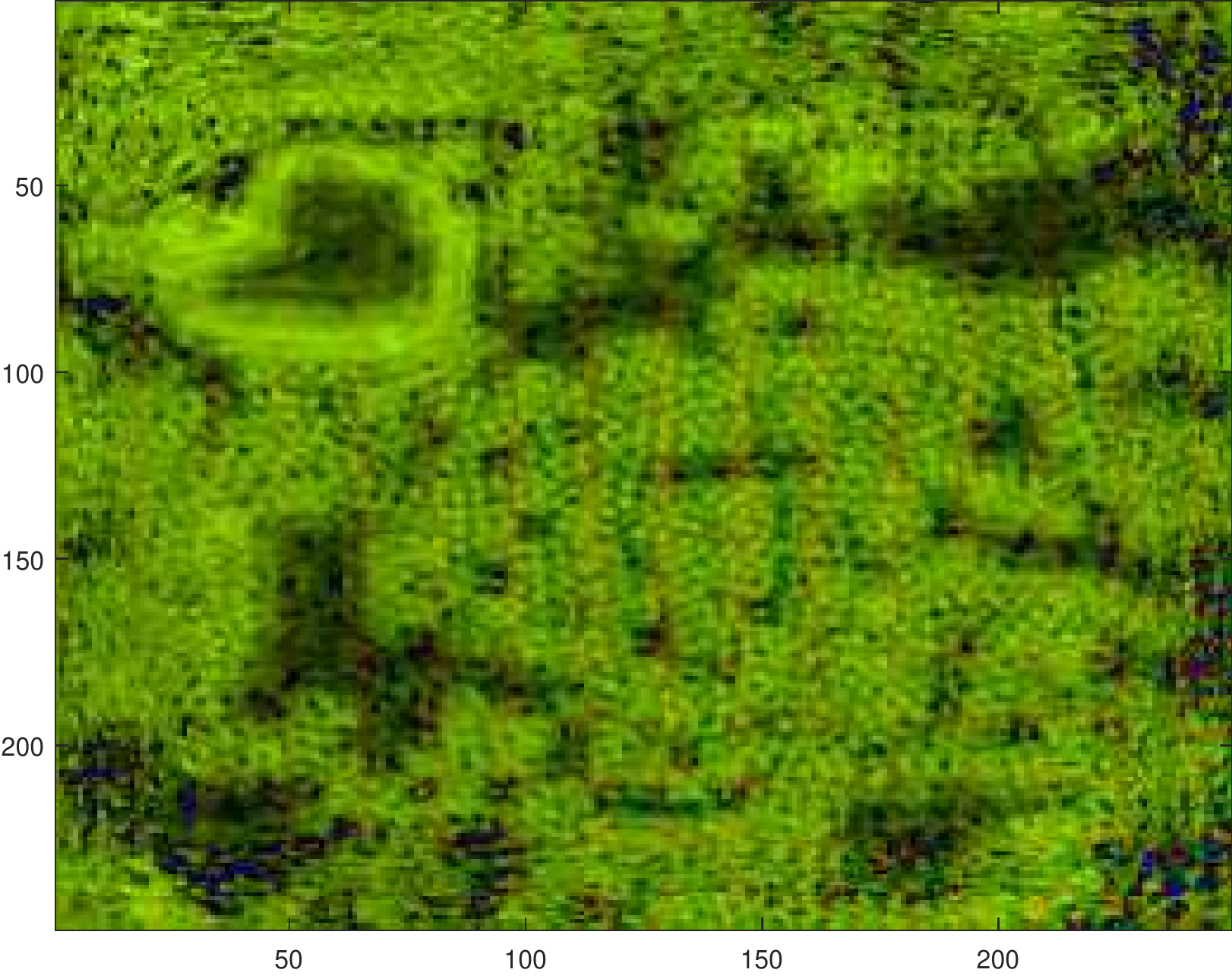}&
\includegraphics[width=.2\columnwidth,height=.2\columnwidth,trim={.6cm .4cm 0cm 0cm},clip]{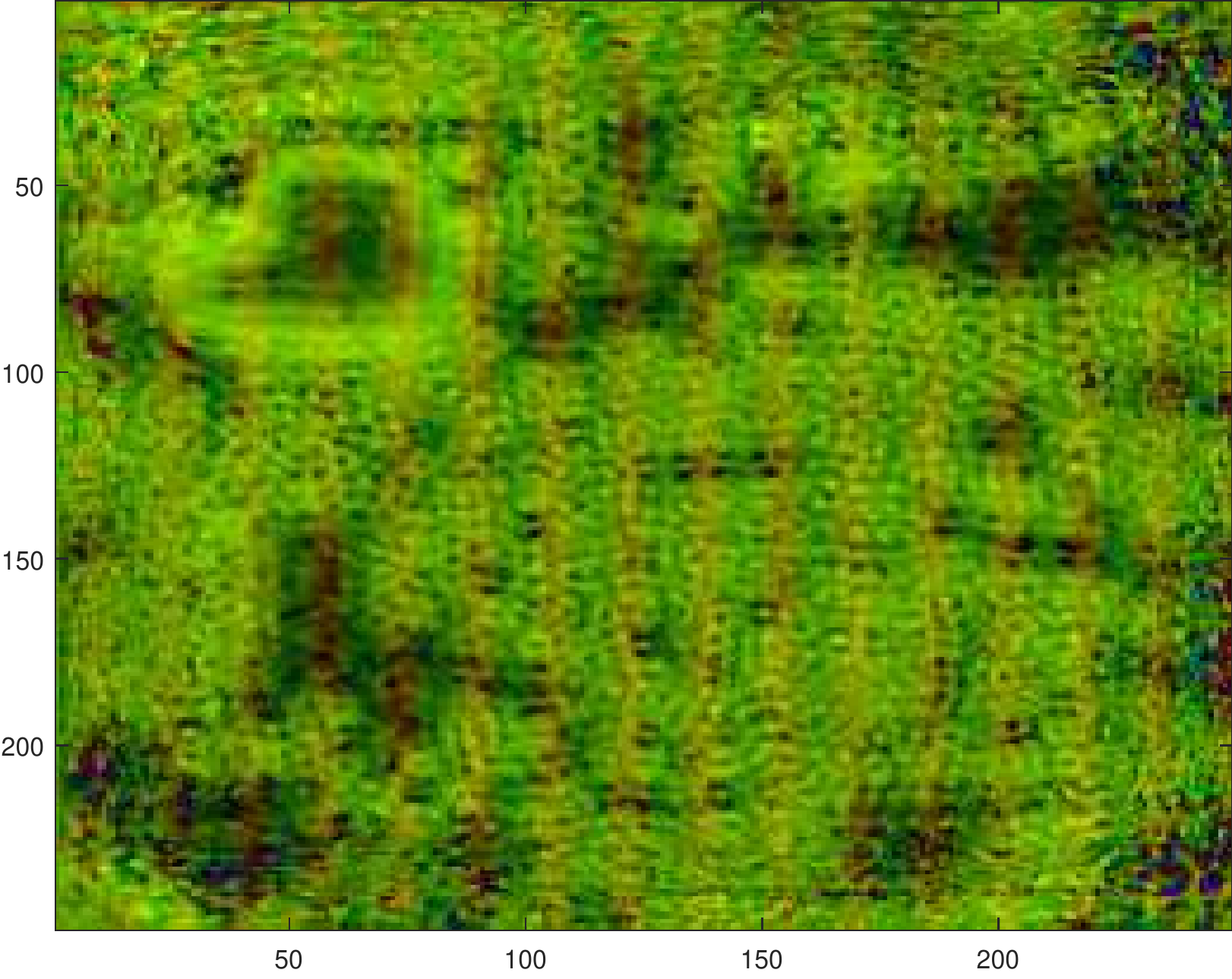}\\
\multicolumn{4}{c}{\tiny{\emph{FC-ADMM}}}\\
\includegraphics[width=.2\columnwidth,height=.2\columnwidth,trim={.6cm .4cm 0cm 0cm},clip]{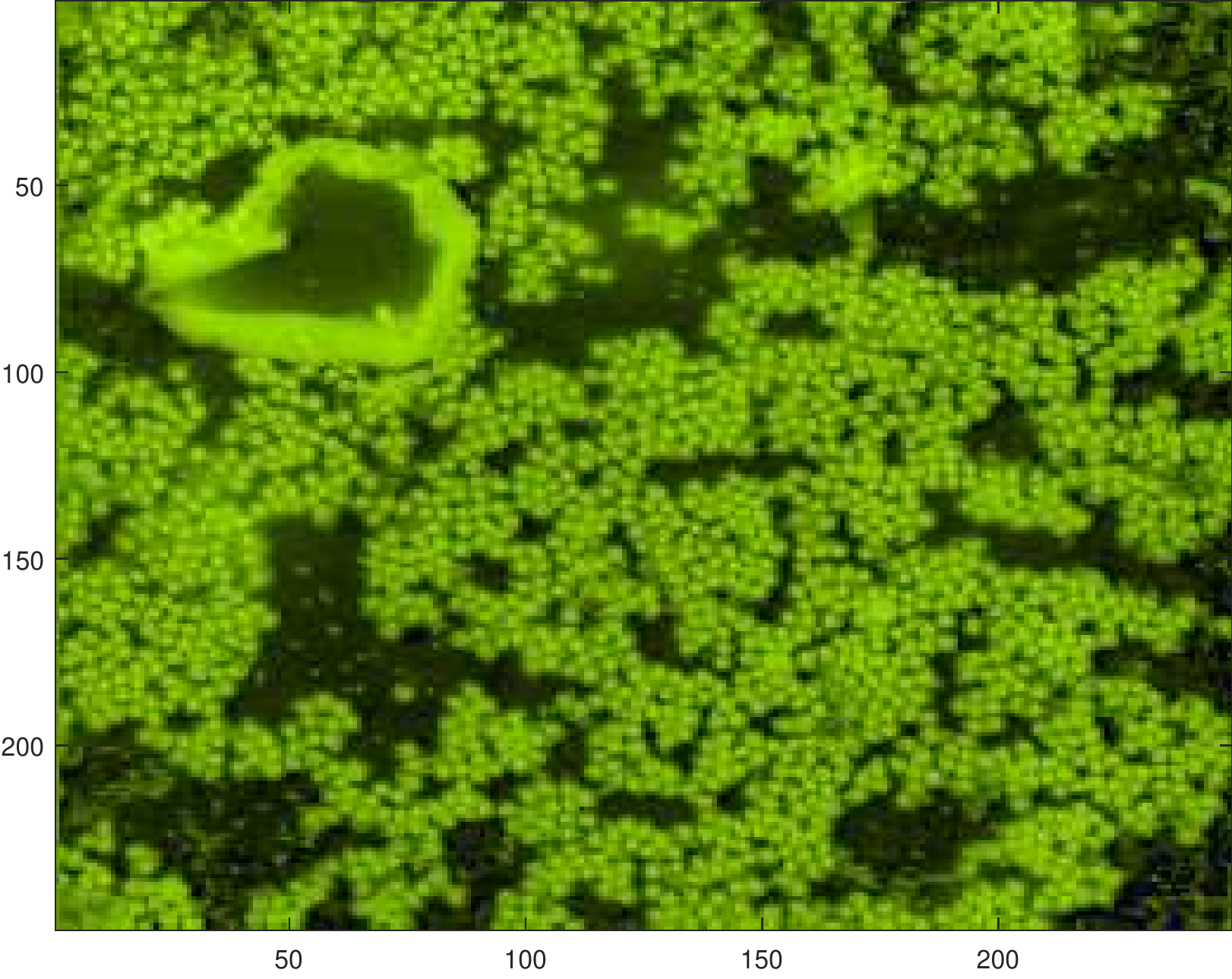}&
\includegraphics[width=.2\columnwidth,height=.2\columnwidth,trim={.6cm .4cm 0cm 0cm},clip]{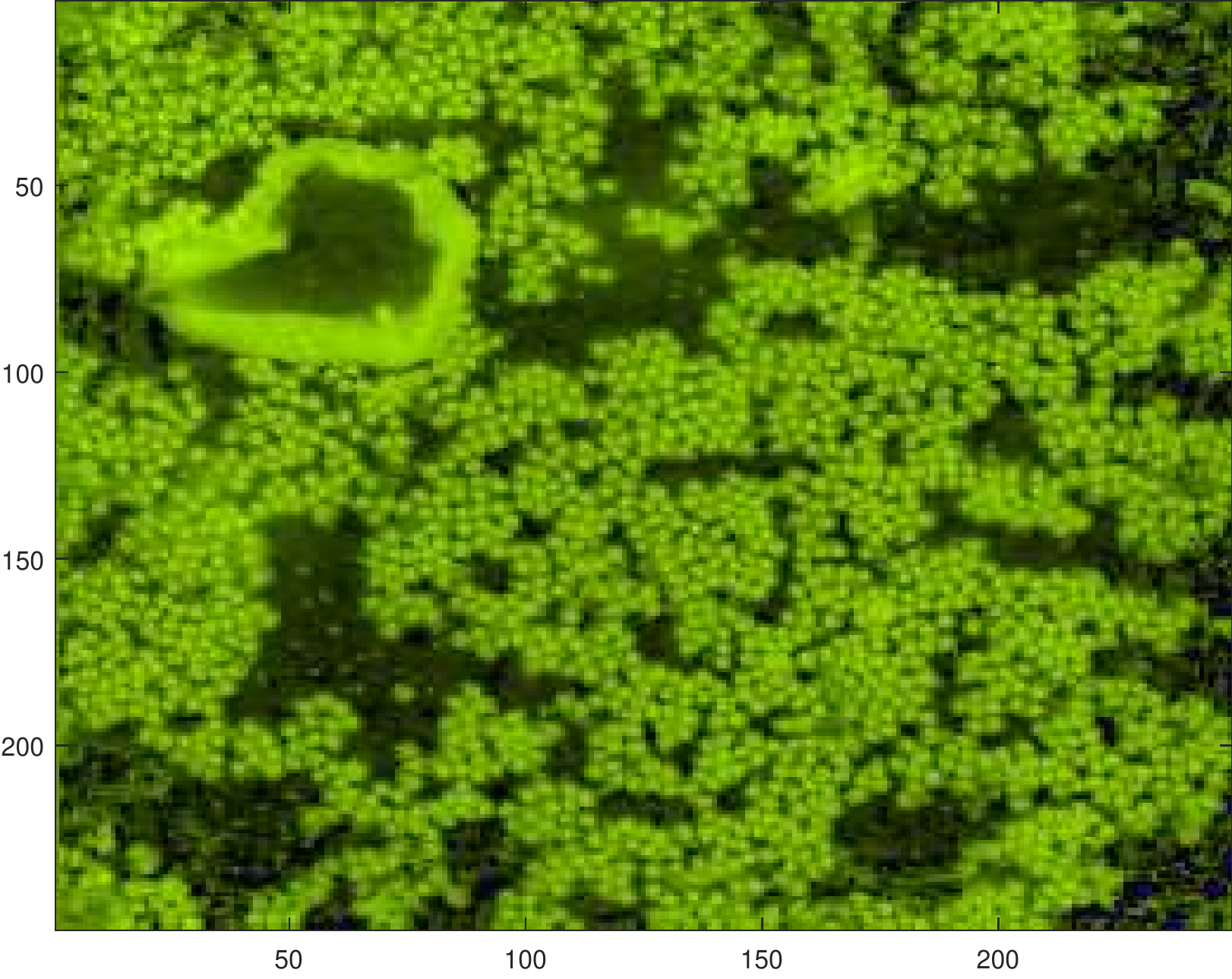}&
\includegraphics[width=.2\columnwidth,height=.2\columnwidth,trim={.6cm .4cm 0cm 0cm},clip]{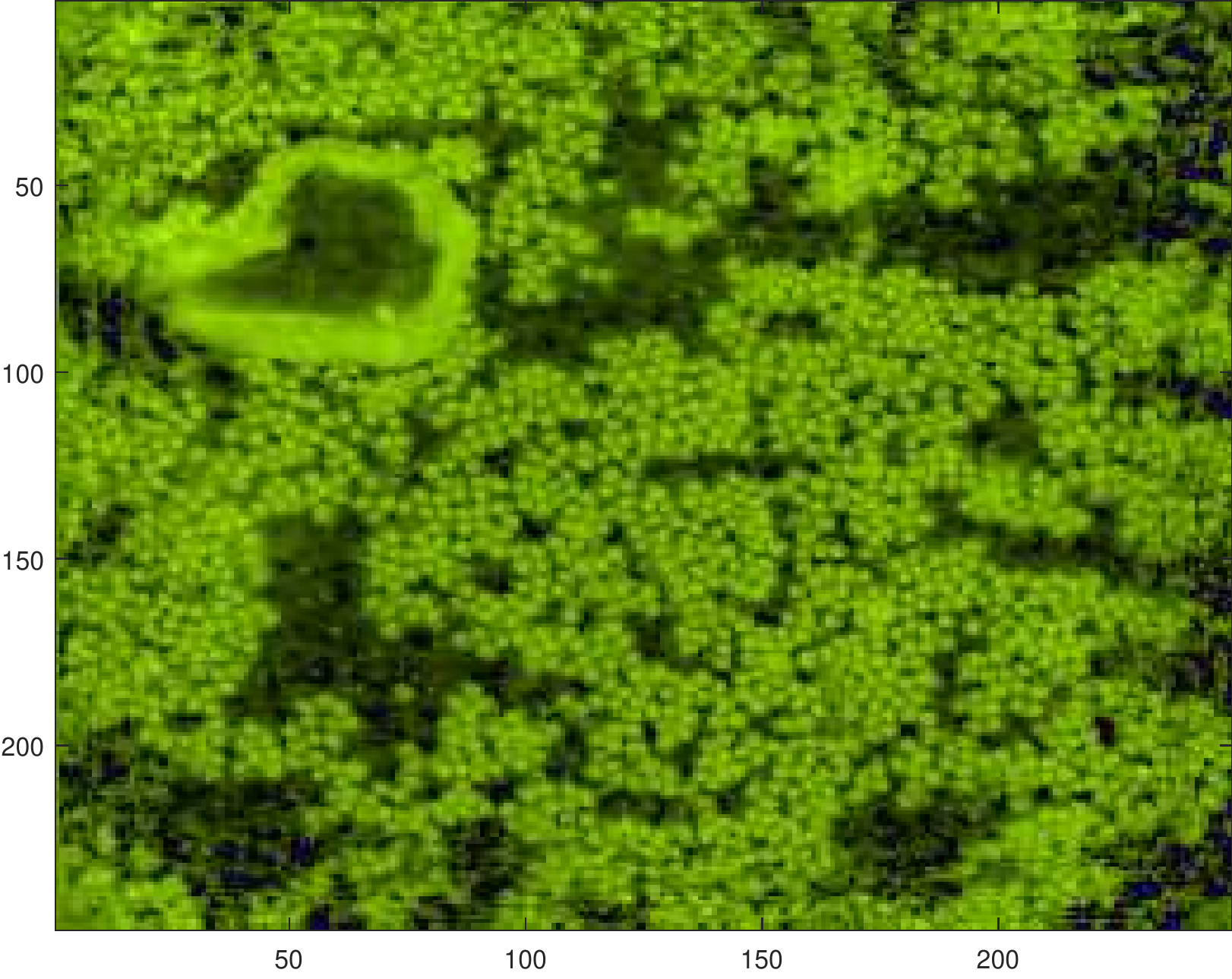}&
\includegraphics[width=.2\columnwidth,height=.2\columnwidth,trim={.6cm .4cm 0cm 0cm},clip]{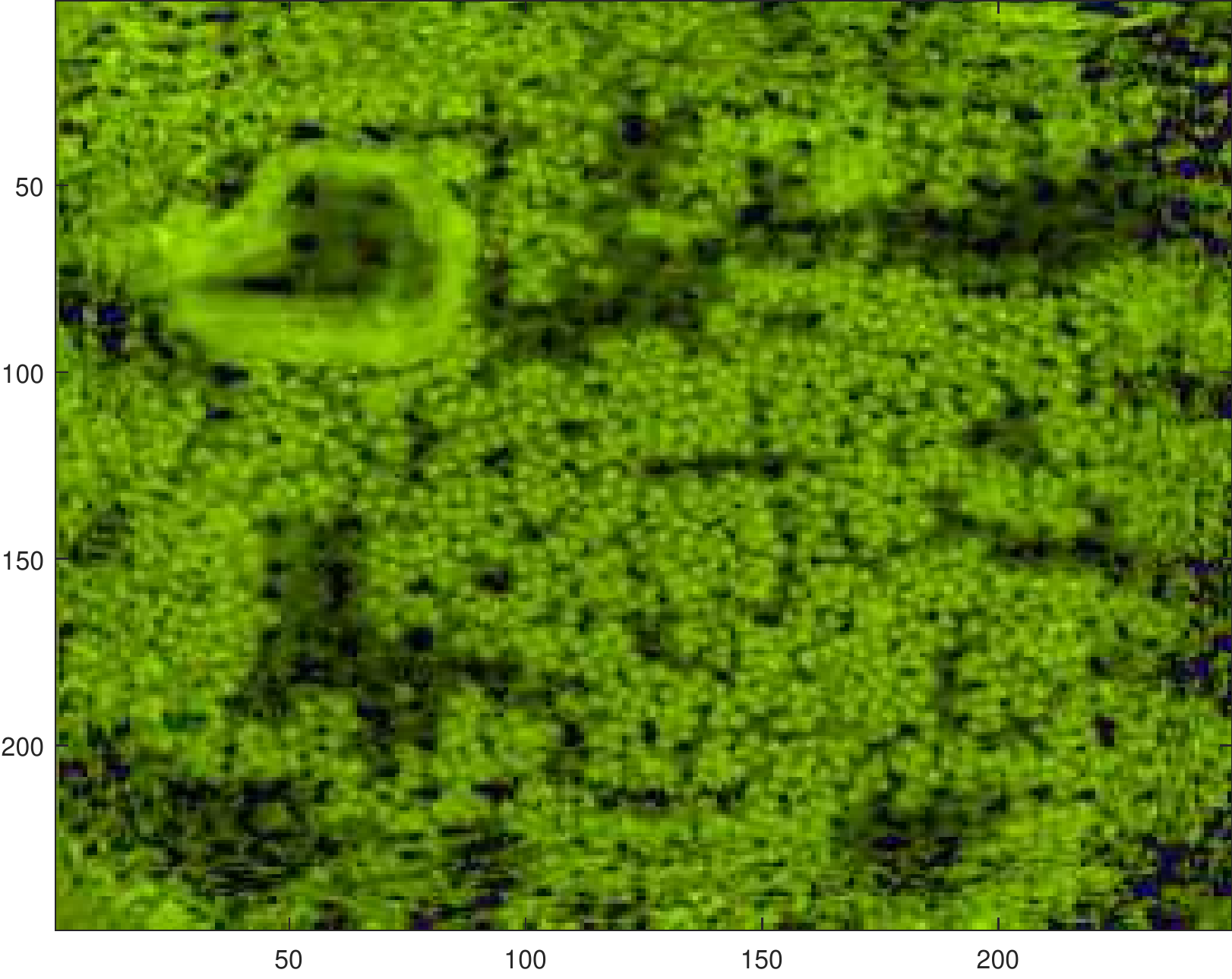}\\
\multicolumn{4}{c}{\tiny{\emph{GDP-ADMM}}}
\\
\cline{1-4}\\
\tiny{$\bm \sigma=(4,0)$}& 
\tiny{$(5,0)$}& 
\tiny{$(6,0)$}& 
\tiny{$(7,0)$}
\\
\includegraphics[width=.2\columnwidth,height=.2\columnwidth,trim={.6cm .4cm 0cm 0cm},clip]{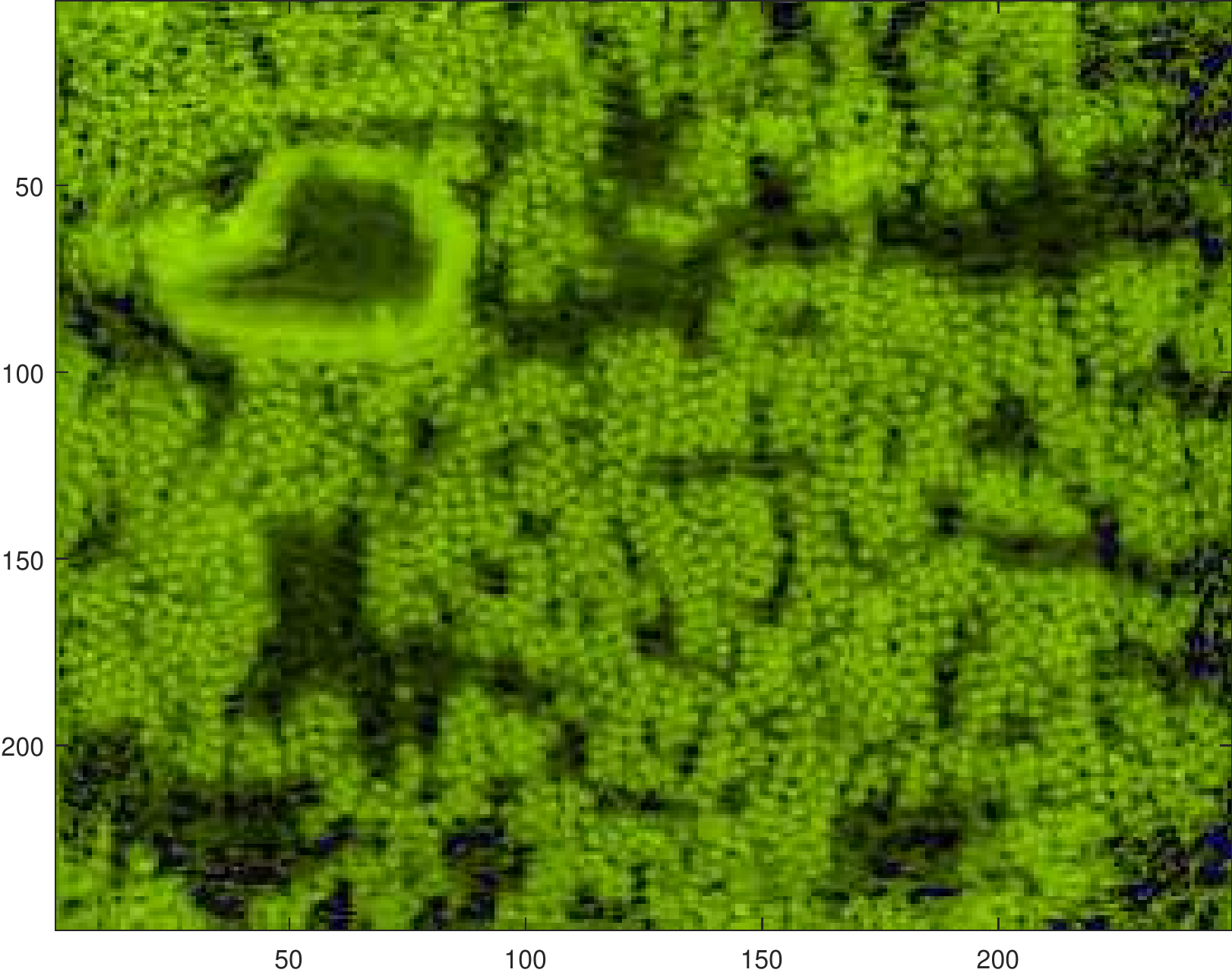}&
\includegraphics[width=.2\columnwidth,height=.2\columnwidth,trim={.6cm .4cm 0cm 0cm},clip]{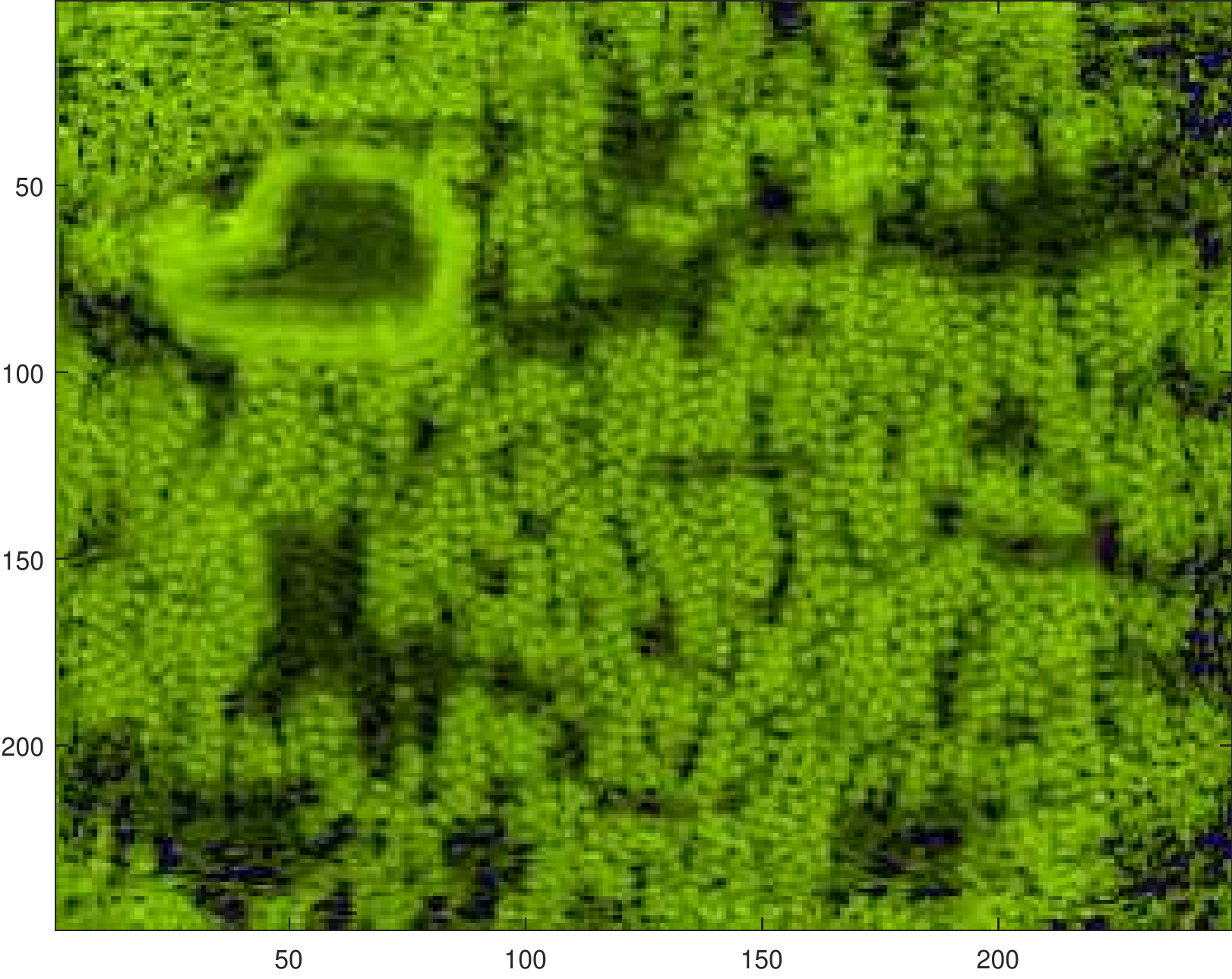}&
\includegraphics[width=.2\columnwidth,height=.2\columnwidth,trim={.6cm .4cm 0cm 0cm},clip]{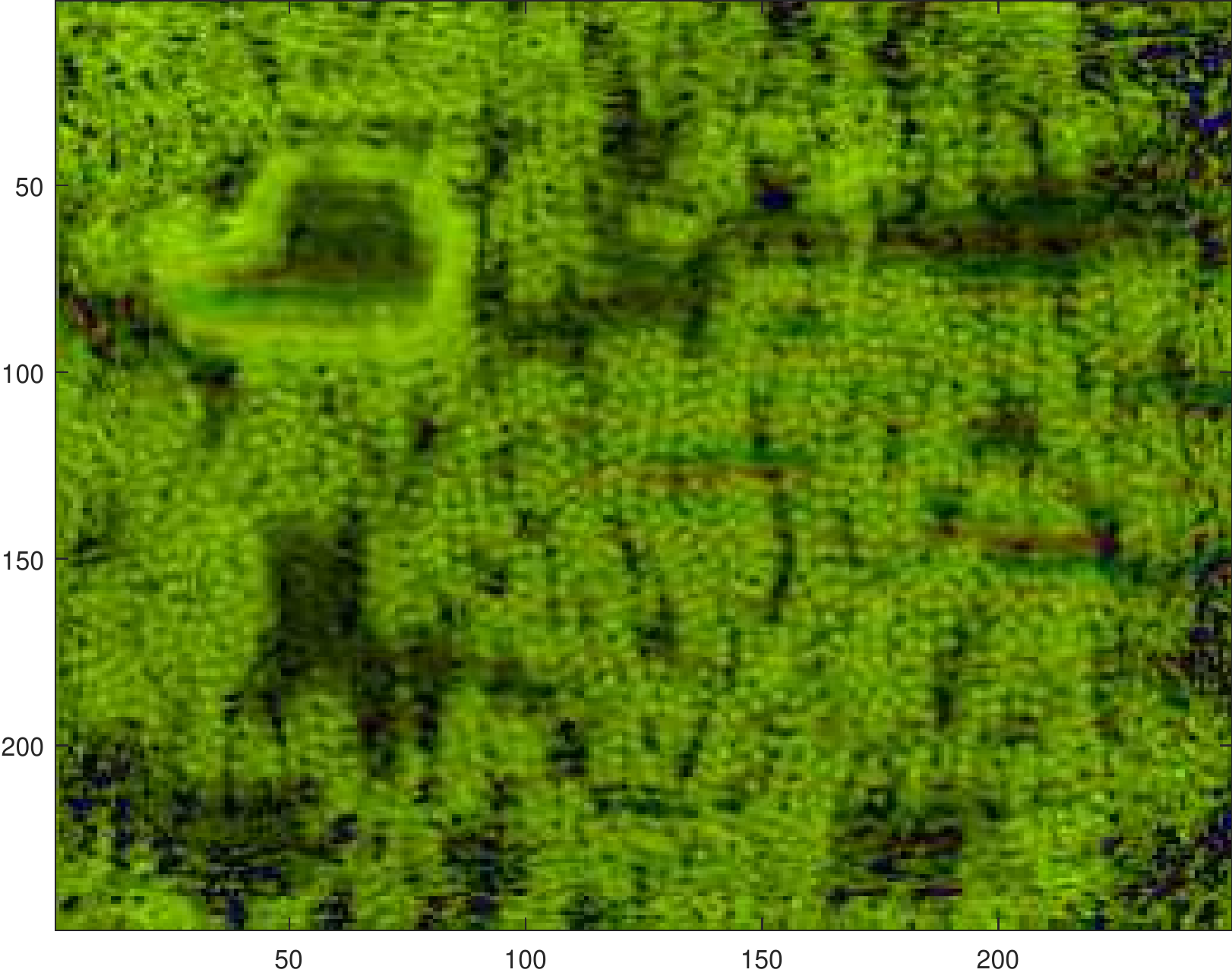}&
\includegraphics[width=.2\columnwidth,height=.2\columnwidth,trim={.6cm .4cm 0cm 0cm},clip]{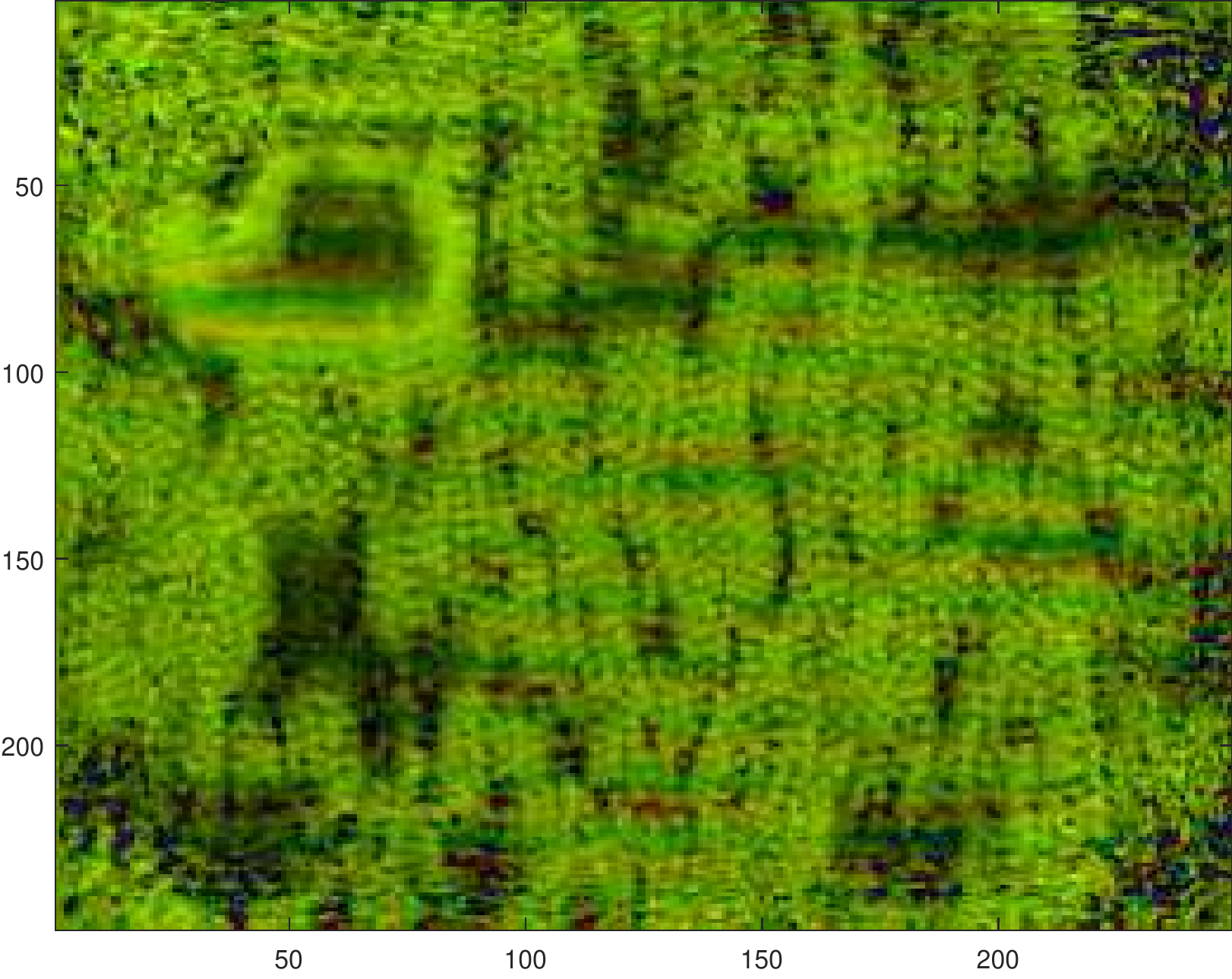} \\
\multicolumn{4}{c}{\tiny{\emph{FC-ADMM}}}\\
\includegraphics[width=.2\columnwidth,height=.2\columnwidth,trim={.6cm .4cm 0cm 0cm},clip]{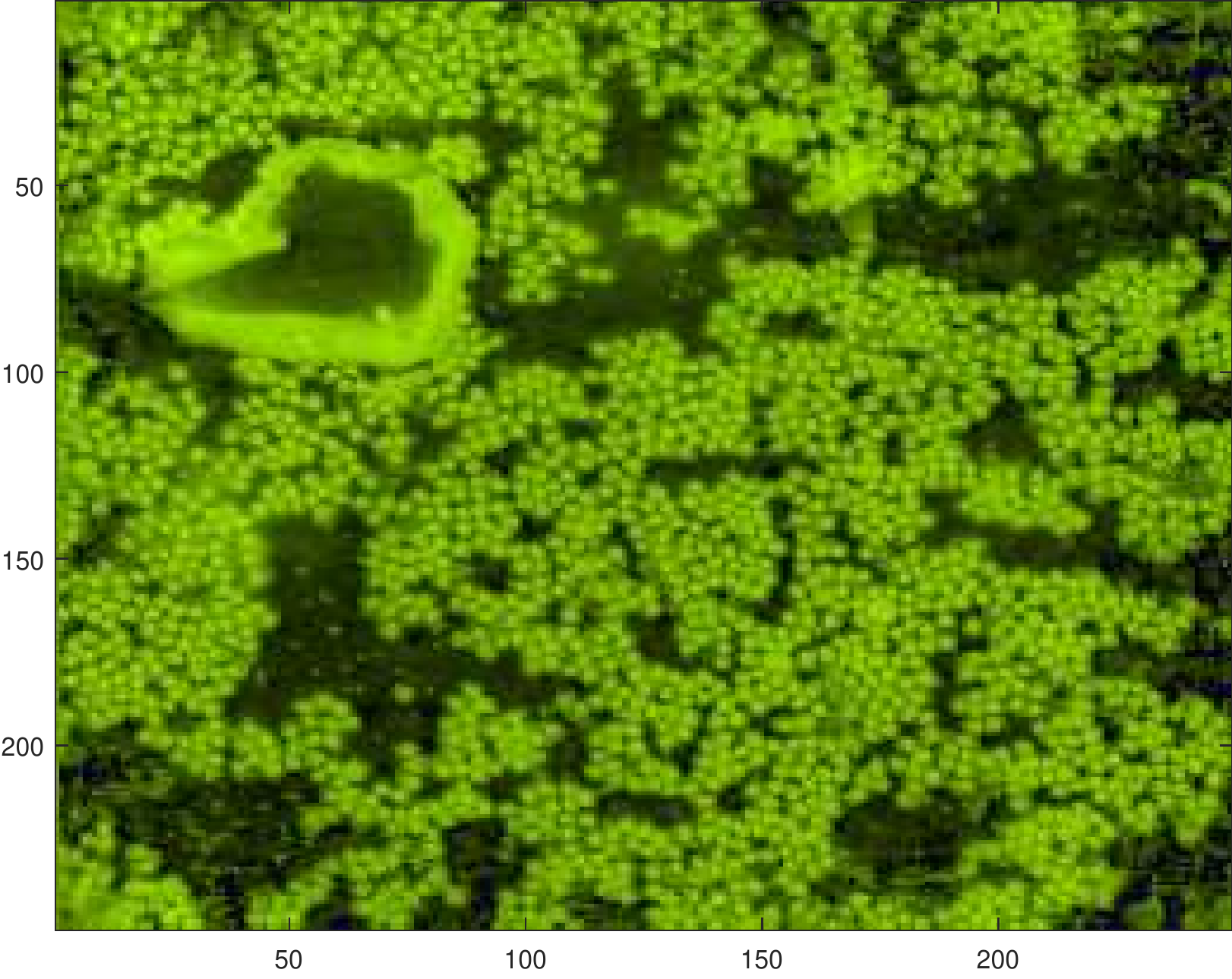}&
\includegraphics[width=.2\columnwidth,height=.2\columnwidth,trim={.6cm .4cm 0cm 0cm},clip]{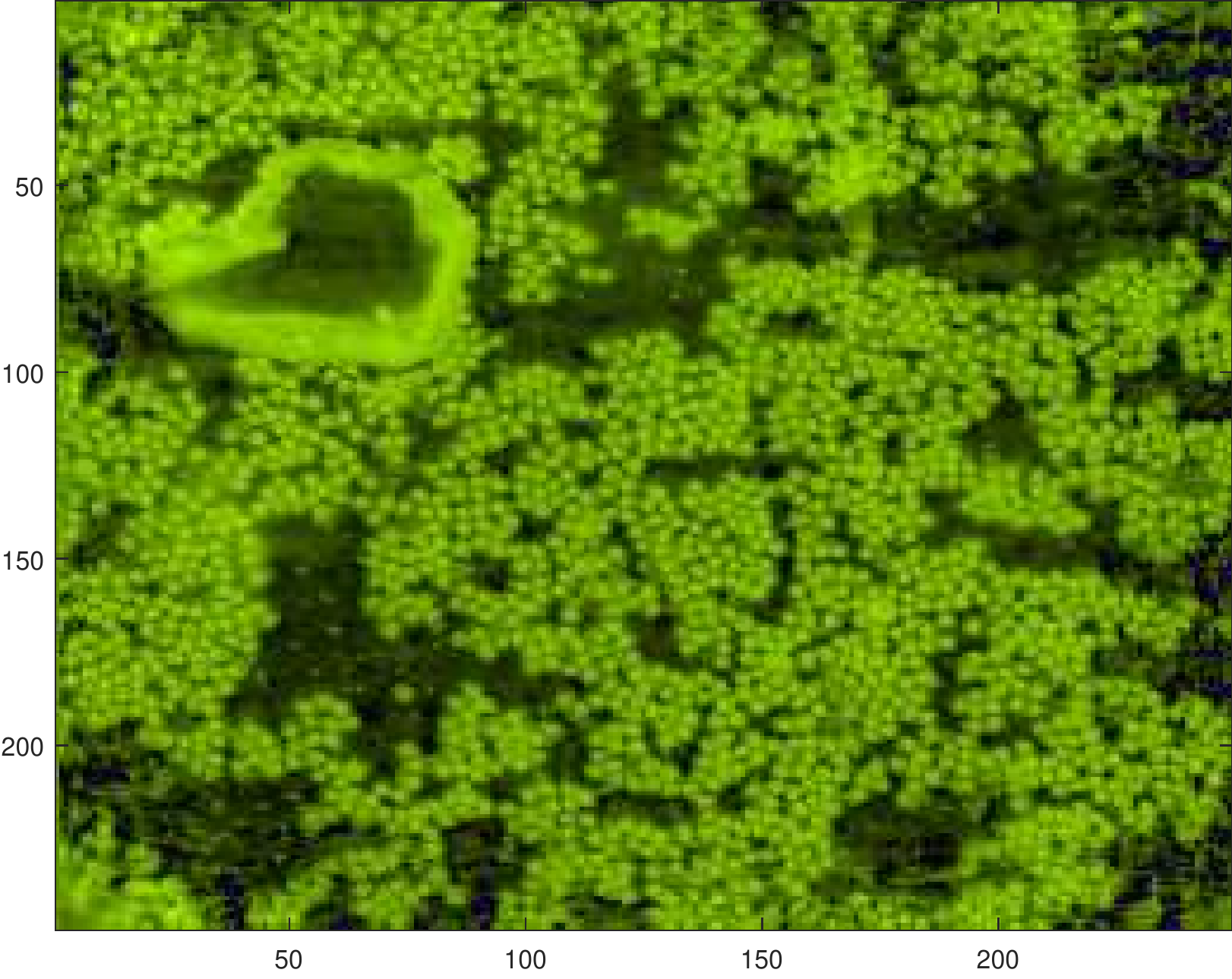}&
\includegraphics[width=.2\columnwidth,height=.2\columnwidth,trim={.6cm .4cm 0cm 0cm},clip]{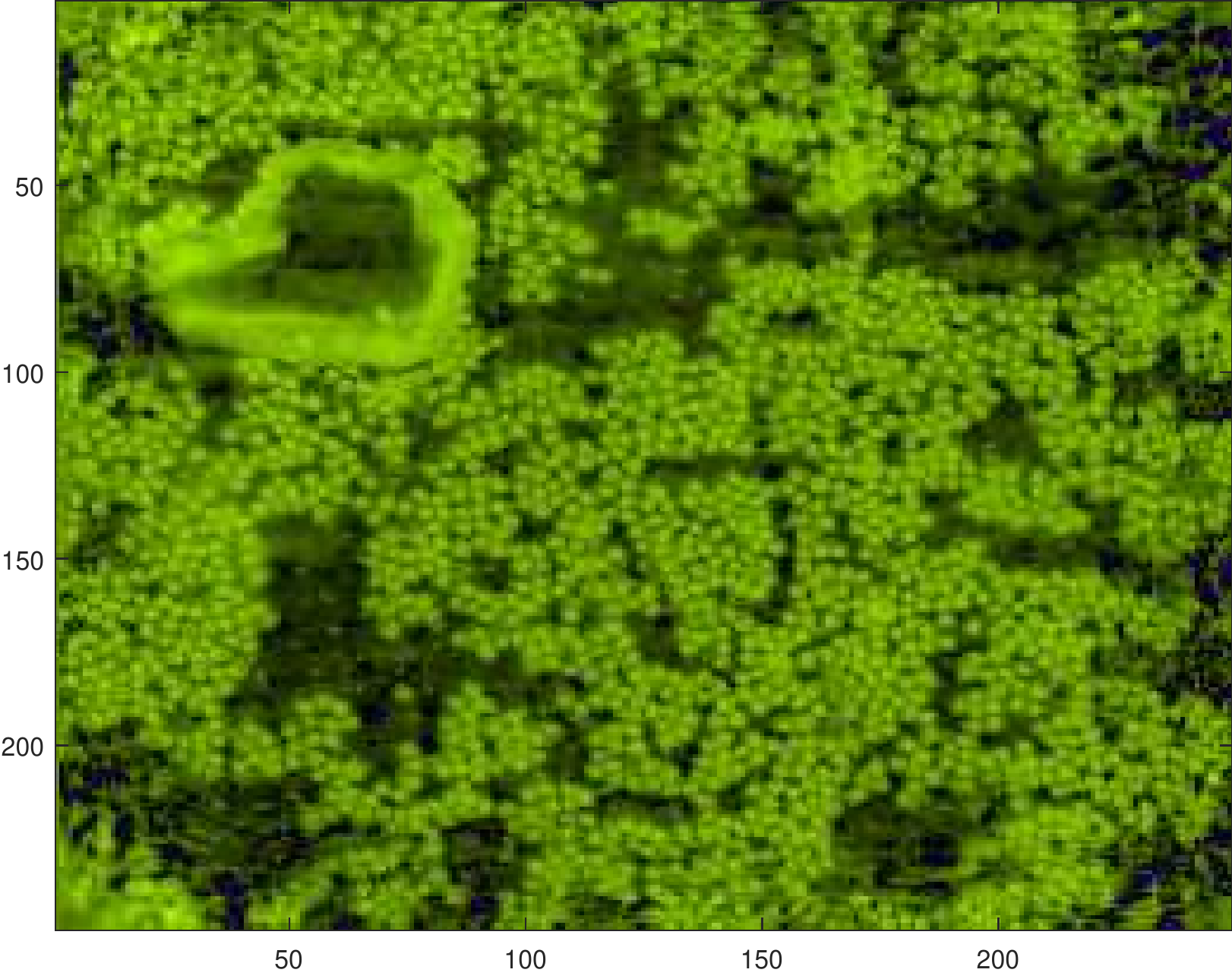}&
\includegraphics[width=.2\columnwidth,height=.2\columnwidth,trim={.6cm .4cm 0cm 0cm},clip]{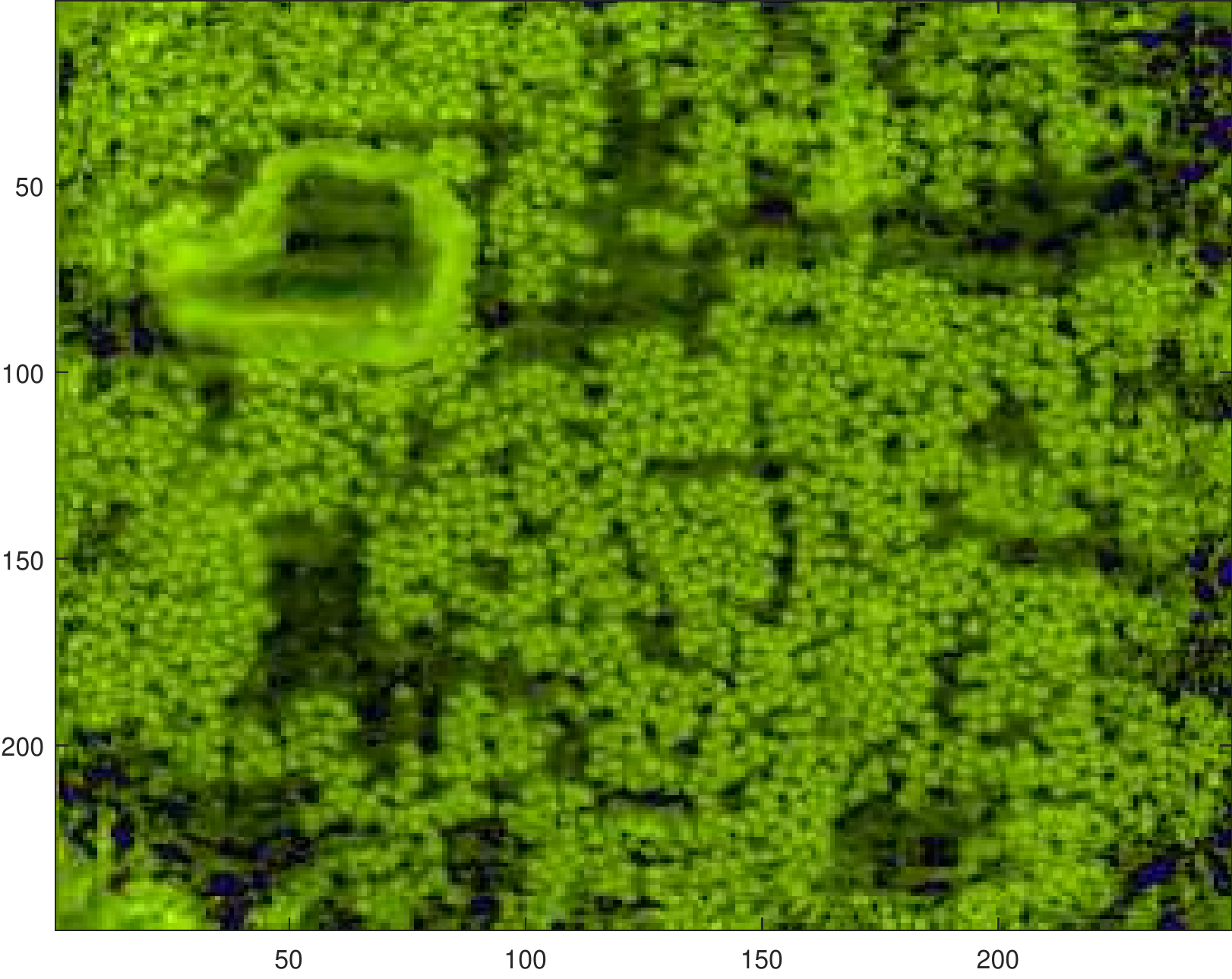}\\
\multicolumn{4}{c}{\tiny{\emph{GDP-ADMM}}}\\
\cline{1-4}
\end{tabular}
\vskip 3mm
\begin{tabular}{ccc}
\includegraphics[width=.25\columnwidth,height=.25\columnwidth,trim={1.6cm 1.4cm 1cm 1cm},clip]{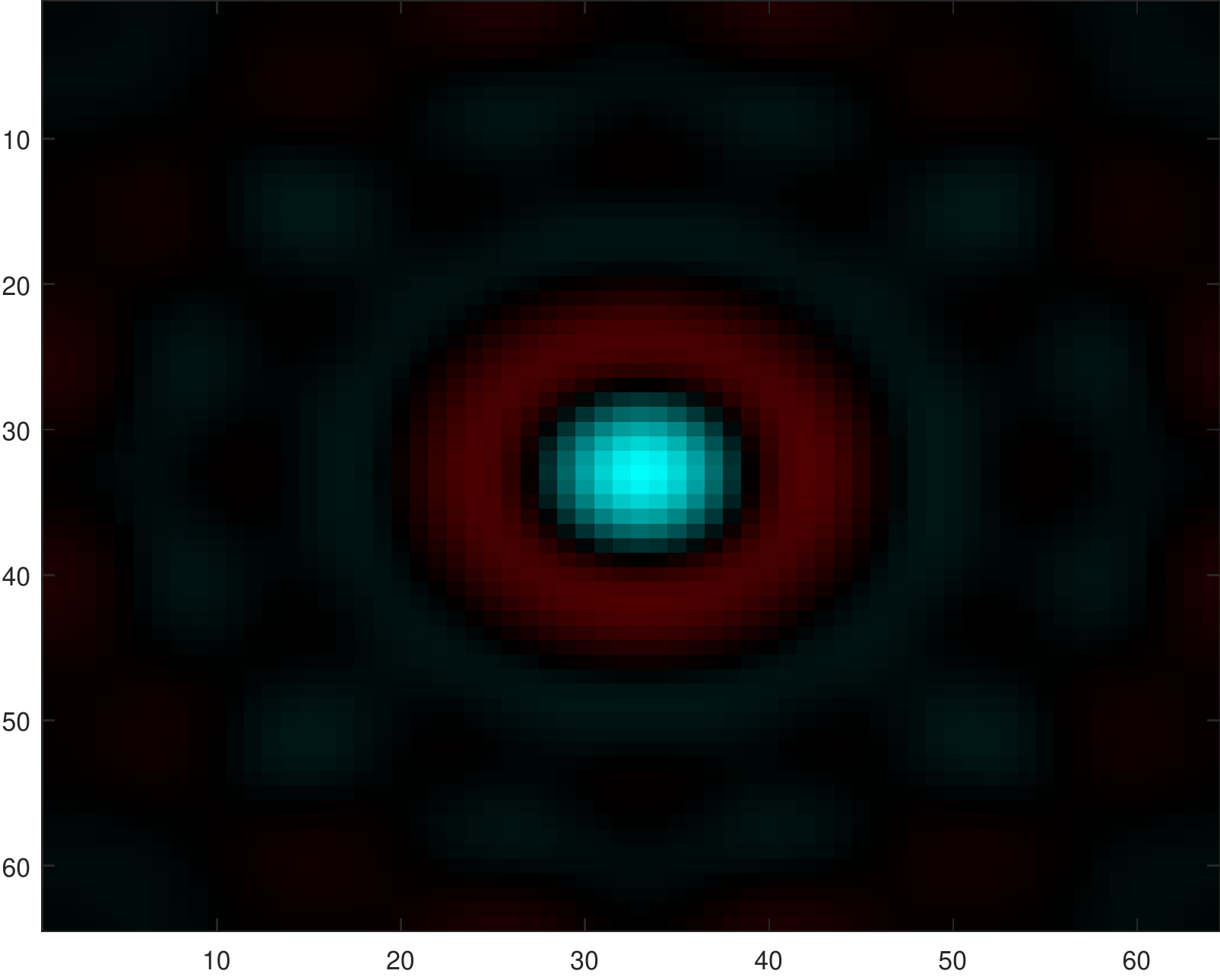}&
\includegraphics[width=.25\columnwidth,height=.25\columnwidth,trim={1.6cm 1.4cm 1cm 1cm},clip]{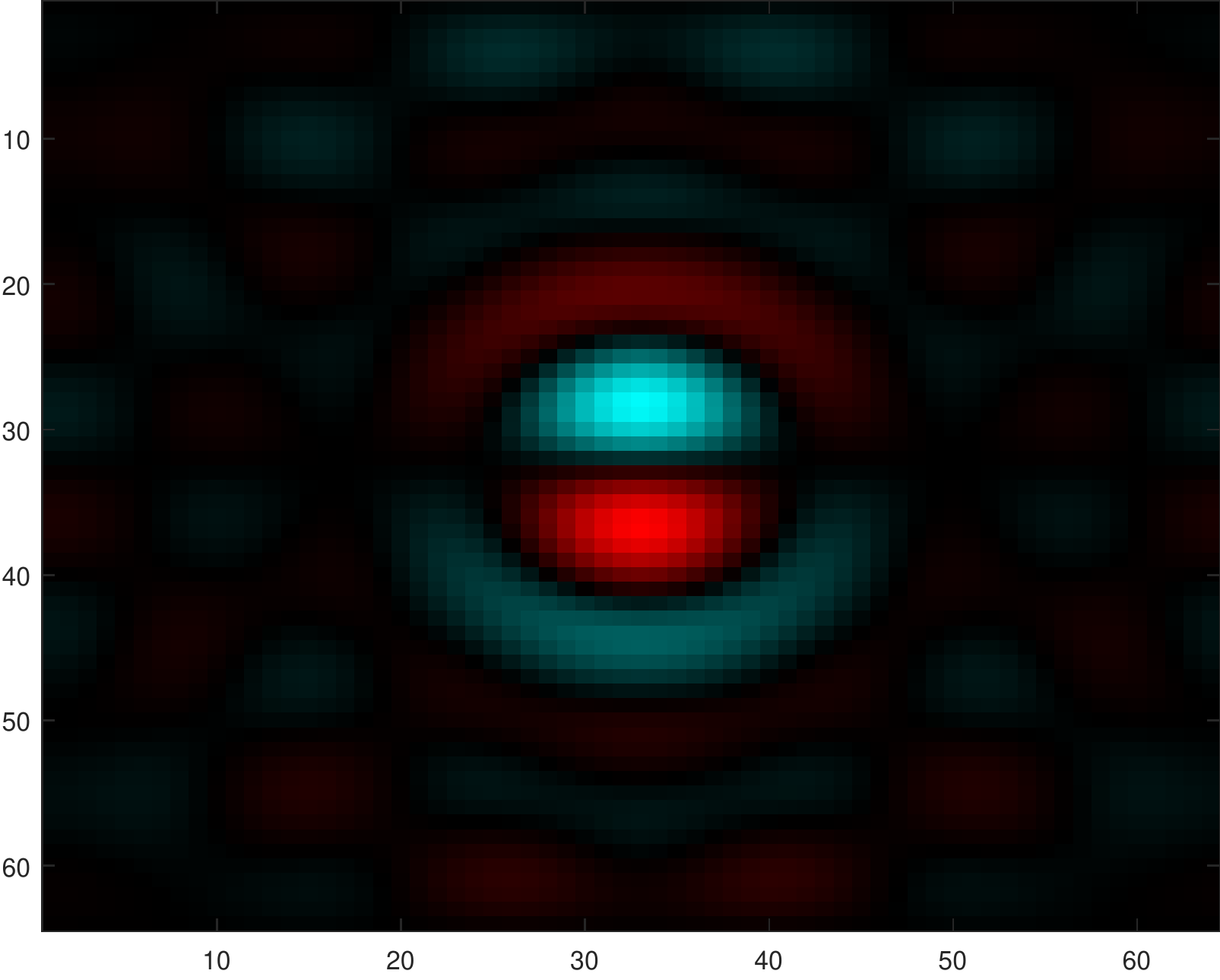}
&
\includegraphics[width=.25\columnwidth,height=.25\columnwidth,trim={1.6cm 1.4cm 1cm 1cm},clip]{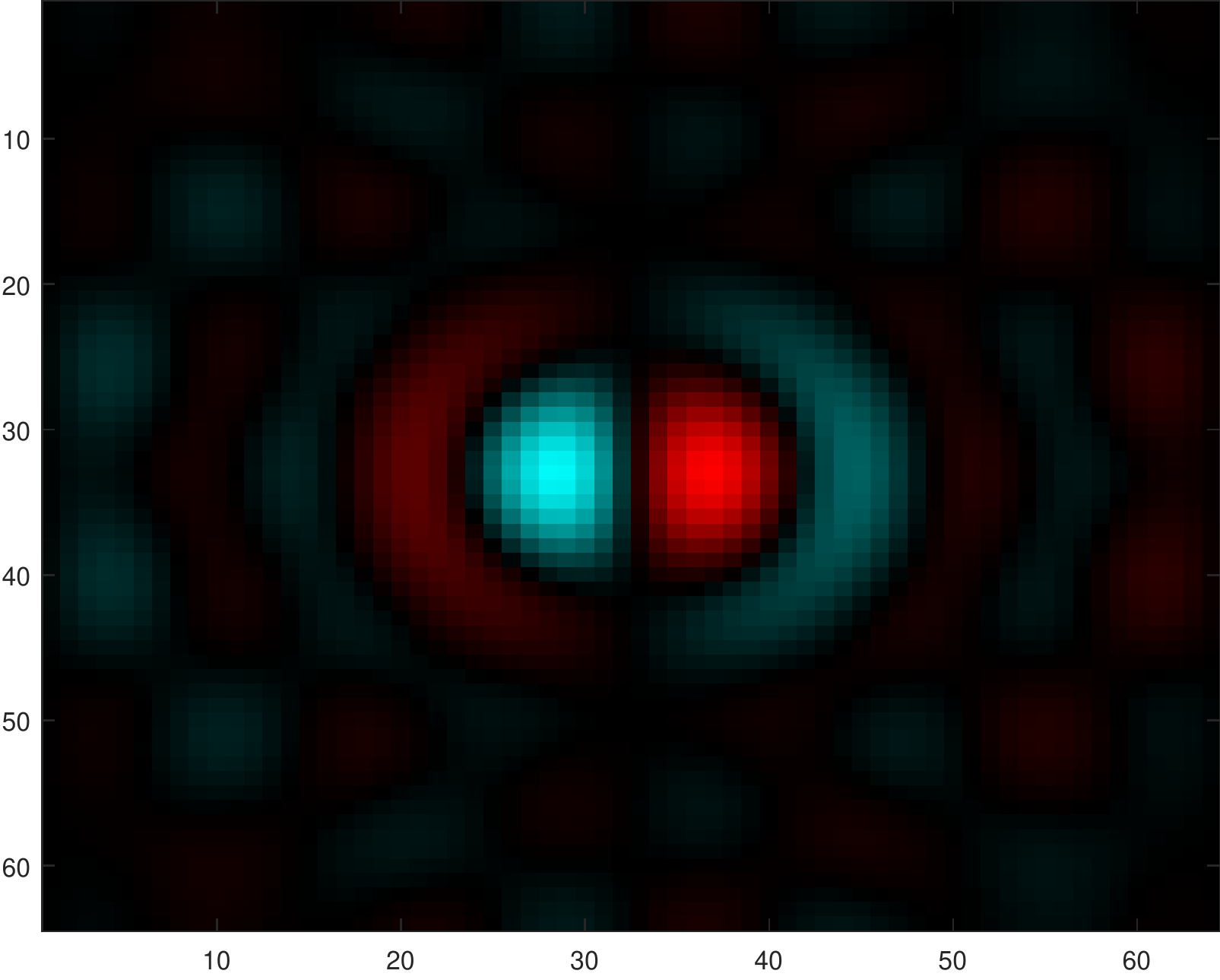}\\
$\tilde\omega$ & $ \nabla_1\tilde\omega$ &$ \nabla_2\tilde\omega$
\end{tabular}
\caption{(First and third rows) reconstructed images of \emph{FC-ADMM}, and (second and fourth rows) \emph{GDP-ADMM} while varying $\bm \sigma$.
  (Bottom row) Probe with its three-modes $\tilde\omega,
  \nabla_1\tilde\omega, \nabla_2\tilde\omega$ from left to right
  recovered from the second row with $\bm \sigma=(4,4)$.
\label{fig3}}
\end{center}
\end{figure}

The reconstructed images can be seen in Figure \ref{fig3},
and the accuracy of the reconstructed results can be seen in Table
\ref{tab1}. Figure \ref{fig3} shows significant improvements when
using \emph{GDP-ADMM} compared to \emph{FC-ADMM}.  When coherence is
very low (4th column), the visual quality of \emph{FC-ADMM} drops, and
small-scale features are completely lost. In comparison, with the same
configuration, \emph{GDP-ADMM} can still produce images containing
significantly sharper large- and small-scale features.  
 However the results by
\emph{GDP-ADMM} seem blurry when the kernel variance $\|\bm \sigma\|$ becomes too large,
as can be seen from the fourth column of Figure (\ref{fig3}).
  Table (\ref{tab1}) shows the enhanced accuracy of
GDP-ADMM compared to the coherent model: residuals are at least 50\%
percent smaller, and images with SNRs about twice as high.  We also
include the results of the recovered probe with its modes in Figure
\ref{fig3}.

\begin{table}
\caption{Performance of \emph{GDP-ADMM}  and \emph{FC-ADMM}.   $e_{pc}$ and $e_{fc}$ are the residuals of \emph{GDP-ADMM} and \emph{FC-ADMM} respectively (the smaller, the better); $\mathrm{SNR}_{pc}$, $\mathrm{SNR}_{fc}$  are the SNRs for  \emph{GDP-ADMM} and \emph{FC-ADMM}, respectively (the larger, the better). }
\begin{center}
\scalebox{.9}{
\begin{tabular}{|c||cccc|}
\cline{1-5} 
$\bm \sigma $& (2,2) & (3,3) & (4,4) & (5,5)\\
    $e_{fc}$&1.36E-1&1.88E-1&2.11E-1&2.20E-1\\
$e_{pc}$&2.60E-2&5.68E-2&8.95E-2&1.14E-1\\
 $\mathrm{SNR}_{fc}$&14.78&9.53&6.14&4.37\\
$\mathrm{SNR}_{pc}$&23.73&18.83&13.02&8.37\\
\cline{1-5} 
$\bm \sigma $& (4,0) & (5,0) & (6,0) & (7,0)\\
 $e_{fc}$&1.70E-1&1.89E-1&2.04E-1&2.14E-1\\
$e_{pc}$&5.62E-2&7.87E-2&9.76E-2&1.12E-1\\
 $\mathrm{SNR}_{fc}$&11.47&9.42&7.61&5.64\\
$\mathrm{SNR}_{pc}$&19.83&16.39&13.46&10.94\\
\cline{1-5}
\end{tabular}
}
\end{center}
\label{tab1}
\end{table}

\begin{figure}
\begin{center}
\begin{tabular}{cccc}
{\includegraphics[width=.22\columnwidth,height=.22\columnwidth,trim={.6cm .4cm 0cm 0cm},clip]{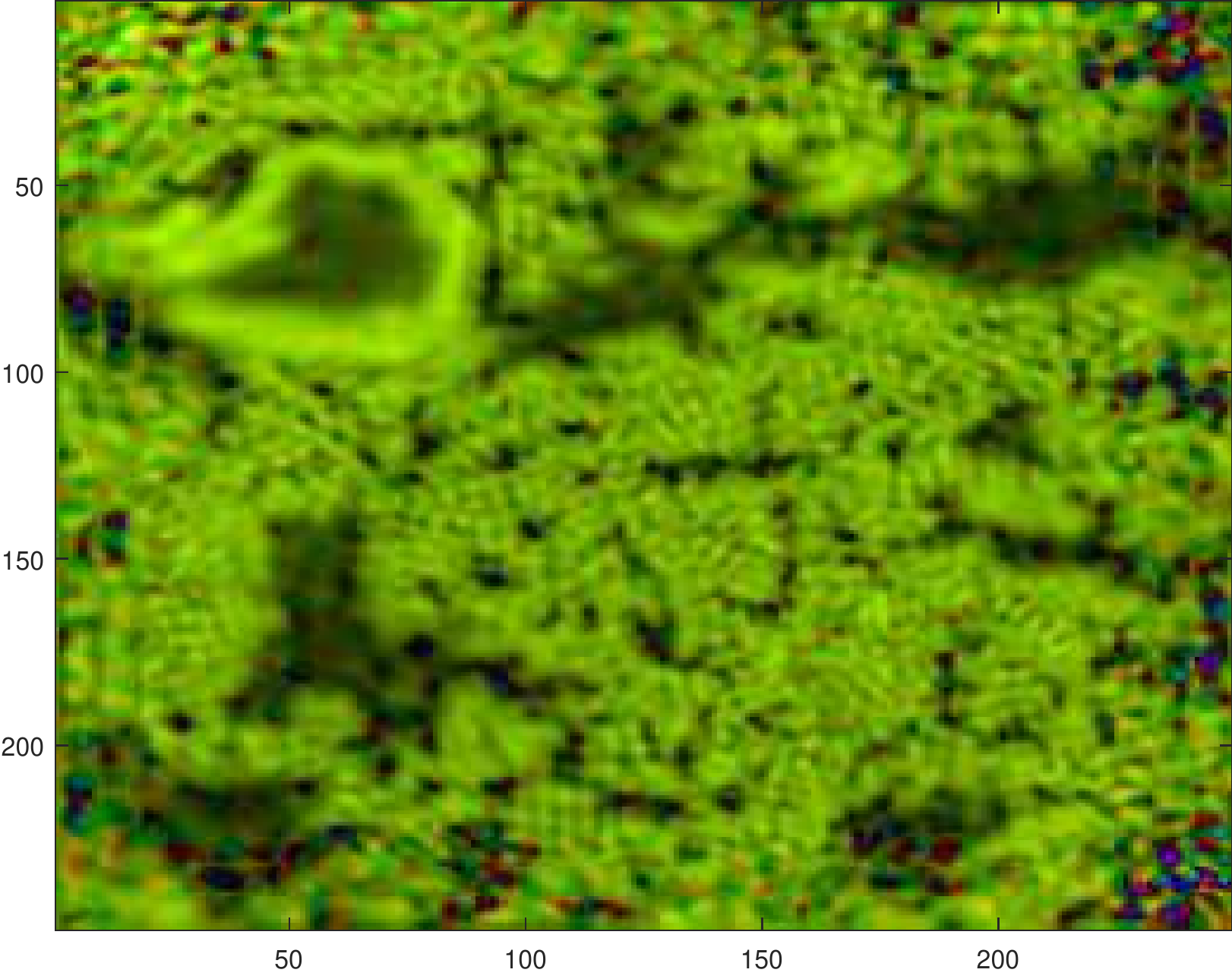}}&
{\includegraphics[width=.22\columnwidth,height=.22\columnwidth,trim={.6cm .4cm 0cm 0cm},clip]{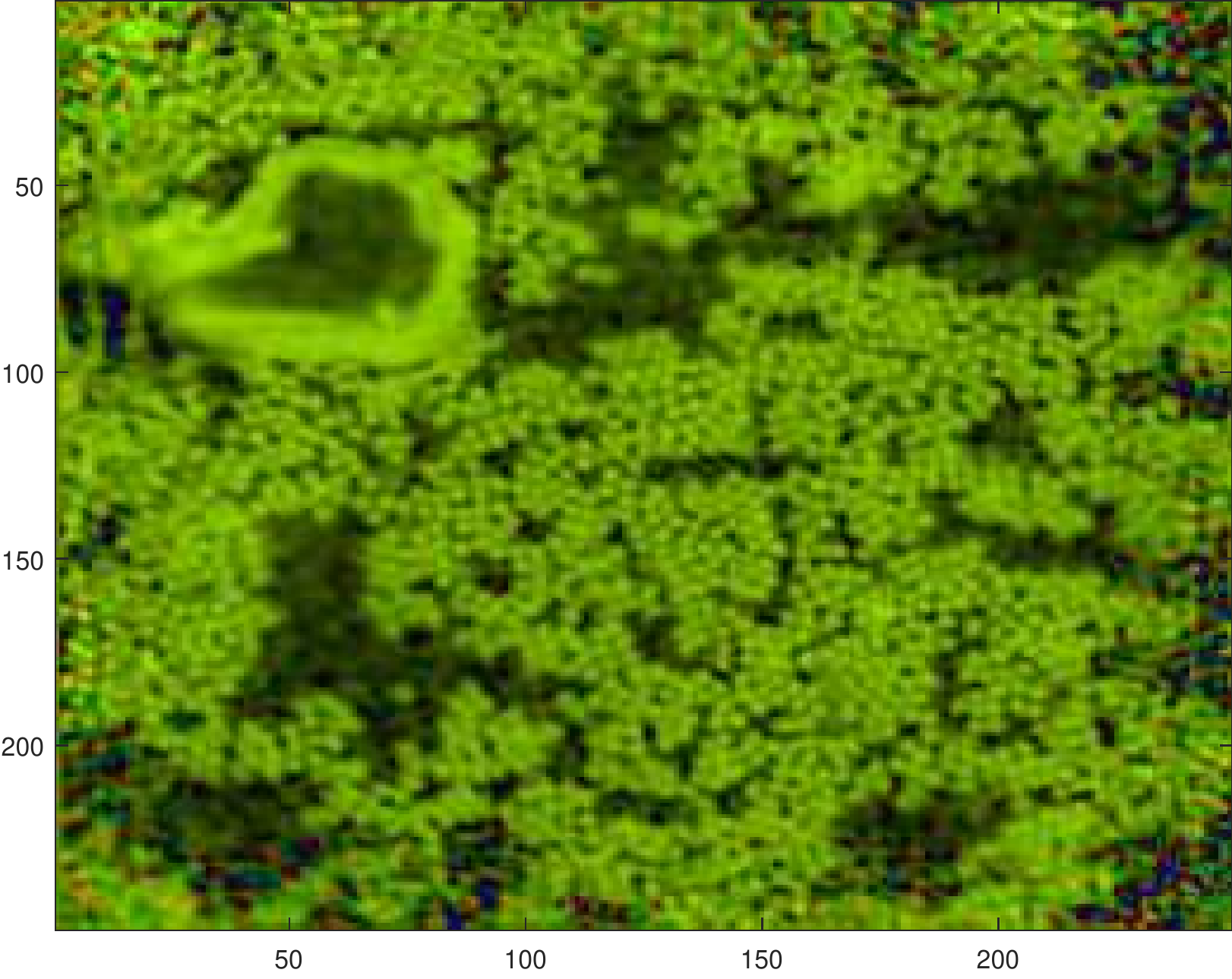}}&
{\includegraphics[width=.22\columnwidth,height=.22\columnwidth,trim={.6cm .4cm 0cm 0cm},clip]{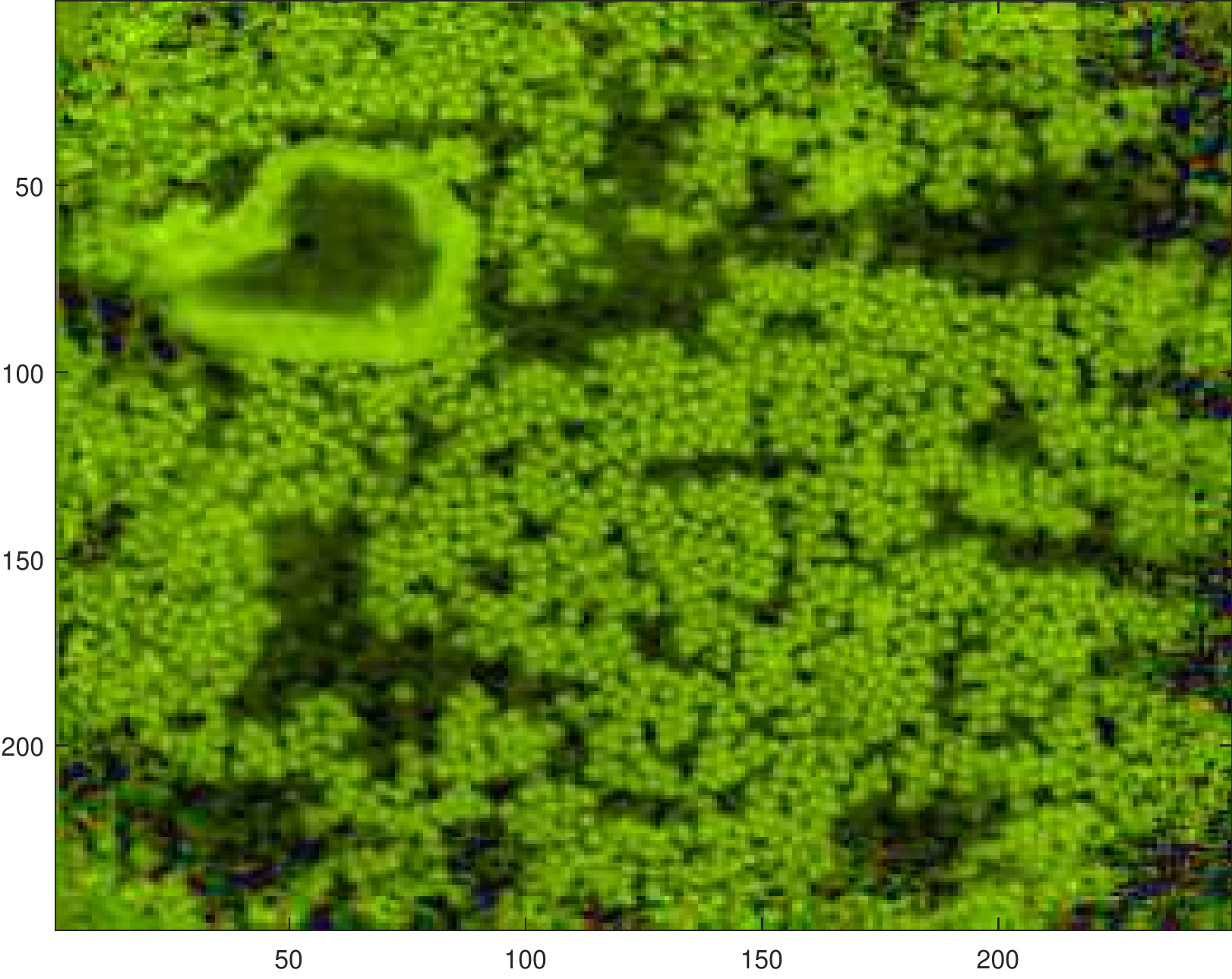}}&
{\includegraphics[width=.22\columnwidth,height=.22\columnwidth,trim={.6cm .4cm 0cm 0cm},clip]{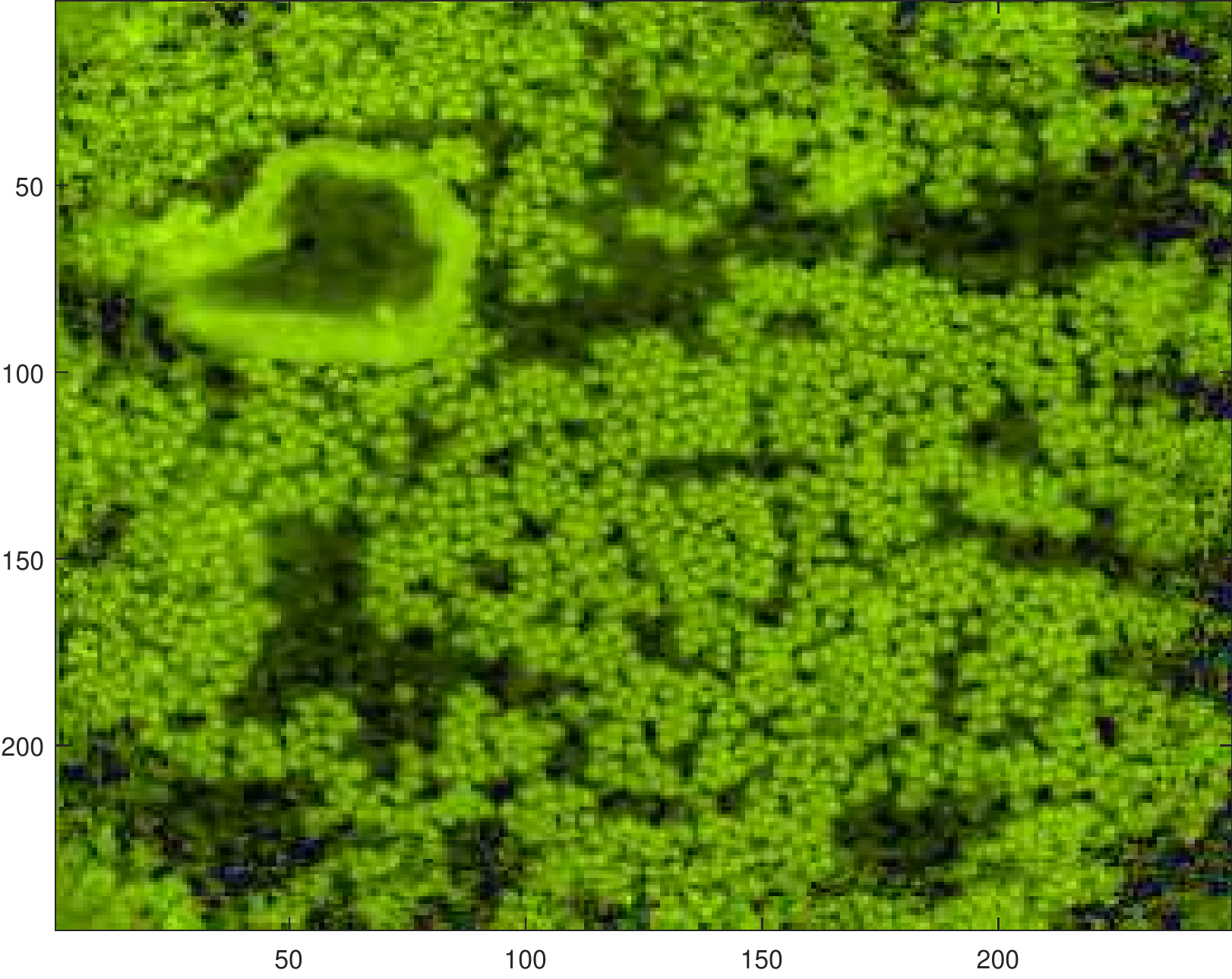}}\\
\tiny{(a) $\mathrm{SNR}$=7.08}&  \tiny{ (b)10.17} &  \tiny{ (c) 11.30}&  \tiny{ (d) 12.35}
\end{tabular}
\end{center}
\caption{Reconstructed images of \emph{GDP-ADMM} while varying the level of Poisson noise, and the SNR with respect to the ground truth.  
From left to right: Noise level $\gamma=0.25,1, 4, 16$,
  (the larger $\gamma$, the weaker the noise). The
  corresponding SNRs of the data, defined by $\mathrm{SNR}(f^{noise},
  f)$, are 23.89, 29.87, 35.91, and 41.07.  }
\label{fig4}
\end{figure}

\paragraph{Noisy data.}
We consider Poisson noisy measurements as
\[
f^{\mathrm{\,noise}}_{pc,j}=\tfrac{1}{\gamma}\mathrm{Poisson}(\gamma f_{pc,j}),
\]
where $\gamma>0$ is used to control the noise level, and $f_{pc,j}$ is
the clean partial coherent intensity data defined by \eqref{intGen}.
Figure \ref{fig4} shows the recovered images at different noise
levels.  It can be seen that, as the noise level decreases (increasing
$\gamma$), the quality of the results increases. We remark that when
$\gamma=0.25$, the image in Figure \ref{fig4} (a) is very blurry,
which implies that it is more challenging to recover high quality
images from partially coherent data contaminated by strong noise.

\begin{figure}
\begin{center}
\begin{tabular}{cc}
\includegraphics[width=.4\columnwidth]{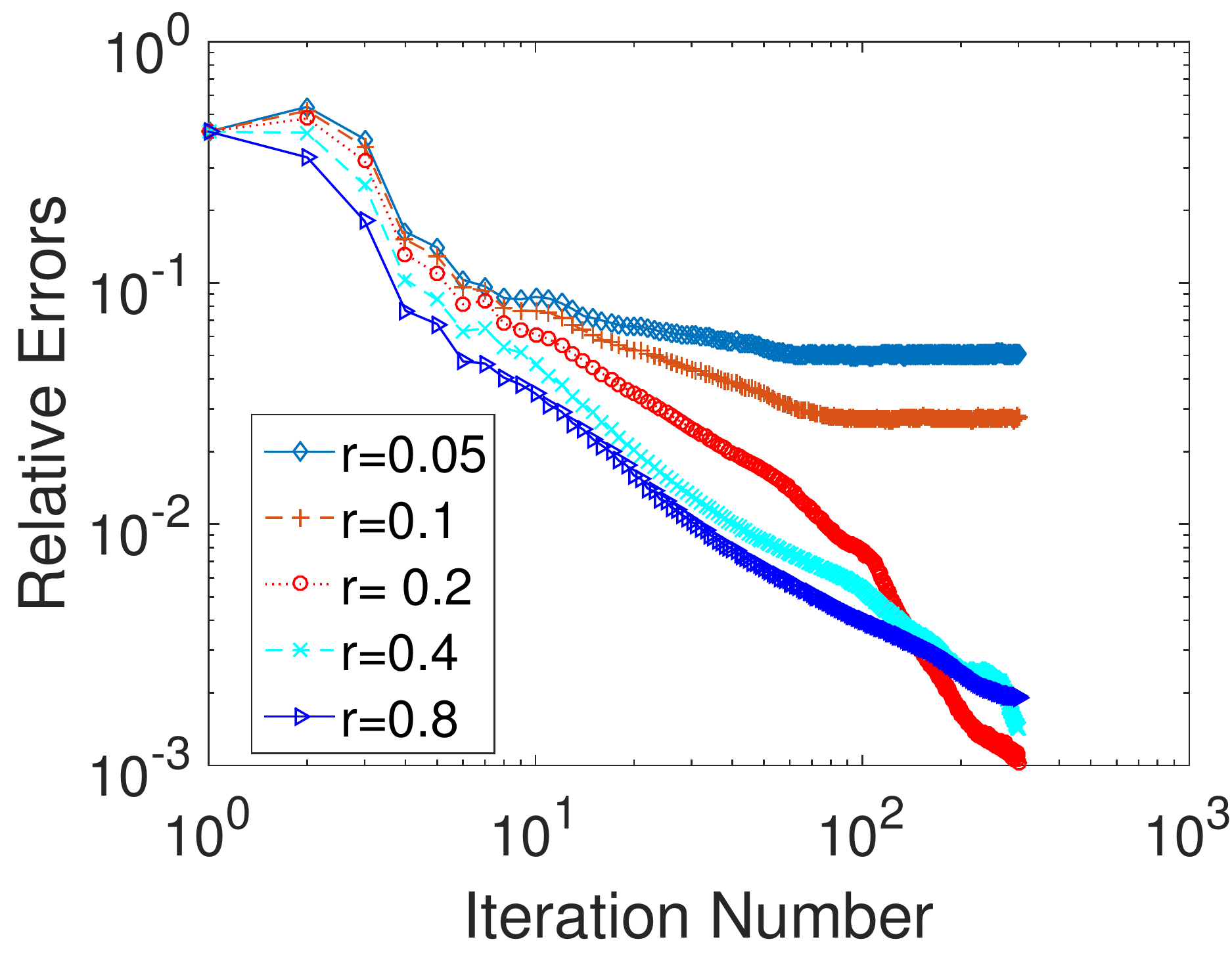}&
\includegraphics[width=.4\columnwidth]{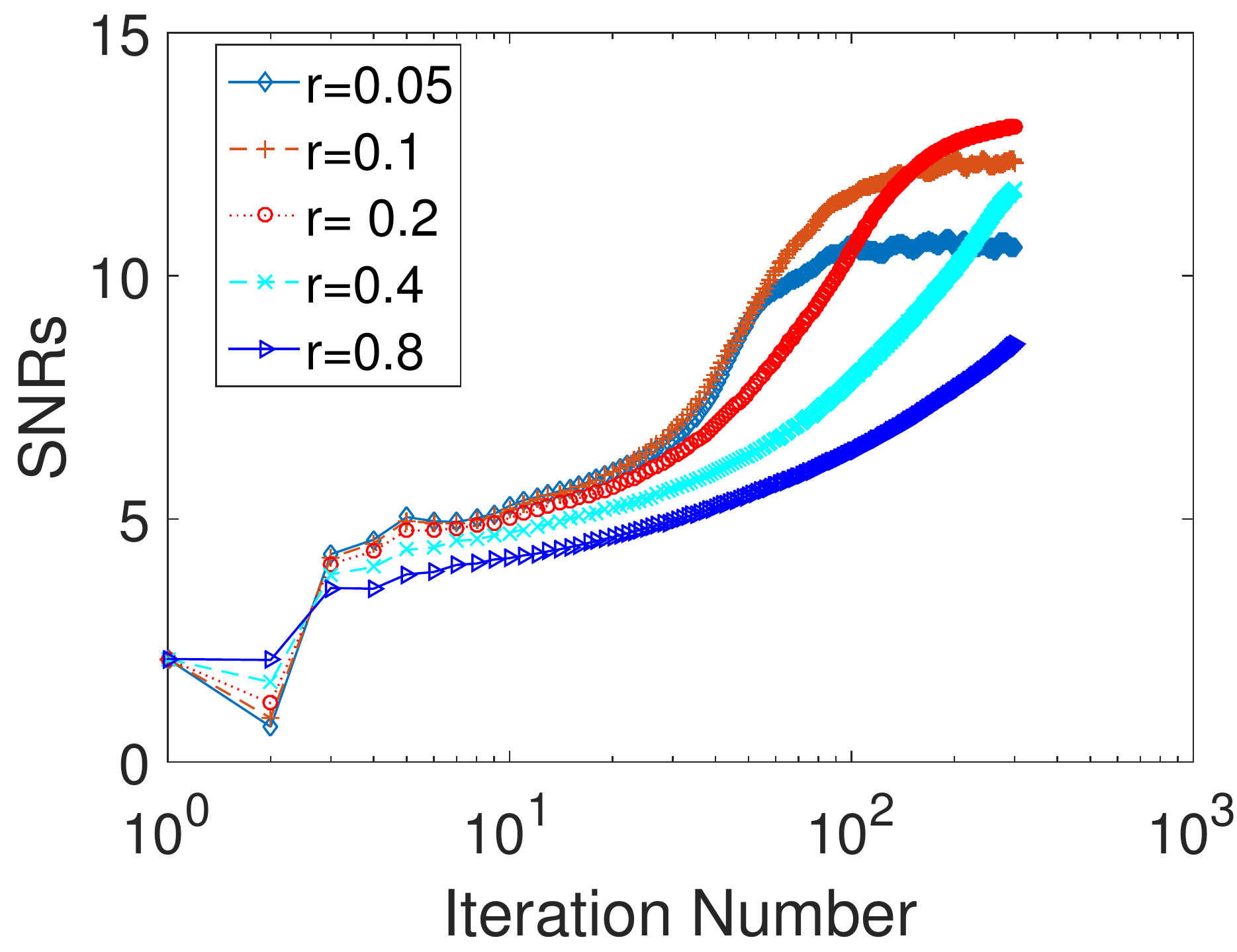}\\
{(a)} & {(b)}
\end{tabular}
\end{center}
\caption{Performance influenced by parameter
  $r\in\{0.05,0.1,0.2,0.4,0.8\}$ of \emph{GDP-ADMM}. (a) Relative
  errors $\tfrac{\|u^{k}-u^{k-1}\|}{\|u^{k}\|}$ of the iterative
  solution with respect to the iteration number; (b) SNRs of iterative
  solutions with respect to the iteration number.}
\label{fig6}
\end{figure}

\paragraph{Parameter $r$ and sliding distances.}
The following experiment is conducted to study the influence of
parameter $r$ by varying $r\in\{0.05,0.1,0.2,0.4,0.8\}$. See the
convergence curve in Figure \ref{fig6}, where the change histories of
the relative errors and SNRs with respect to iteration numbers are
reported. One can readily see that with smaller parameter $r$, the
algorithm tends to be unstable (see Figure \ref{fig6} (b)). It is
consistent with the condition for convergence guarantee in
\cite{chang2017blind}, where the parameter $r$ should be sufficiently
large to make the augmented Lagrangian monotonically decrease. If the
parameter is too big, the iterative solution could be trapped into the
unsatisfactory local minima. Therefore, to gain optimal performance, a
moderate $r$ should be used.  We remark that although the parameter
$r$ is selected manually, it can be applied to diverse cases with
different degrees of partial coherence and sliding distances when
fixing the kernel function.

\begin{figure}
\begin{center}
\begin{tabular}{cccc}
{\includegraphics[width=.22\columnwidth,height=.22\columnwidth,trim={.6cm .4cm 0cm 0cm},clip]{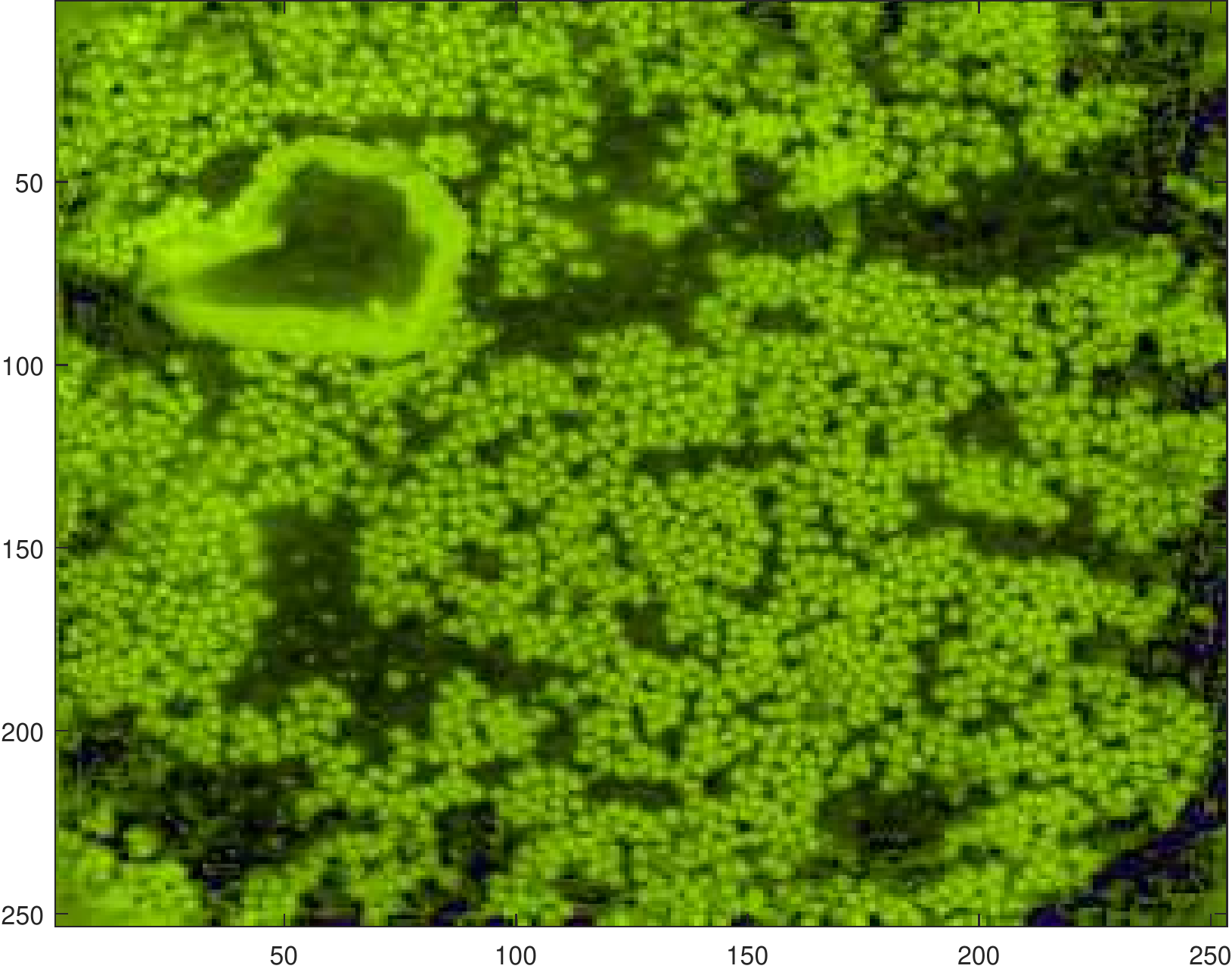}} &
{\includegraphics[width=.22\columnwidth,height=.22\columnwidth,trim={.6cm .4cm 0cm 0cm},clip]{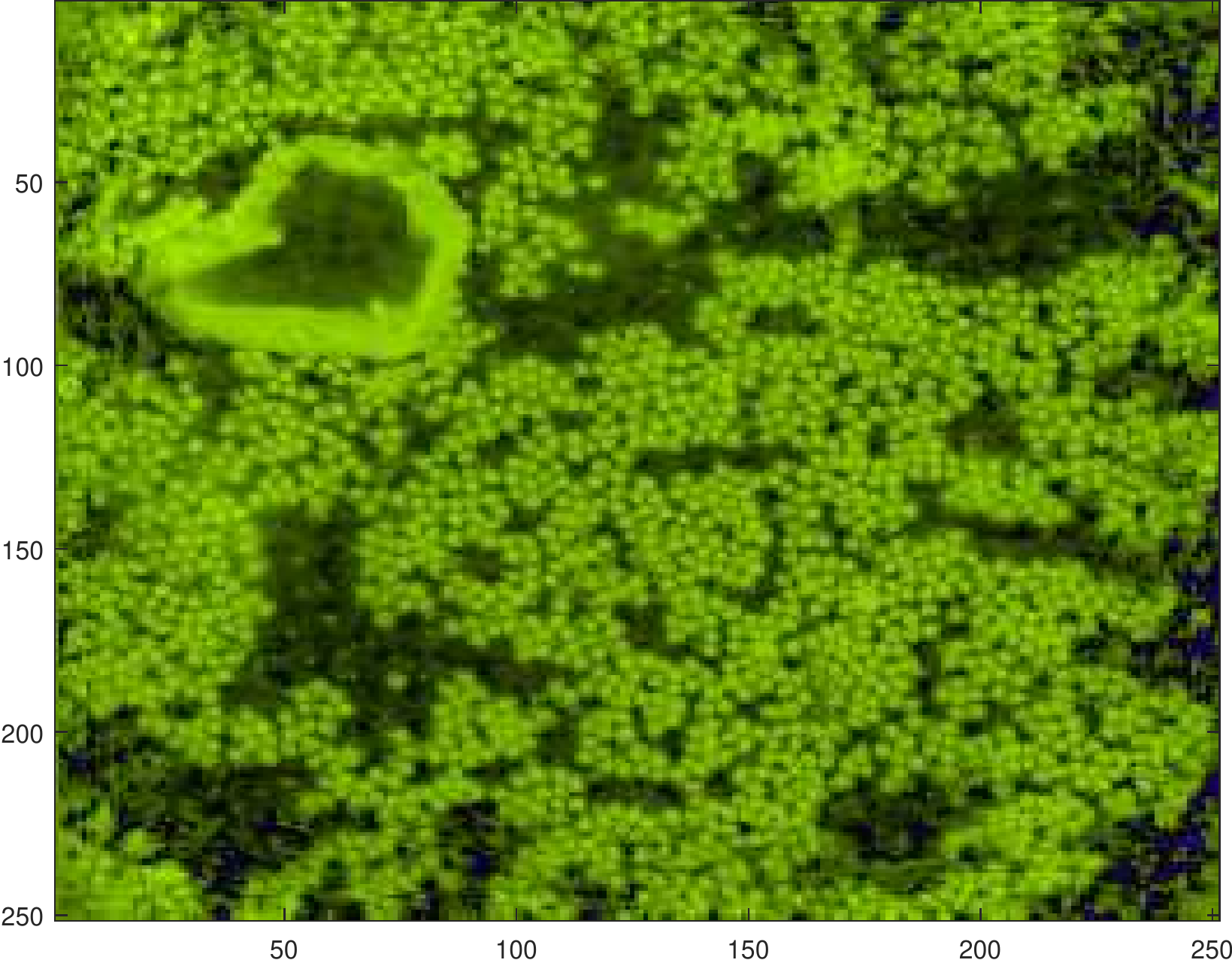}} &
{\includegraphics[width=.22\columnwidth,height=.22\columnwidth,trim={.6cm .4cm 0cm 0cm},clip]{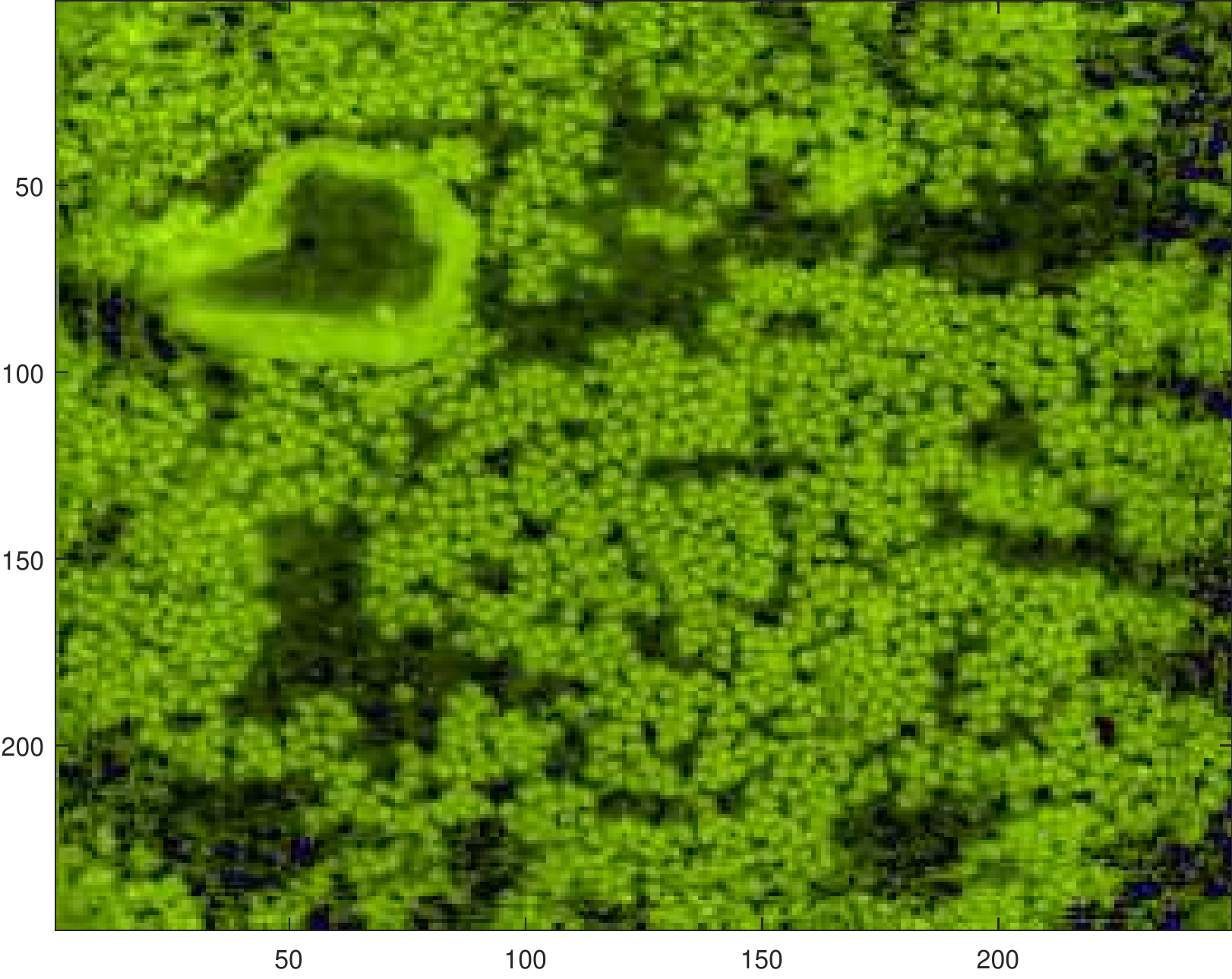}} &
{\includegraphics[width=.22\columnwidth,height=.22\columnwidth,trim={.6cm .4cm 0cm 0cm},clip]{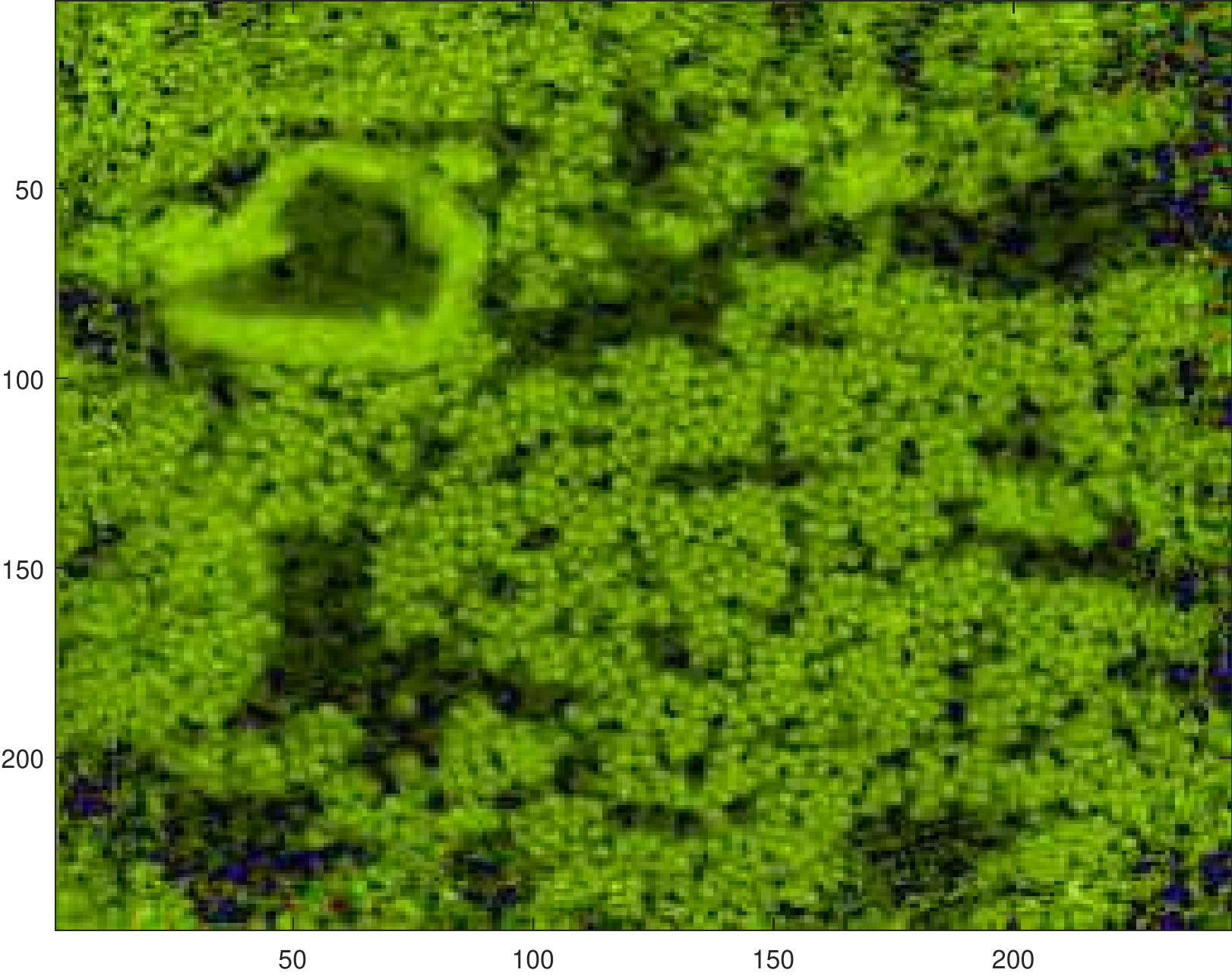}}\\
\tiny{$\mathrm{Dist}$=4} & \tiny{6} & \tiny{8} & \tiny{12}\\
 \tiny{SNR=15.93} & \tiny{14.54} & \tiny{13.02} &  \tiny{8.96} 
\end{tabular}
\end{center}
\caption{Reconstructed images using \emph{GDP-ADMM} with different
  sliding distances $\mathrm{Dist}$, 
 Gaussian kernel function ($\bm \sigma=(4,4)$), and beam width $D=7$. 
}
\label{fig7}
\end{figure}

In Figure \ref{fig7} we conduct the tests with different sliding
distances $\mathrm{Dist}\in\{4, 6, 8,12\}$, which determine the
redundancy levels of the data. On one hand, inferred from the reported
results, scanning with a smaller step size or $\mathrm{Dist}$ will
help to increase the quality of the images. On the other hand, when
$\mathrm{Dist}=12$ ( almost twice as large as the beam width), \emph{GDP-ADMM}
can still produce satisfactory results with clear small-scale
features, showing the robustness of \emph{GDP-ADMM} with respect to  the
redundancy of the measured data.
%
%

\paragraph{Proximal term.}

\begin{figure}
\begin{center}
\begin{tabular}{cc}
{\includegraphics[width=.43\columnwidth,trim={.6cm .4cm 0cm 0cm},clip]{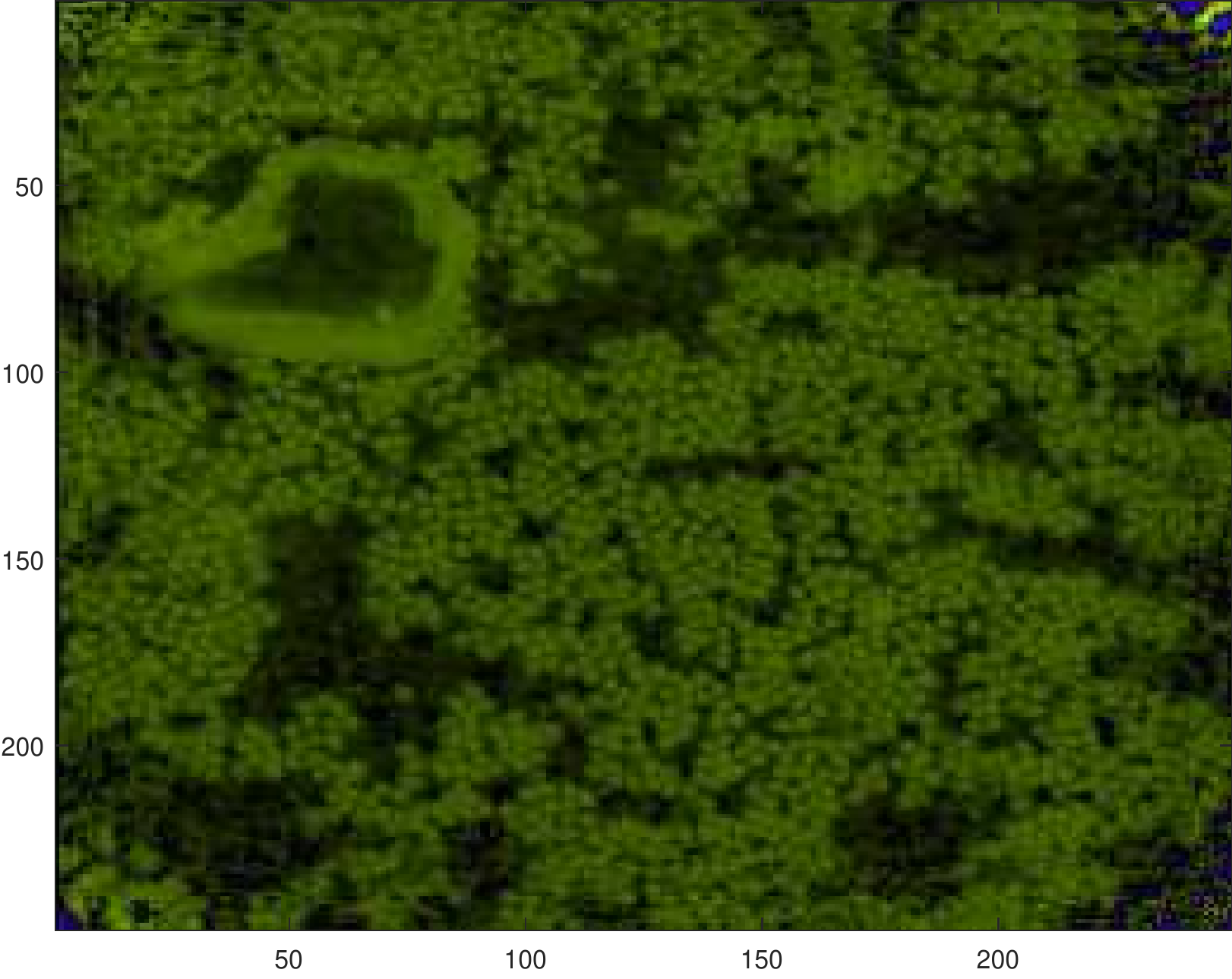}}
&
{\includegraphics[width=.43\columnwidth,trim={.6cm .4cm 0cm 0cm},clip]{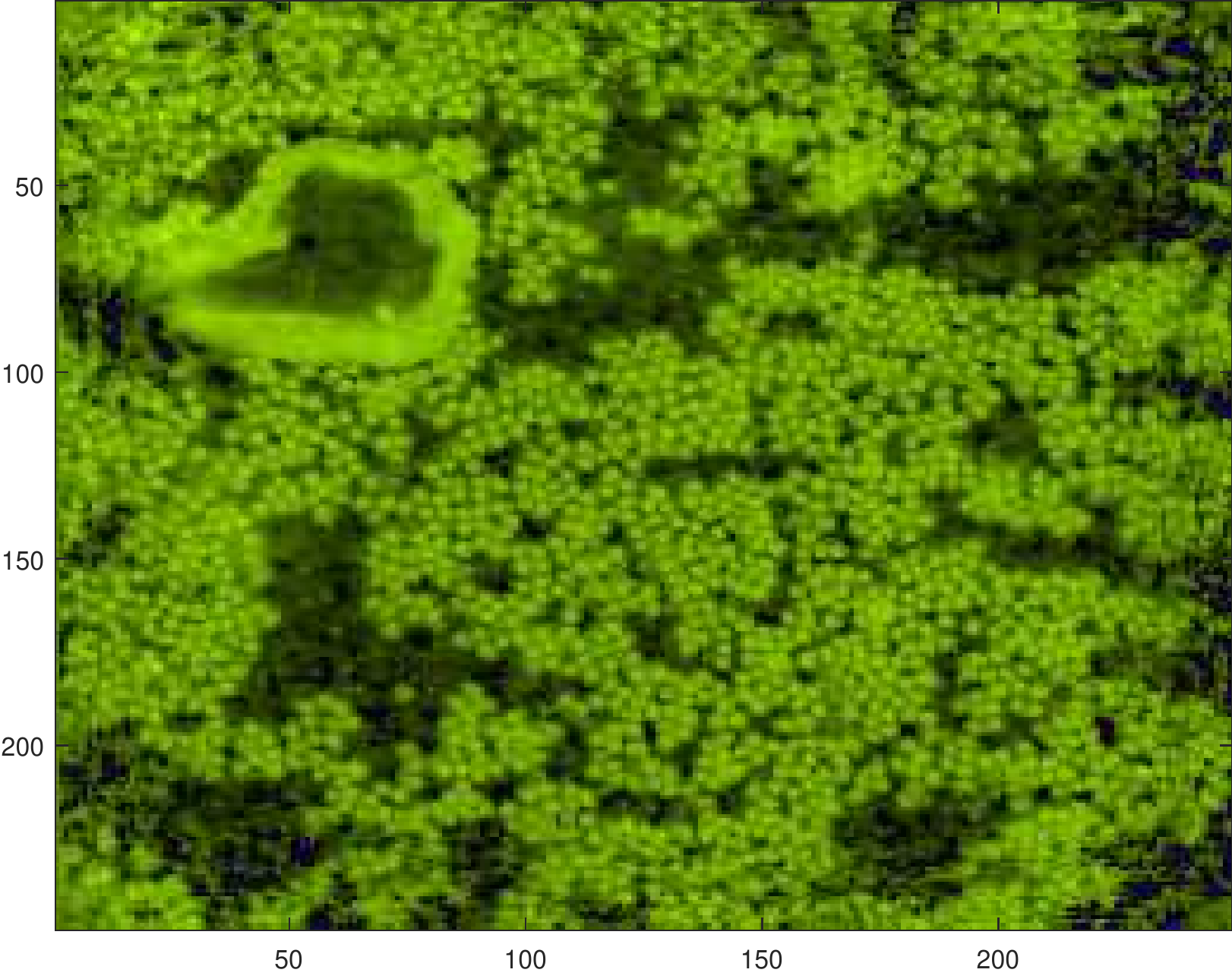}}\\
(a) & (b)\\
{\includegraphics[width=.43\columnwidth,trim={.2cm .2cm 0cm 0cm},clip]{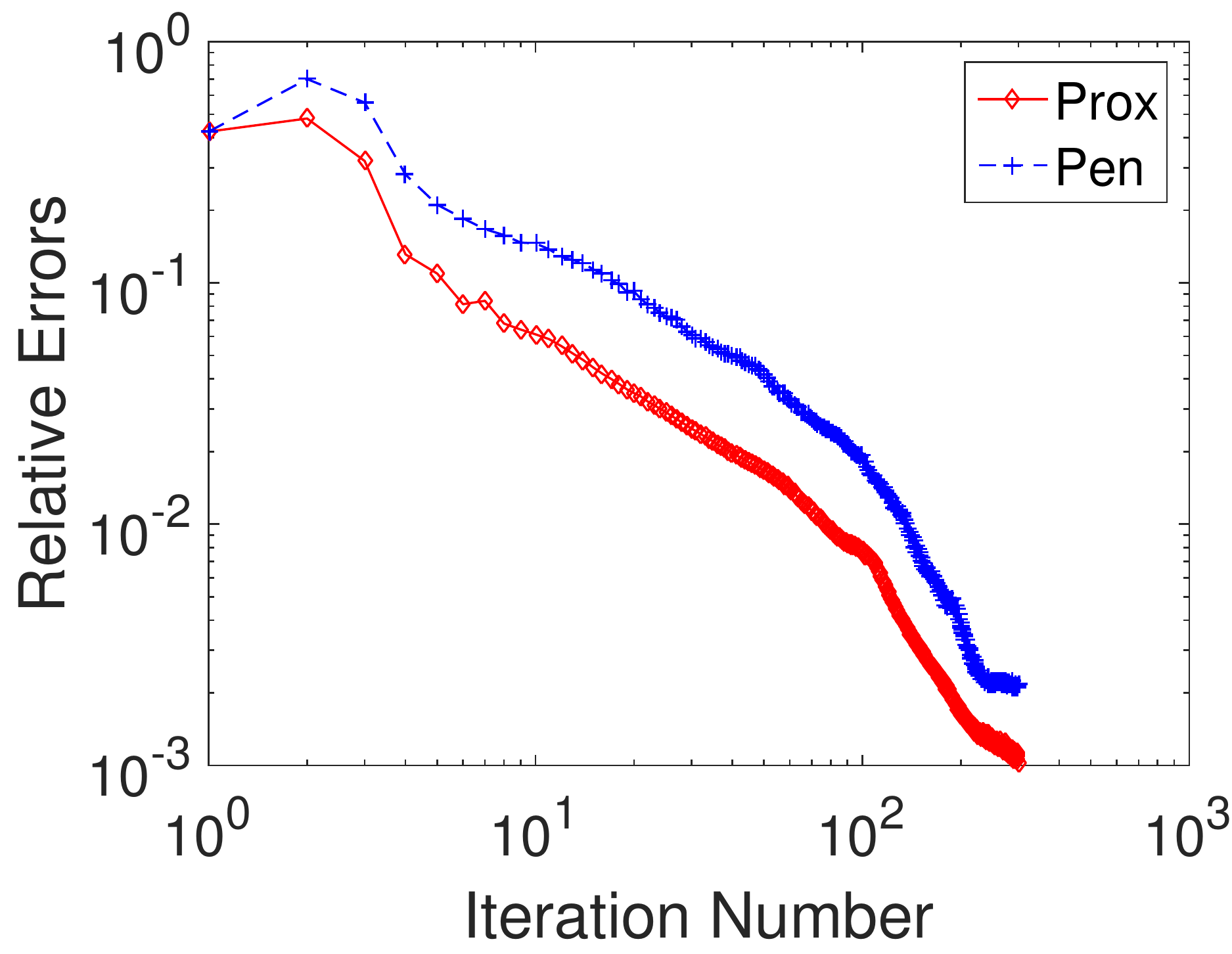}}
&
{\includegraphics[width=.43\columnwidth,trim={.2cm .2cm 0cm 0cm},clip]{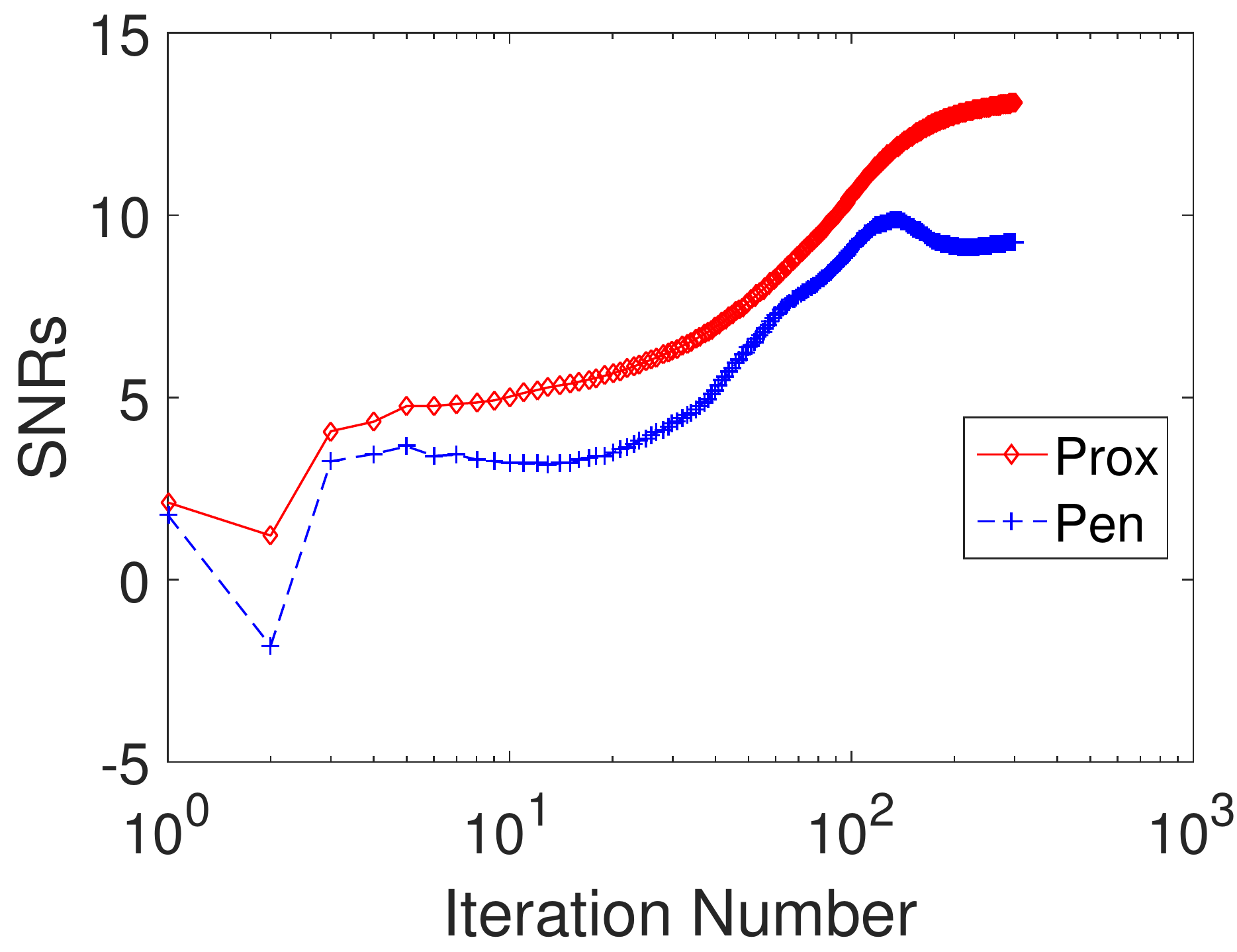}}\\
(c) & (d)\\
\end{tabular}
\end{center}
\caption{GDP-ADMM variant with penalization (``Pen'', (a)), versus 
\emph{GDP-ADMM} baseline (``Prox'', (b)). The relative
  errors and SNRs are reported in (c) and (d), respectively.
  $\mathrm{Dist}=8$, $\bm \sigma=(4,4)$, beam width $D=7$.}
\label{fig8}
\end{figure}

In the following experiment we show the effect of the proximal term.
We conduct tests comparing \emph{GDP-ADMM} baseline with a variant that replaces
 the proximal term with a penalization term $\tfrac{\tau}{2}\|u\|^2$, i.e.
\[
u^{k+1}=(N_2^{k}+\tfrac{\tau}{r}\mathbf{I})^{-1}b^{k},
\]
with identity matrix ${\mathbf I}$ and $\tau=1\times
10^{-5}\times\|\mathrm{diag}(N_2^{k})\|_\infty \times r$ when solving
Eq. \eqref{eqH}, Appendix \ref{apdx-2}. See the results in Figure
\ref{fig8}. One can readily observe that with the help of the proximal
term, the recovered images have cleaner boundaries (Figure \ref{fig8}
(b)) and higher SNR values (Figure \ref{fig8} (d)). Moreover,
\emph{GDP-ADMM} is speeded up when using the proximal term, as shown
by the convergence of the relative errors in Figure \ref{fig8} (c).

\begin{figure}
\begin{center}
\begin{tabular}{cccc}
\includegraphics[width=.22\columnwidth,height=.22\columnwidth,trim={.6cm .4cm 0cm 0cm},clip]{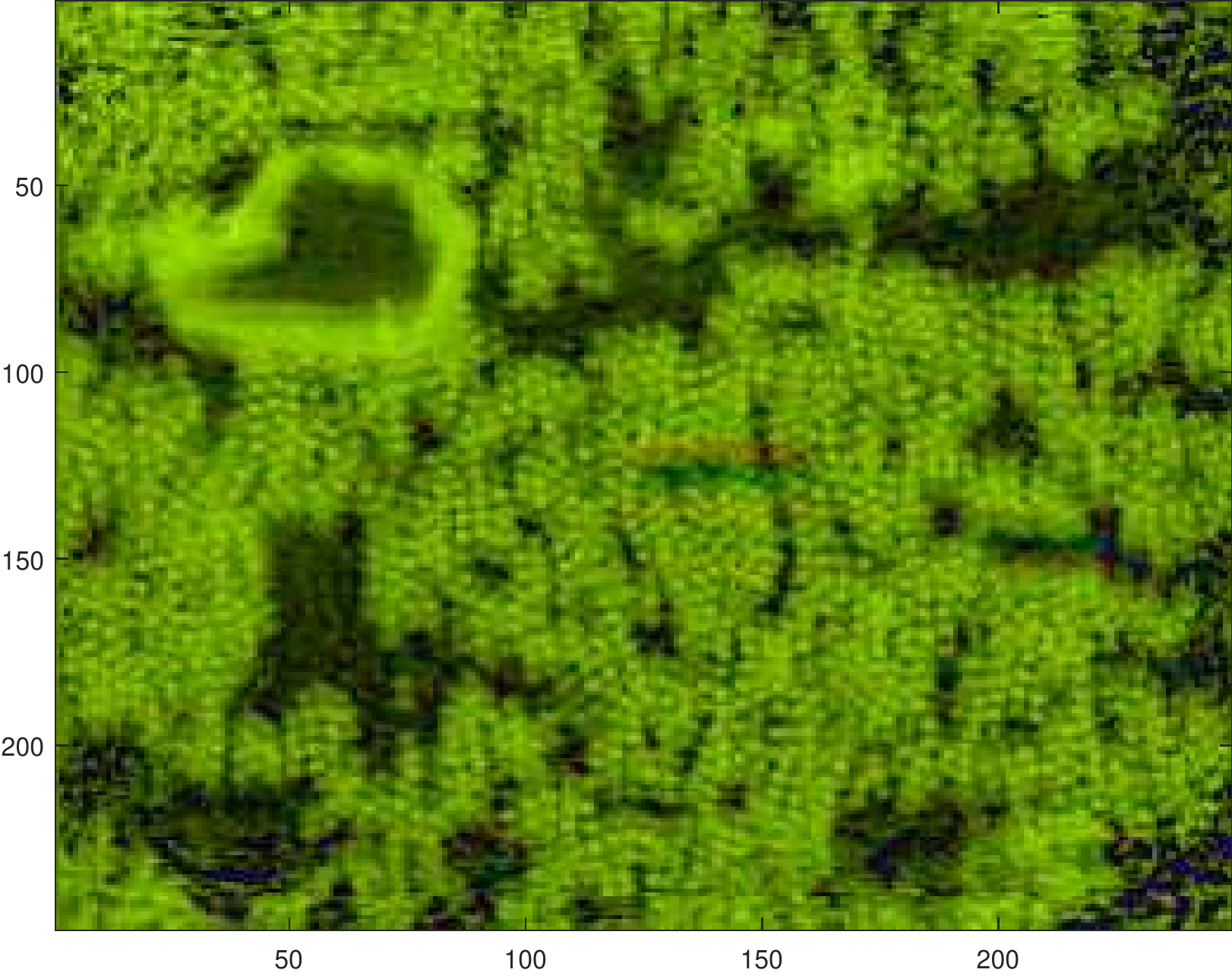}&
\includegraphics[width=.22\columnwidth,height=.22\columnwidth,trim={.6cm .4cm 0cm 0cm},clip]{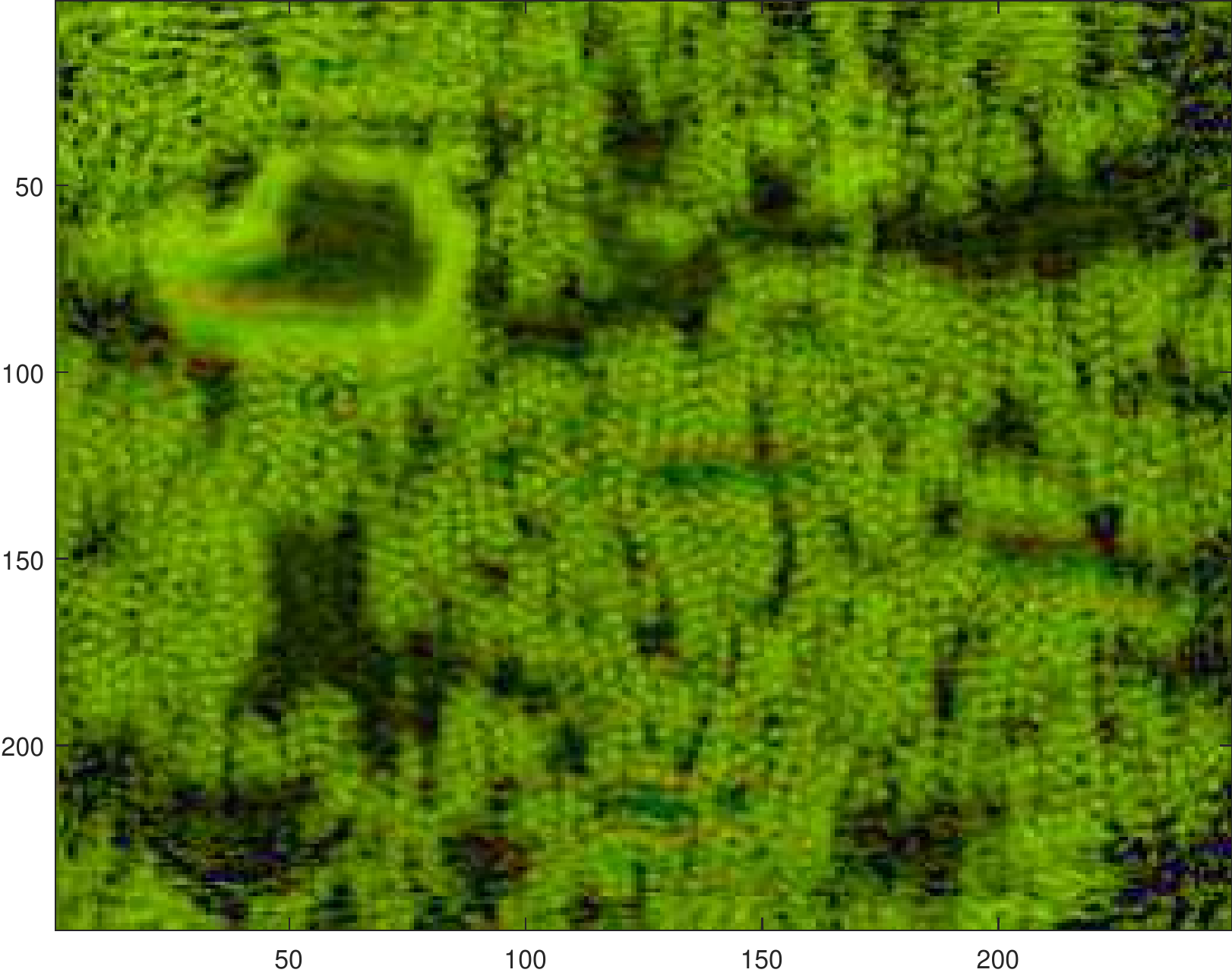}&
\includegraphics[width=.22\columnwidth,height=.22\columnwidth,trim={.6cm .4cm 0cm 0cm},clip]{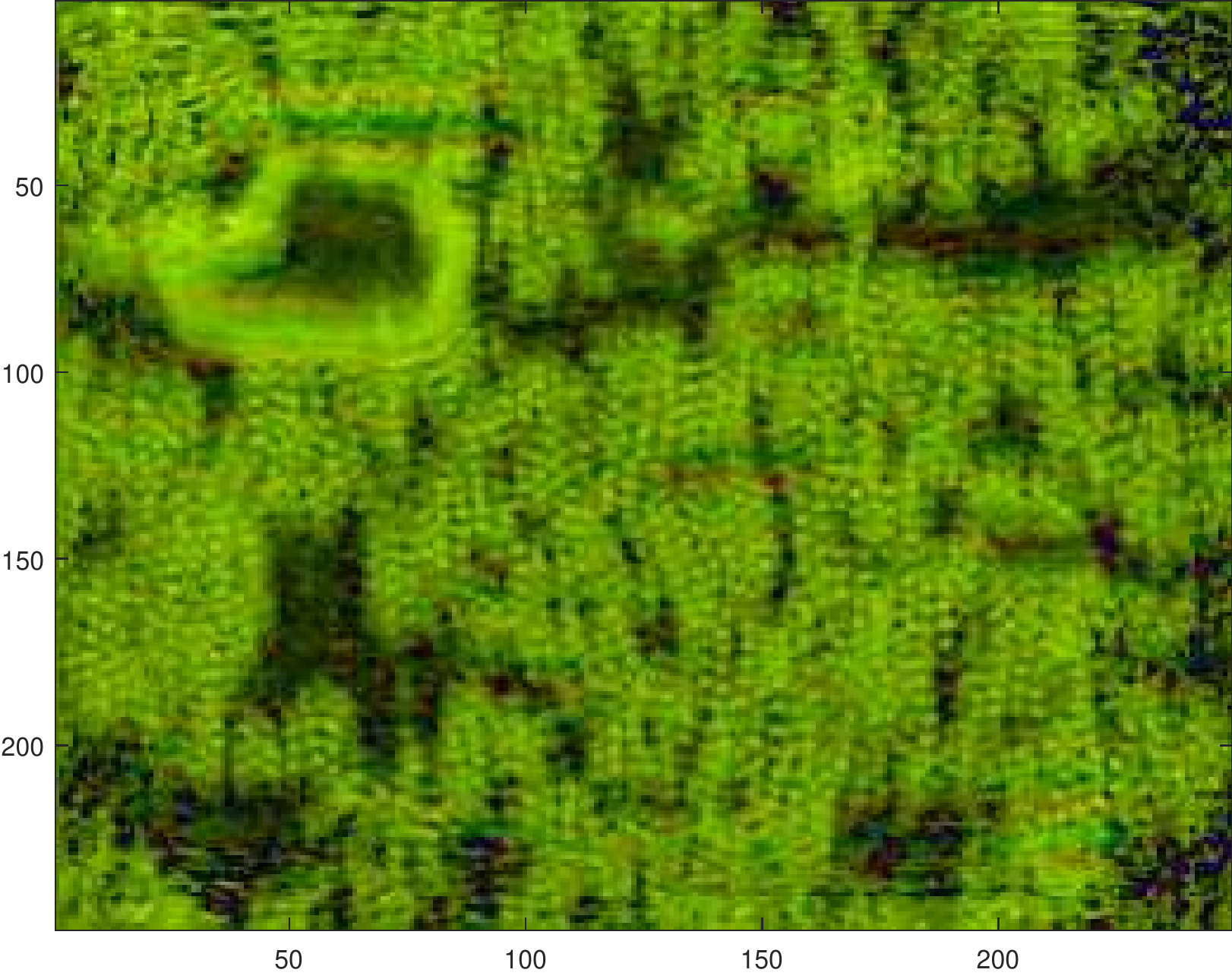}&
\includegraphics[width=.22\columnwidth,height=.22\columnwidth,trim={.6cm .4cm 0cm 0cm},clip]{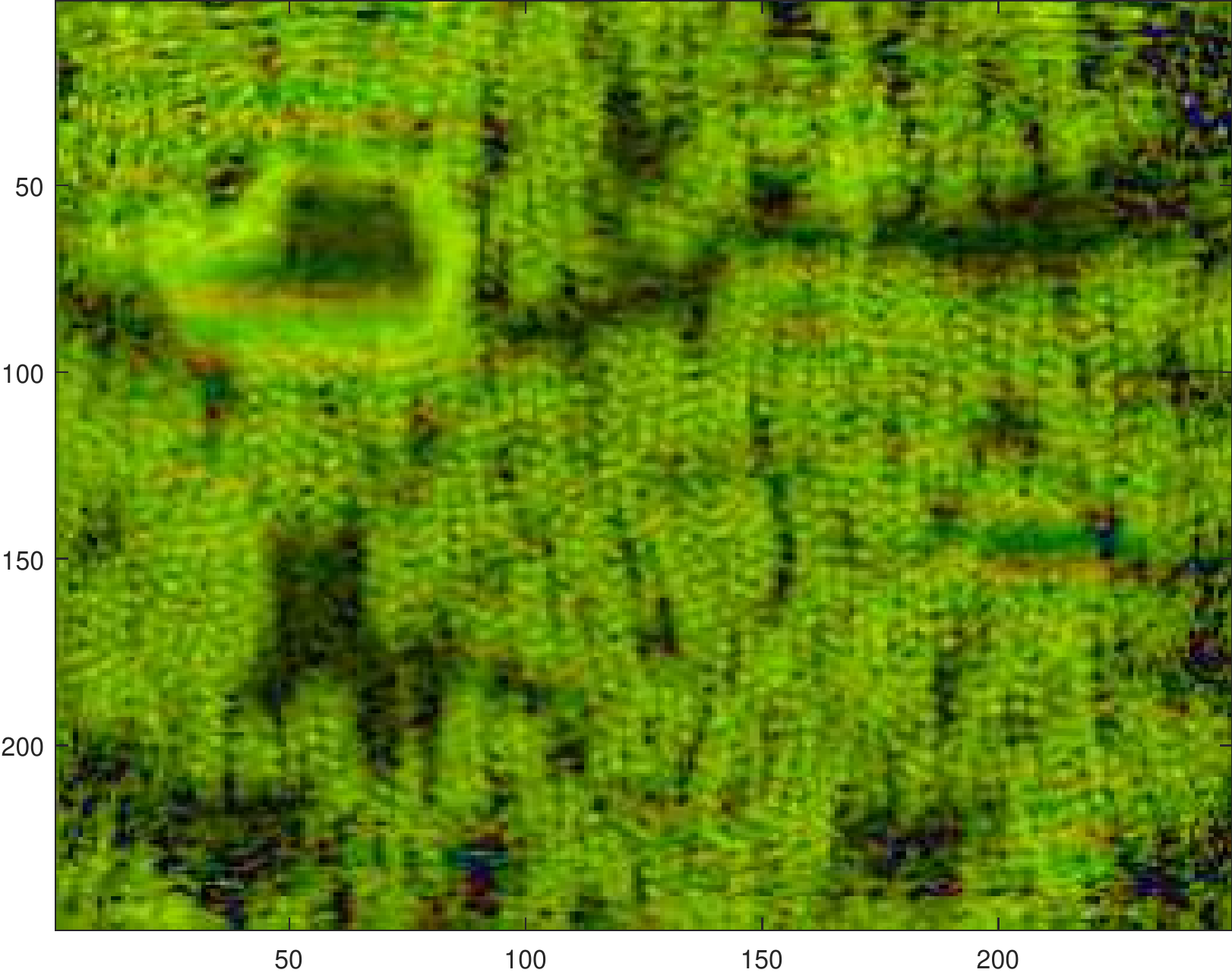}
\\
\multicolumn{4}{c}{\tiny{\emph{FC-ADMM}}}\\
\\
\includegraphics[width=.22\columnwidth,height=.22\columnwidth,trim={.6cm .4cm 0cm 0cm},clip]{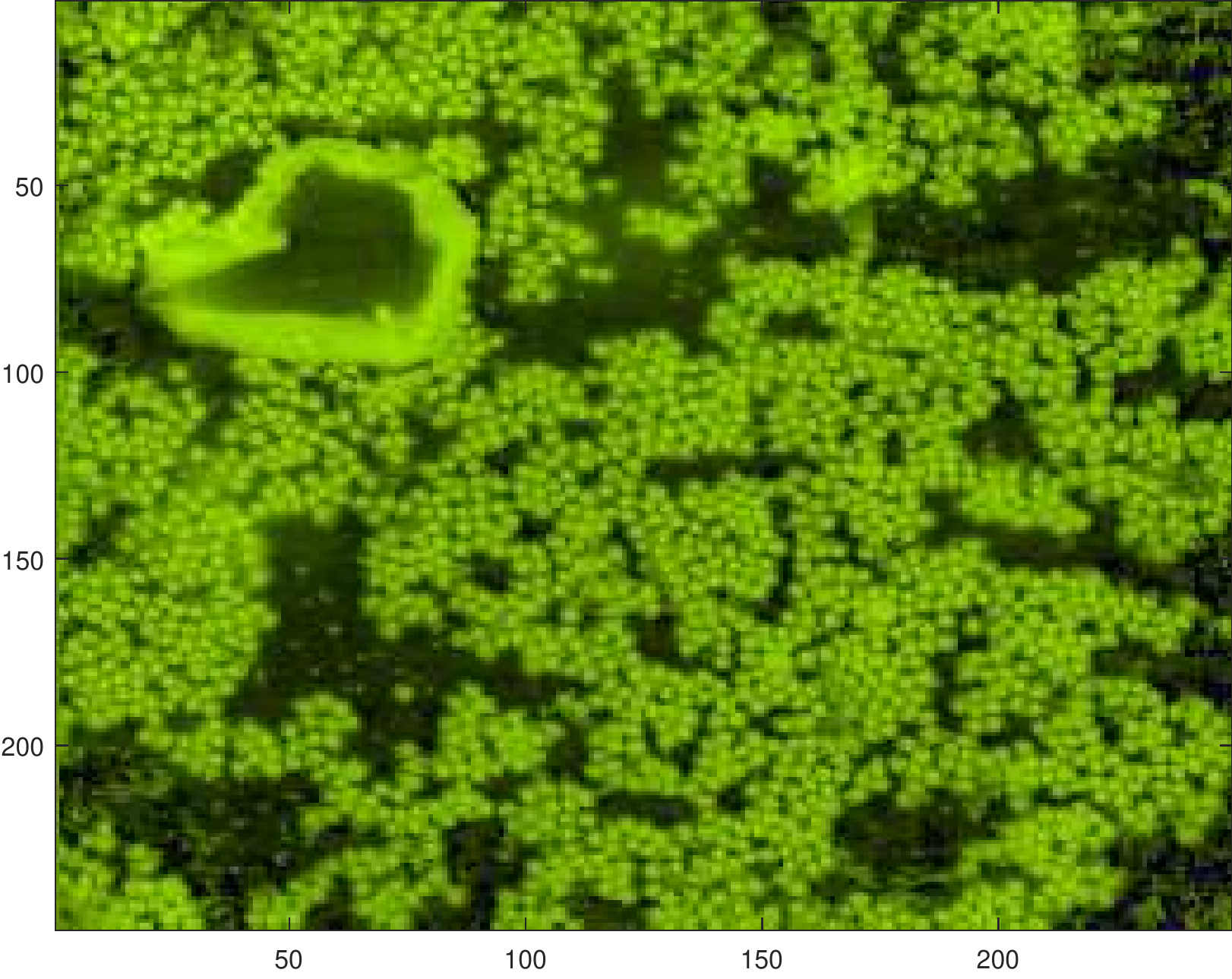}&
\includegraphics[width=.22\columnwidth,height=.22\columnwidth,trim={.6cm .4cm 0cm 0cm},clip]{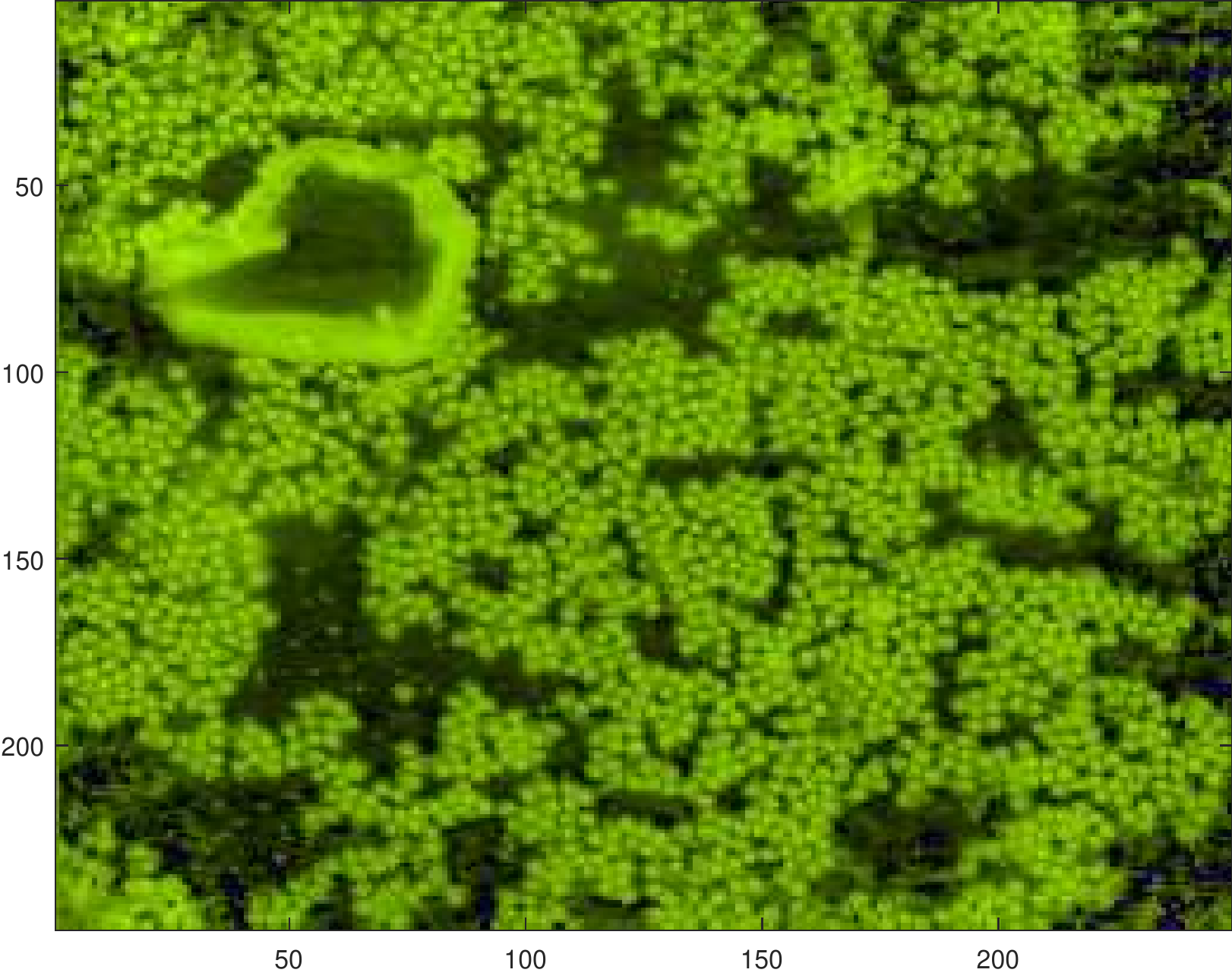}&
\includegraphics[width=.22\columnwidth,height=.22\columnwidth,trim={.6cm .4cm 0cm 0cm},clip]{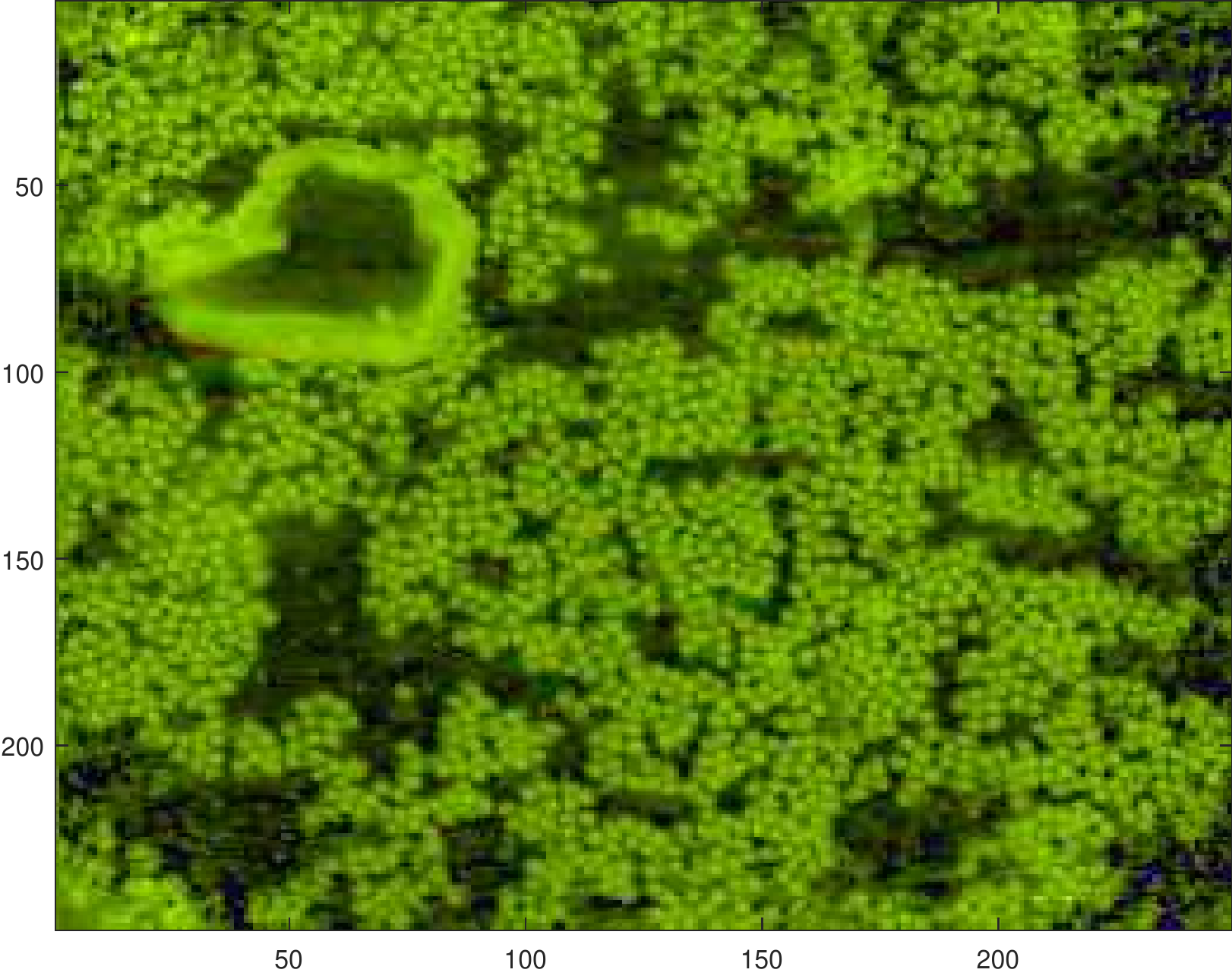}&
\includegraphics[width=.22\columnwidth,height=.22\columnwidth,trim={.6cm .4cm 0cm 0cm},clip]{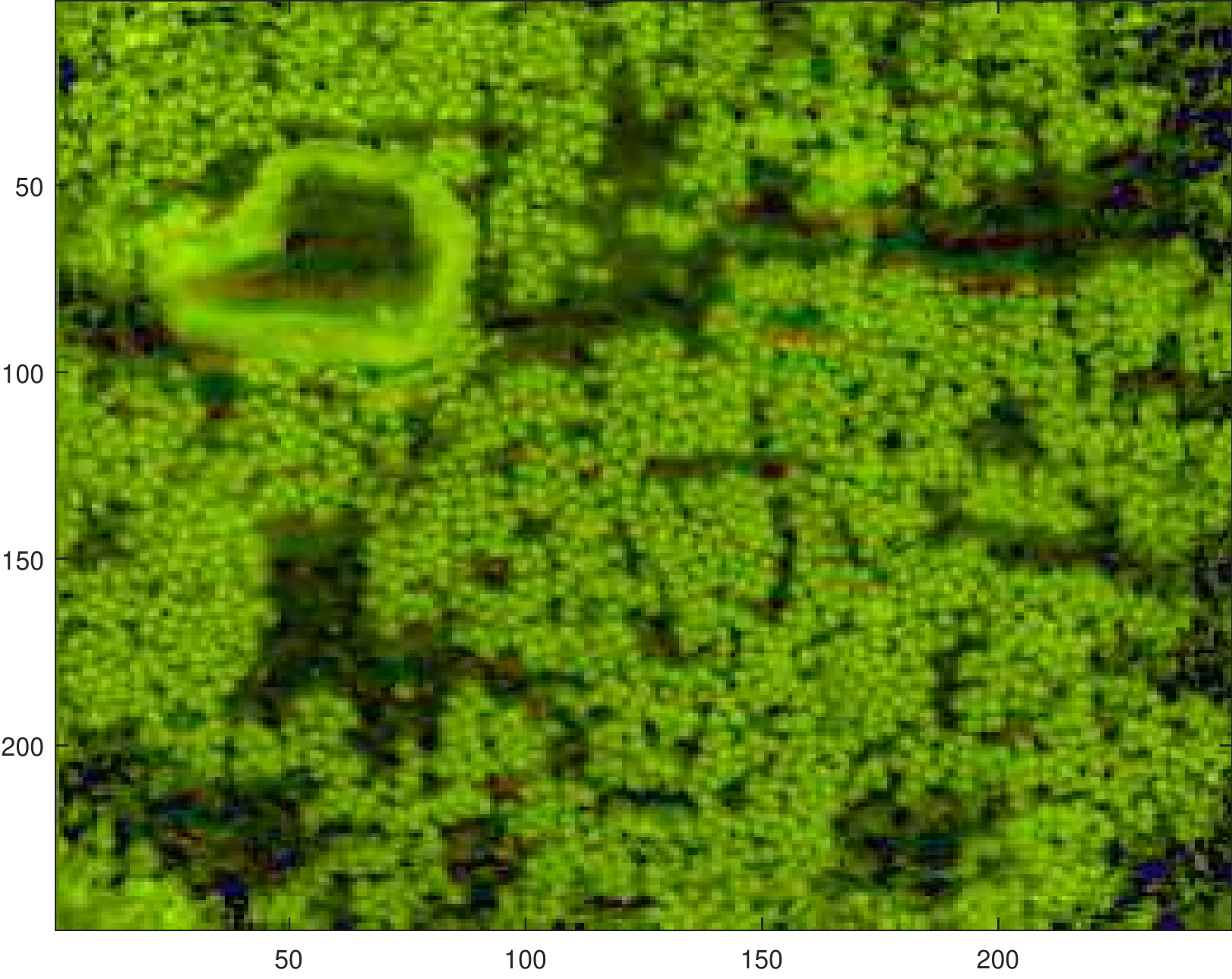}\\
\multicolumn{4}{c}{\tiny{\emph{GDP-ADMM}}}\\
\tiny{$d=11$}& \tiny{13}  & \tiny{15} & \tiny{17 }
\end{tabular}
\end{center}
\caption{Reconstructed images by \emph{FC-ADMM} (top row) and
  \emph{GDP-ADMM} (bottom) for {\textit{fly-scan}} ptychography with
  varying kernel width $d$, beam width $D=7$, sliding distance $\mathrm{Dist}=8$. }
\label{fig5}
\end{figure}

\paragraph{Different kernels.}
Previous experiments reported in this section are based on the
Gaussian kernel function.  Here we assess the
performance of \emph{GDP-ADMM} when using a binary kernel function for
{\textit{fly-scan}} ptychography \cite{deng2015continuous}.  Figure
\ref{fig5} shows the reconstructed images obtained by varying the
kernel size $d$.  Similarly to the case of Gaussian kernel
function, \emph{GDP-ADMM} displays a significant increase in visual
quality. In the first row of Figure \ref{fig5}, the results produced
by \emph{FC-ADMM} are completely blurry. On the other hand,
\emph{GDP-ADMM} (second row of Figure \ref{fig5}) achieves much
sharper overall results.  One can also see the obvious decrease of the
residual and increase of SNRs by \emph{GDP-ADMM} in Table \ref{tab2}.

Similar improvements by \emph{GDP-ADMM} can be obtained with other
motion blur type kernel functions, however, due to page limitations,
we do not provide further results. These results show that
\emph{GDP-ADMM} can be applied to partial coherence problems with more
general kernel function.

\begin{table}
\caption{{\textit{Fly-scan}}: Performance of \emph{GDP-ADMM}  and \emph{FC-ADMM}. 
Here $d$ is the size of the binary kernel, and the beam width is $D=7$ pixels. }
\begin{center}
\scalebox{.9}{
\begin{tabular}{|c||cccc|}
\cline{1-5} 
$d$ &   11& 13 & 15 & 17 \\
\cline{1-5} 
\cline{1-5} 
$e_{fc}$& 2.00E-1&2.16E-1&2.29E-1&2.37E-1\\
\cline{1-5} 
$e_{pc}$&4.47E-2&6.46E-2&9.04E-2&1.16E-1\\
\cline{1-5} 
\cline{1-5} 
$\mathrm{SNR}_{fc}$&6.19&5.60&7.15&6.71\\
\cline{1-5} 
$\mathrm{SNR}_{pc}$&21.15&18.07&14.63&11.58\\
\cline{1-5} 
\end{tabular}
}
\end{center}
\label{tab2}
\end{table}

\paragraph{Runtime and memory performance}
Finally we report the the computational performance 
 of the \emph{GDP-ADMM}
algorithm on a machine with an Intel i7-5600U CPU and 16G RAM using
\textsc{Matlab}. \emph{GDP-ADMM} requires  two additional modes and four additional
variables $z_{1}, z_{2}$ and $\Lambda_1, \Lambda_2$ compared with \emph{FC-ADMM} . Because of this, the memory cost and runtime of \emph{GDP-ADMM} are in theory about three
times as large as those of \emph{FC-ADMM} per iteration.
When computing the image in Figure \ref{fig2},  \emph{GDP-ADMM} requires  an average of 873MB of RAM and takes 655 seconds to compute, whereas \emph{FC-ADMM} requires 344MB ad takes 218 seconds. These results are  consistent with the previous
theoretical estimate. 

\begin{figure}
\begin{center}
\includegraphics[width=.6\columnwidth]{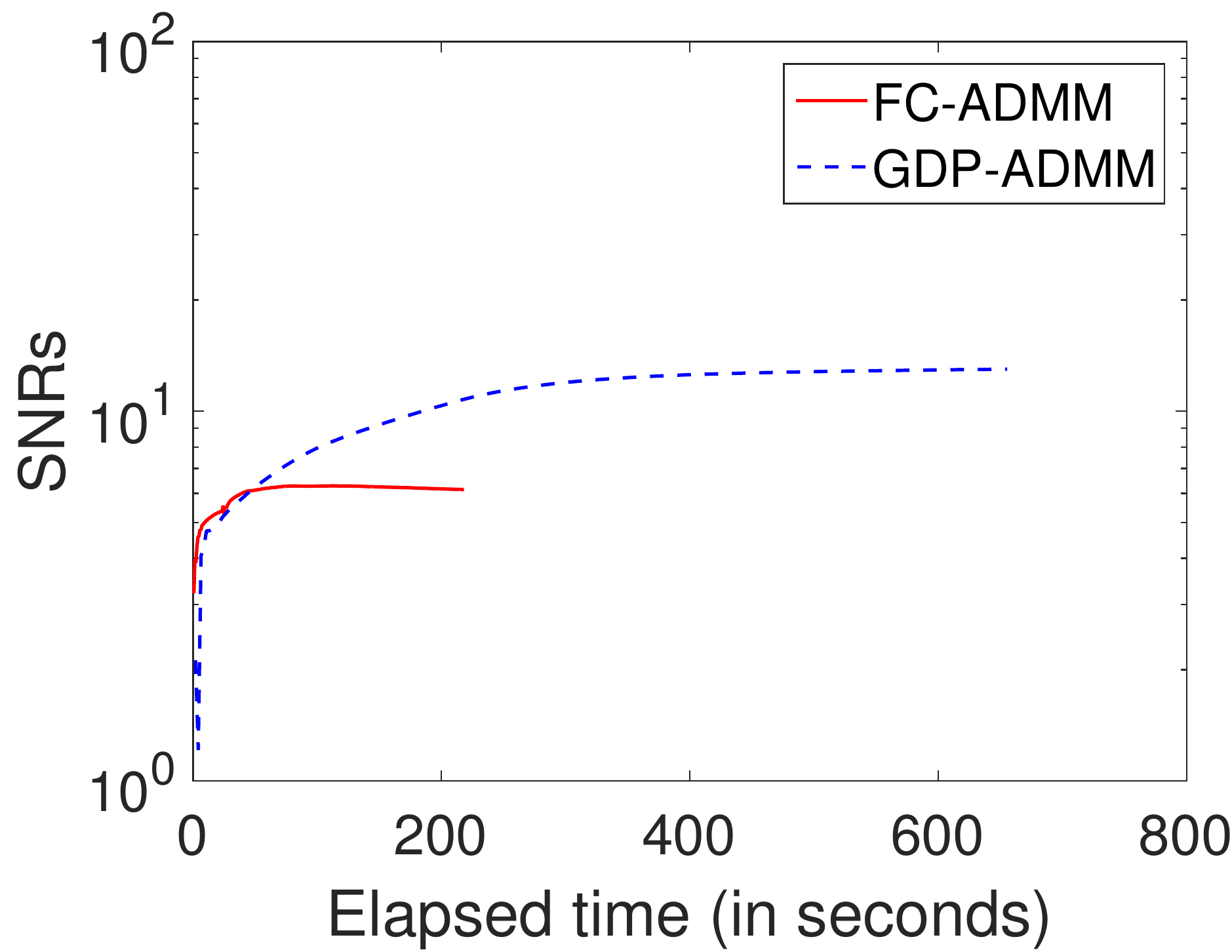}
\end{center}
\caption{ Histories of the SNRs changes for \emph{FC-ADMM} and \emph{GDP-ADMM}  with respect to the elapsed time in seconds. $\sigma_1=\sigma_2=4$, Dist=8.  }
\label{fig9}
\end{figure}

We further investigate the change histories of
the SNRs for \emph{FC-ADMM} and \emph{GDP-ADMM} with respect to
elapsed the time, and report the results in Figure \ref{fig9}. { The SNR histories show
  that,  \emph{FC-ADMM} can recover
  better images in the first 50 seconds,} but \emph{GDP-ADMM} improves the image further after  that. Hence, in order to accelerate the
\emph{GDP-ADMM} algorithm, we could use the iterative solution of
\emph{FC-ADMM} as the initialization for \emph{GDP-ADMM}. We also
emphasize that the runtime and memory requirements \emph{GDP-ADMM} are insensitive to the variances and the support sizes of the kernel functions.  It is important to note that, if we solve the problem directly following
\cite{marchesini2013augmented}, the probe and the weights in
\eqref{modelMX} in the case of the setting in Figure \ref{fig3} (a)-(b),
at most $(4\sigma_1+1)\times (4\sigma_1+1)=441$ translated probes for
$\sigma_1=\sigma_2=5$ should be introduced, which requires much more
memory and computation costs.

\section{Conclusions}

\label{sec5}
In this paper, we propose  Gradient Decomposition of
the Probe (\emph{GDP}), an efficient model that exploits translational
kernel separability. The \emph{GDP} model increases the approximation
accuracy compared with the coherent model, and it holds for a general
partial coherent source. We derive an optimization model coupling the
variances of the kernel with the transverse coherence widths, and a
fast and memory efficient proximal first order \emph{GDP-ADMM} algorithm to
solve the nonlinear optimization problem based on \emph{GDP}.  Numerical
experiments demonstrate the effectiveness of such approximation in the
case of Gaussian kernel function, and binary functions in fly-scanning
schemes, showing that the proposed methods are suitable for 
general partially coherent sources.

 However the results by
\emph{GDP-ADMM} seem blurry when inferred from the fourth column of Figure
\ref{fig3}, and further improvement may be achieved by considering sparse
prior, or using piecewise Taylor expansion series or other
  integration schemes. This will be subject of future work.
Additional research directions to extend this work are: (i) consider partial coherence with
framewise different variances $\bm \sigma_j$ when using scan geometries on irregular
grids or spirals, (ii) combine translational blur at the sample with detector blur in          Eq.  \eqref{modelPC} for near field ptychography.
(ii)   incorporate longitudinal coherence when using
broadband illumination with chromatic
aberrations of the probe, dispersion of the diffraction pattern \cite{enders2014ptychography}, and spectral Kramers-Kronig dispersion across resonances
\cite{hirose2017use}.



\section{Acknowledgments}

 This work was partially funded by the Center for Applied Mathematics for
Energy Research Applications, a joint ASCR-BES funded project within the
Office of Science, US Department of Energy, under contract number
DOE-DE-AC03-76SF00098, and by the Advanced Light Source, which is a DOE Office of Science User Facility under contract no. DE-AC02-05CH11231.

%



\bibliographystyle{siam}


\bibliography{rD}

\appendix

\section{\emph{GDP-ADMM}}\label{apdx-2}

 Once the variables $\omega, u, z, \Lambda, \sigma$ are set,
in the $(k+1)^{th}$ iteration the 
augmented Lagrangian is modified by adding a proximal term as follows:
$$
\tilde{ \mathcal L}^{k}_{r}(\tilde\omega,u, z,\bm \sigma ,\Lambda)=\mathcal L_{r}(\tilde\omega,u,z,\bm \sigma ,\Lambda)+\tfrac{\tau}{2}\|u-u^{k}\|^2_{M^{k}}\,,
$$
where for simplicity we removed superscripts and subscripts, and
$\|u\|_{M^{k}}^2=\langle M^{k} u, u\rangle$ is a positive
semi-definite matrix $M^{k},$ defined by the approximated solution
$\left(\tilde\omega^{k+1}, u^{k}, \{z_l^{k}\}, \{\sigma_l^{k}\},
  \{\Lambda_{l}^{k}\}\right)$ in the $k^{th}$ iteration. We remark
that with the help of an additional proximal term
$\tfrac{\tau}{2}\|u-u^{k}\|^2_{M^{k}}$, the subproblem in Step 2 
admits a unique solution. Below we demonstrate how to solve each
subproblem. 
\begin{equation}\label{eqADMM-S-PC}
\begin{aligned}
&\mbox{Step 1:~} \tilde\omega^{k+1}=\arg\min\limits_{\tilde\omega} \mathcal L_{r}(\tilde\omega,u^{k},\{z^{k}_{l}\},\bm \sigma^{k},\{\Lambda^{k}_{l}\});\\
&\mbox{Step 2:~} u^{k+1}=\arg\min\limits_u \tilde{ \mathcal L}_{r}^{k}\left (\tilde\omega^{k+1},u,\{z^{k}_{l}\},\bm \sigma^{k},\{\Lambda^{k}_{l}\} \right );\\
&\mbox{Step 4:~} \{z_{l}^{k+1}\}_{l=0}^2=\arg\min\limits_{\{z_l\}} \mathcal L_{r}(\tilde\omega^{k+1},u^{k+1},\{z_{l}\},\bm \sigma^{k},\{\Lambda^{k}_{l}\});\\
&\mbox{Step 5:~} \bm \sigma^{k+1}=\arg\min_{\bm \sigma} \mathcal L_{r} (\tilde\omega^{k+1},u^{k+1},\{z^{k+1}_{l}\},\bm \sigma,\{\Lambda^{k}_{l}\});\\
&\mbox{Step 6:~}\Lambda_{0}^{k+1}=\Lambda_{0}^{k}+(z_{0}^{k+1}-\mathcal A(\tilde\omega^{k+1},u^{k+1}));\\
&\mbox{~~~~~~~\quad}\Lambda_{1}^{k+1}=\Lambda_{1}^{k}+(z_{1}^{k+1}-\mathcal A(\sigma^{k+1}_1\nabla_1\tilde\omega^{k+1},u^{k+1}));\\
&\mbox{~~~~~~~\quad}\Lambda_{2}^{k+1}=\Lambda_{2}^{k}+(z_{2}^{k+1}-\mathcal A(\sigma^{k+1}_2\nabla_2\tilde\omega^{k+1},u^{k+1}));
\end{aligned}
\end{equation}

Step 1 involves finding the first order stationary point $\omega^{k+1}$
of the augmented Lagrangian which has the following form:
\begin{equation}
\begin{split}\label{eq:K}
 N_{1}^{k} \tilde \omega^{k+1}-c^{k}=0,
\end{split}
\end{equation}
with symmetric sparse matrix $N_1^k$ defined as:
\begin{equation}
\label{eqK}
\begin{split}
N_1^k:=&\mathrm{diag}\left (\sum\limits_j \left |\mathcal S_j u^{k} \right |^2\right )\\
&+\sum\limits_{l=1}^2 (\sigma^{k}_l)^2\nabla_l^T \left (\mathrm{diag}\left(\sum\limits_j|\mathcal S_j u^{k}|^2\right)\nabla_l \right ),\\
c^k:=&{\sum\limits_j\left(\mathcal S_j (u^{k})^*\circ\bar z^{k}_{0,j}+\sum\limits_{l=1}^2\sigma^{k}_l\nabla_l^T\left(\mathcal S_j(u^{k})^*\circ \bar z^{k}_{l,j}\right) \right),}
\end{split}
\end{equation}
and
$\bar z^{k}_{l,j}:= \mathcal F^*(z^{k}_{l,j}+\Lambda^{k}_{l,j}).$
In this section, the gradient operator is given in a discrete setting
for simplicity, where $\nabla_1, \nabla_2$ represent the forward {
  finite} difference operator with respect to $x,y$ directions.

Step 2 is similar to Step 1, the closed form solution can be expressed as:
\begin{equation}\label{eq:H}
\left ( N^k_2 +\tfrac{\tau}{r} M^{k} \right )u^{k+1}-\left (b^k +\tfrac{\tau}{r}M^{k} u^{k} \right )=0,
 \end{equation}
 with diagonal matrix $N_2^k$:
\begin{equation}\label{eqH}
\begin{split}
N_2^{k}:=&\mathrm{diag}\left (\sum\limits_j \mathcal S_j^T\left( |\tilde\omega^{k+1}|^2+\sum\limits_{l=1}^2 \sigma^2_l|\nabla_l \tilde\omega^{k+1}|^2\right)\right ),\\
b^{k}:=&\sum\nolimits_j\mathcal S_j^T\left((\tilde\omega^{k+1})^*\circ \bar z^{k}_{0,j}+\sum\limits_{l=1}^2 \sigma^{k}_l \nabla_l(\tilde\omega^{k+1})^*\circ \bar z^{k}_{l,j} \right).
\end{split}
\end{equation}
Empirically, in order to avoid divisions by 0 when solving
\eqref{eq:H}, set the matrix $M^{k}$ to a diagonal matrix with
diagonal elements:
\begin{equation}
\label{eq:MK}
M^{k}(l,l):=\left\{
\begin{split}
&0,\qquad\qquad\qquad\, \text{if~} N_2^{k}(l,l)\geq \tfrac{1}{10}\| \mathrm{diag}(N_2^{k})\|_\infty; \\
&\|\mathrm{diag}(N_2^{k})\|_\infty-\tfrac{r}{\tau} N_2^{k}(l,l),\, \text{otherwise.}
\end{split}
\right.
\end{equation}

Step 3 can be expressed as the following problem:
\[
\begin{split}
\min_{\{z_{l,j}\}_{0\leq l\leq 2}^{0\leq j\leq J-1}}&\sum\nolimits_j\tfrac12\left\|\sqrt{f_{pc,j}}- \sqrt{\sum\nolimits_l| z_{l,j}|^2}\right\|^2\\
&+\tfrac{r}{2}  (\|z_0+\Lambda_0-\mathcal A(\tilde\omega,u)\|^2\\
 &+\|z_1+\Lambda_1-\mathcal A(\sigma_1\nabla_1\tilde\omega,u)\|^2\\
 &+\|z_2+\Lambda_2-\mathcal A(\sigma_2\nabla_2\tilde\omega,u)\|^2),
\end{split}
\]
which can be solved  independently with respect to each frame, hence one needs to consider:
\[\begin{split}
\min_{\{z_{l,j}\}_{l=0}^{2}} &\tfrac12\left\|\sqrt{f_{pc,j}}- \sqrt{\sum\nolimits_l| z_{l,j}|^2}\right\|^2\\
+&\tfrac{r}{2}\sum\nolimits_{l}\|z_{l,j}-\tilde z_{l,j}\|^2~\forall 0\leq j\leq J-1,
\end{split}
\]
where $\tilde z_{0,j}=\mathcal A_j(\tilde\omega,u)-\Lambda_{0,j},
\tilde z_{1,j}=\sigma_1\mathcal
A_j(\nabla_1\tilde\omega,u)-\Lambda_{1,j},\tilde
z_{2,j}=\sigma_2\mathcal A_j(\nabla_2\tilde\omega,u)-\Lambda_{2,j}.$
Similarly, it also has a closed form solution, see Eq. \eqref{eqSubZ}, Section \ref{sec3}. 

Step 4 requires solving a linear least square problem whose closed form solution is:
\[
\begin{split}
&\sigma_l=\frac{\Re(\langle z_l+\Lambda_l,\mathcal A(\nabla_l \tilde\omega,u) \rangle)}{\|\mathcal A(\nabla_l \tilde\omega,u)\|^2}.
\end{split}
\]




\section{{Derivation of \eqref{eqM-1}}}\label{apdx-1}

\begin{widetext}

\begin{equation}
{
\begin{split}
\!f_{pc,j}(q)\!= &\! \int\!\! \left | \mathcal F_{x\rightarrow q}\left (\mathcal S_j u(x)\left ( \omega(x)-y^T\nabla \omega(x)+\tfrac{1}{2}y^T\nabla^2\omega(x)y \right )\right )\right |^2 \kappa(y)\mathrm{d}y+\mathcal O\left (\int |y|^3\kappa(y)\mathrm{d}y\right)\\
=&\!\int |\mathcal F_{x\rightarrow q}(\mathcal S_j u(x)( \omega(x)\!+\tfrac{1}{2}y^T\nabla^2\omega(x)y))|^2 \kappa(y)\mathrm{d}y\!+\!\int |\mathcal F_{x\rightarrow q}(\mathcal S_j u(x)y^T\nabla \omega(x))|^2 \kappa(y)\mathrm{d}y\! \!\\
&+\mathcal O\left (\int |y|^3\kappa(y)\mathrm{d}y\right )\\
\!\!&\!=\!\int\! |\mathcal A_j(\omega,u)+\tfrac12y^T\mathcal A_j(\nabla^2\omega,u)y|^2 \kappa(y)\mathrm{d}y\!+\int \!\! |y^T\mathcal A_j(\nabla \omega,u)|^2 \kappa(y)\mathrm{d}y \!+\!\mathcal O(\int |y|^3\kappa(y)\mathrm{d}y)\!\!\\
=& |\mathcal A_j(\omega,u)|^2+2\Re\left(\mathcal A^*_j(\omega,u)\left(\tfrac12\sigma_1^2\mathcal A_j(\nabla_{11}\omega,u)+\tfrac12\sigma_2^2\mathcal A_j(\nabla_{22}\omega,u)+\sigma_{12}\mathcal A_j(\nabla_{12}\omega,u)\right)\right)\\
&+\mathcal O\left (\int |y|^4 \kappa(y)\mathrm{d}y\right)+\sigma_1^2|\mathcal A_j(\nabla_1 \omega, u)|^2+
\sigma_2^2|\mathcal A_j(\nabla_2 \omega, u)|^2+\mathcal O\left (\int |y|^3\kappa(y)\mathrm{d}y\right )\\
=&|\mathcal A_j(\omega,u)|^2\!+\!2\Re\left(\mathcal A^*_j(\omega,u)\mathcal A_j(\tfrac{1}{2}(\sigma_1^2\nabla_{11}\omega\!+\sigma_2^2\nabla_{22}\omega\!+2\sigma_{12}\nabla_{12}\omega),u)\right)\!\!\\
&+\sigma_1^2|\mathcal A_j(\nabla_1 \omega), u)|^2+
\sigma_2^2|\mathcal A_j(\nabla_2 \omega, u)|^2+\mathcal O\left (\int |y|^4 \kappa(y)\mathrm{d}y\right )+\mathcal O\left (\int |y|^3\kappa(y)\mathrm{d}y\right )\\
=&|\mathcal A_j(\omega+\tfrac{1}{2}(\sigma_1^2\nabla_{11}\omega+\sigma_2^2\nabla_{22}\omega+2\sigma_{12}\nabla_{12}\omega),u)|^2\\
&+\sigma_1^2|\mathcal A_j(\nabla_1 \omega, u)|^2
+\sigma_2^2|\mathcal A_j(\nabla_2 \omega, u)|^2+
\mathcal O(\int |y|^3\kappa(y)\mathrm{d}y),
\end{split}
}
\label{eqM-1-1}
\end{equation}

\noindent
where the first equality is based on:
\[
\begin{split}
&\int \left |\mathcal F_{x\rightarrow q}(\mathcal S_j u(x)( \omega(x)+\tfrac{1}{2}y^T\nabla^2\omega(x)y))-\mathcal F_{x\rightarrow q}(\mathcal S_j u(x)y^T\nabla \omega(x)) \right |^2 \kappa(y)\mathrm{d}y\\
=&\int \left( \left |\mathcal F_{x\rightarrow q}(\mathcal S_j u(x)( \omega(x)+\tfrac{1}{2}y^T\nabla^2\omega(x)y))\right |^2+ \left |\mathcal F_{x\rightarrow q}(\mathcal S_j u(x)y^T\nabla \omega(x)) \right |^2 \right )\kappa(y)\mathrm{d}y\\
&-2 \int_y \Re \left ( (\mathcal F_{x\rightarrow q}(\mathcal S_j u(x)( \omega(x)+\tfrac{1}{2}y^T\nabla^2\omega(x)y)))^* \mathcal F_{x\rightarrow q}(\mathcal S_j u(x)y^T\nabla \omega(x))\right ) \kappa(y)\mathrm{d}y)\\
=&\int \left ( |\mathcal F_{x\rightarrow q}(\mathcal S_j u(x)( \omega(x)+\tfrac{1}{2}y^T\nabla^2\omega(x)y))|^2 +|\mathcal F_{x\rightarrow q}(\mathcal S_j u(x)y^T\nabla \omega(x)) |^2 \right ) \kappa(y)\mathrm{d}y\\
&-2\Re\left (\int_y(\mathcal A_j(\omega,u)+y^T\mathcal A_j(\nabla^2 \omega,u)y)^*(y^T\mathcal A_j(\nabla \omega,u))\kappa(y)\mathrm{d}y\right )\\
=&\int |\mathcal F_{x\rightarrow q}(\mathcal S_j u(x)( \omega(x)+\tfrac{1}{2}y^T\nabla^2\omega(x)y))|^2\mathrm{d}y+|\mathcal F_{x\rightarrow q}(\mathcal S_j u(x)y^T\nabla \omega(x)) |^2 \kappa(y)\mathrm{d}y\\
&-2\Re((\mathcal A_j(\omega,u))^*\mathcal A_j(\nabla \omega,u)\overbrace{\int_y y^T\kappa(y)\mathrm{d}y}^{\mathop{=0}{ \eqref{thirdMZ}}}\\
&+\sum\limits_{1\leq i_1,i_2,i_3\leq 2}\mathcal A^*_j(\nabla_{i_1,i_2} \omega,u)\mathcal A_j(\nabla_{i_3} \omega,u)\overbrace{\int_y y_{i_1}y_{i_2}y_{i_3}\kappa(y)\mathrm{d}y}^{\mathop{=0}{ \eqref{thirdMZ}}})\\
=&\int \left (|\mathcal F_{x\rightarrow q}(\mathcal S_j u(x)( \omega(x)+\tfrac{1}{2}y^T\nabla^2\omega(x)y))|^2+|\mathcal F_{x\rightarrow q}(\mathcal S_j u(x)y^T\nabla \omega(x)) |^2 \right ) \kappa(y)\mathrm{d}y,
\end{split}
\]
 similarly, the first two terms of the fourth equality are derived by:
 \[{
 \begin{split}
 &\int |\mathcal A_j(\omega,u)+\tfrac12y^T\mathcal A_j(\nabla^2\omega,u)y|^2 \kappa(y)\mathrm{d}y\\
 &=\int |\mathcal A_j(\omega,u)|^2 \kappa(y)\mathrm{d}y+\Re \left(\mathcal A^*_j(\omega,u)\int y^T\mathcal A_j(\nabla^2\omega,u)y \kappa(y)\mathrm{d}y\right )\\
 &\qquad+\int |\tfrac12y^T\mathcal A_j(\nabla^2\omega,u)y|^2 \kappa(y)\mathrm{d}y\\
 &=\int |\mathcal A_j(\omega,u)|^2 \kappa(y)\mathrm{d}y+\Re(\mathcal A^*_j(\omega,u)(\sigma_1^2\mathcal A_j(\nabla_{11}\omega,u)\\
 &\qquad+\sigma_2^2\mathcal A_j(\nabla_{22}\omega,u) +2\sigma_{12}\mathcal A_j(\nabla_{12}\omega,u)))
 +\mathcal O\left (\int |y|^4 \kappa(y)\mathrm{d}y \right ),
 \end{split}
}
 \]

 \noindent
 and  the third term of the fourth equality is derived by:
\begin{equation}\nonumber
{
\begin{split}
\int |y^T\mathcal A_j(\nabla \omega,u)|^2 \kappa(y)\mathrm{d}y\mathop{=}^{\eqref{thirdMZ}}\sigma_1^2|\mathcal A_j(\nabla_1 \omega, u)|^2+
\sigma_2^2|\mathcal A_j(\nabla_2 \omega, u)|^2.
\end{split}
}
\end{equation}

\end{widetext}


\end{document}